\documentclass[onecolumn,showkeys,aps,prd]{revtex4}
\usepackage{amsfonts}
\usepackage{amsmath}
\usepackage{amssymb}
\usepackage{graphicx}
\usepackage{hyperref}
\usepackage{bbm}
\usepackage{floatrow}
\usepackage[utf8]{inputenc} 
\usepackage{xcolor}
\usepackage{xcoffins}
\NewCoffin\tablecoffin
\usepackage{ulem}
\usepackage{verbatim}
\usepackage{graphicx}

\usepackage[T1]{fontenc}    
\usepackage{hyperref}       
\usepackage{url}            
\usepackage{booktabs}       
\usepackage{amsfonts}       
\usepackage{nicefrac}       
\usepackage{microtype}      
\usepackage{lipsum}
\usepackage{graphicx}
\graphicspath{ {./images/} }
\usepackage{lettrine}
\usepackage{stmaryrd}
\usepackage[utf8]{inputenc}
\usepackage{fancyhdr}
\usepackage[T1]{fontenc} 
\usepackage{latexsym}
\usepackage{amsmath,amssymb}
\usepackage{array}
\usepackage{manfnt}
\usepackage{xcolor}
\usepackage{tikz}
\usepackage[makeroom]{cancel}
\usepackage{graphicx}
\usepackage{hyperref}
\usepackage{color}
\usepackage{bbm}
\usepackage{float}

\newcommand{\R}{\mathbb{R}}
\def\g{{\mathfrak{g}}}

\newcommand{\be}{\begin{equation}}
\newcommand{\ee}{\end{equation}}
\newcommand{\bea}{\begin{eqnarray}}
\newcommand{\eea}{\end{eqnarray}}

\usepackage[ps,matrix,arrow,curve]{xy}

\usepackage{amsthm}
\theoremstyle{definition}
\newtheorem{theorem}{Theorem}[section]
\newtheorem{lemma}[theorem]{Lemma}

\newtheorem{definition}[theorem]{Definition}

\def\emph#1{{\sl #1\/}}

\def\tr{\mathop{\rm tr}\nolimits}

\def\g{\mathfrak{g}}

\newcommand{\ds}{\displaystyle}
\newcommand{\del}{\partial}
\newcommand{\itGamma}{\mit{\Gamma}}
\newcommand{\D}{\mathrm{d}}

\newcommand{\rmd}{\mathrm{d}}

\newcommand{\killing}[2]{ {\langle #1 , #2 \rangle } }

\newcommand{\realni}{\ensuremath{\mathbb{R}}}

\newcommand{\cA}{{\cal A}}

\newcommand{\cF}{{\cal F}}
\newcommand{\cG}{{\cal G}}
\newcommand{\cH}{{\cal H}}

\newcommand{\cM}{{\cal M}}

\def\g{\mathfrak{g}}

\def\l{\mathfrak{l}}

\def\ie{{\sl i.e.\/}}
\def\dd{{\cal D}}

\usepackage{xcoffins}
\usepackage{ulem}

\begin{document} 
\title{Topological invariant of $4$-manifolds based on a $3$-group}

\author{Tijana Radenkovi\'c}
 \email{rtijana@ipb.ac.rs}
\affiliation{Institute of Physics, University of Belgrade, Pregrevica 118, 11080 Belgrade, Serbia}

\author{Marko Vojinovi\'c}
 \email{vmarko@ipb.ac.rs}
\affiliation{Institute of Physics, University of Belgrade, Pregrevica 118, 11080 Belgrade, Serbia}

\begin{abstract}
    {We study a generalization of $4$-dimensional $BF$-theory in
the context of higher gauge theory. We construct a triangulation independent topological state sum $Z$, based on the classical $3BF$ action for a general $3$-group and a $4$-dimensional spacetime manifold $\cM_4$. This state sum coincides with Porter’s TQFT for $d=4$ and $n=3$. In order to verify that the constructed state sum is a topological invariant of the underlying $4$-dimensional manifold, its behavior under Pachner moves is analyzed, and it is obtained that the state sum $Z$ remains the same. This paper is the generalization of the work done by Girelli, Pfeiffer, and Popescu for the case of state sum based on the classical $2BF$ action with the underlying $2$-group structure. }
\end{abstract}

\keywords{higher gauge theory, state sum model, $3$-gauge theory, $3$-groups}
\pacs{04.60.-m; 
  04.60.Nc; 
  04.60.Pp; 
  02.20.-a; 
  02.40.Sf  
  }

\maketitle

%
\section{Introduction}
%
Within the Loop Quantum Gravity framework, one studies the nonperturbative quantization of gravity, both canonically and covariantly, see \cite{RovelliBook, zakopane, RovelliVidottoBook, Thiemann2007} for an overview and a comprehensive introduction. The covariant approach focuses on defining of the path integral for the gravitational field by considering a triangulation of a spacetime manifold and specifying the path integral as a discrete state sum of the gravitational field configurations living on the simplices in the triangulation. This quantization technique is usually referred to as the \emph{spinfoam quantization method}, and it can be divided into three major steps:
\begin{enumerate}
\item first, one writes the classical action $S[g]$ as a topological $BF$-like action plus simplicity constraints,

\item then one uses the algebraic structure underlying the topological sector of the action to define a topological state sum $Z$,

\item and finally, one deforms the topological state sum by imposing simplicity constraints, thus promoting it into a path integral for a physical theory.
\end{enumerate}
Spinfoam models for gravity are usually constructed by constraining the topological gauge theory known as $BF$ theory, obtaining the Plebanski formulation of general relativity \cite{plebanski1977}. For example, in $3$ dimensions, the prototype spinfoam model is known as the Ponzano-Regge model \cite{PonzanoRegge1968}. In $4$ dimensions there are multiple models, such as the Barrett-Crane model \cite{BarrettCrane,BarrettCrane1}, the Ooguri model \cite{Ooguri}, and the most sophisticated EPRL/FK model \cite{EPRL,FK} (see also \cite{EPRLFK, MV2013, MV2015}). All these models aim to define a viable theory of a quantum gravitational field alone, without matter fields. The attempts to include matter fields have had limited success \cite{RovelliSpinfoamFermions}, mainly because the mass terms cannot be expressed in the theory due to the absence of the tetrad fields from the topological $BF$ sector of the theory.

In order to overcome this problem, a new approach has been developed within the framework of {\emph{higher gauge theory}} (for a review of higher gauge theory, see \cite{BaezHuerta2011, Zimu}, and for its applications in physics see \cite{Wolf, Nitta, Nitta2, Nitta3, Nitta4, Jurco, Jurco2, Jurco3, Song, Song2, Wolf2014, Wolf2017}). Within higher gauge theory formalism, one generalizes the $BF$ action, which is based on some Lie group, to a $2BF$ action based on the $2$-group structure. Within this approach \cite{MikovicVojinovic2012}, one rewrites the action for general relativity as a constrained $2BF$ action, such that the tetrad fields are present in the topological sector. This result opened up the possibility to couple all matter fields to gravity in a straightforward way. Nevertheless, the matter fields could not be naturally expressed using the underlying algebraic structure of a $2$-group, rendering the spinfoam quantization method only half-implementable, since the matter sector of the classical action could not be expressed as a topological term plus a simplicity constraint, which means that the steps 2 and 3 above could not be performed for the matter sector of the action.

This final issue has recently been resolved in \cite{Radenkovic2019}, where one more step in the categorical ladder is performed in order to generalize the underlying algebraic structure from a $2$-group to a $3$-group (see also \cite{MV2020} for the $4$-group formulation). This generalization then naturally gives rise to the so-called $3BF$ action, which proves to be suitable for a unified description of both gravity and matter fields. The first step of the spinfoam quantization program is carried out in \cite{Radenkovic2019} where the suitable gauge $3$-groups have been specified, and the corresponding constrained $3BF$ actions constructed so that the desired classical dynamics of the gravitational and matter fields is obtained. A reader interested in the construction of the constrained $2BF$ actions describing the Yang-Mills field and Einstein-Cartan gravity, and $3BF$ actions describing the Klein-Gordon, Dirac, Weyl, and Majorana fields, each coupled to gravity in the standard way, is referred to \cite{MikovicVojinovic2012, Radenkovic2019}. 

In this paper, we focus our attention on the ssecond step of the spinfoam quantization program: we will construct a triangulation independent topological state sum $Z$, based on the classical $3BF$ action for a general $3$-group and a $4$-dimensional spacetime manifold $\cM_4$. This state sum  coincides with the Porter’s TQFT \cite{Porter98, Porter96} for $d = 4$ and $n = 3$. In order to verify that the constructed state sum is topological, we analyze its behavior under Pachner moves \cite{Pachner}. Pachner moves are local changes of a triangulation that preserve topology, such that any two triangulations of the same manifold are connected by a finite number of Pachner moves. In $4$ dimensions, there are five different Pachner moves: the $3-3$ move, $4-2$ move, and $5-1$ move, and their inverses. After defining the state sum, we calculate its behavior under these Pachner moves. We obtain that the state sum $Z$ remains the same, proving that it is a topological invariant of the underlying 4-dimensional manifold. This construction thus completes the second step of the quantization procedure. Our result paves the way for the third step of the covariant quantization procedure and a formulation of a quantum theory of gravity and matter by imposing the simplicity constraints on the state sum. We leave the third step for future work.

The layout of the paper is as follows. In Section \ref{sec:lagrange} we review the pure and the constrained $nBF$ theories describing some of the physically relevant models -- the constrained $2BF$ actions describing the Yang-Mills field and Einstein-Cartan gravity, and constrained $3BF$ actions describing the Klein-Gordon and Dirac fields coupled to Yang-Mills fields and gravity in the standard way. In Section~\ref{sec:preliminaries}, we review the relevant algebraic tools involved in the description of higher gauge theory,  $2$-crossed modules, and $3$-gauge theory. Starting from the notion of Lie $3$-groups, we generalize the integral picture of gauge theory to a $3$-gauge theory that involves curves, surfaces, and volumes labeled with elements of non-Abelian groups. In Section~\ref{sec:discrete}, we define the discrete state sum model of topological higher gauge theory in dimension $d=4$. The proof that the state sum is invariant under the Pachner moves and thus independent of the chosen triangulation is presented in Appendix~\ref{app:pachner}. 

Notations and conventions throughout the paper are as follows. If $G$ is a finite group, $ \int_G dg=1/|G| \sum_{g \in  G}$ denotes the normalized sum over all group elements, while $\delta_G $ denotes
corresponding $ \delta$-distribution on $G$. The $\delta$-function is defined for every $g \in G$ such that
$\delta_G(g) = |G|$ if $g = e$, \ie\,, if $g$ is the unit element of the group, and $\delta_G(g) = 0$ if $g\neq e$. If $G$ is
a Lie group, $\int_G dg$ and $\delta_G$ denote the Haar
measure and the $\delta$-distribution on $G$, respectively. We define our model for any closed and oriented combinatorial manifold
$\mathcal{M}_4$ of dimension $d=4$. The set of all $k$-simplices, $0\leq k\leq d$, is denoted by $\Lambda_k$. The set of vertices
$\Lambda_0$ is finite and ordered, and every $k$-simplex is labeled by $(k+1)$-tuples of vertices $(i_0 \ldots i_k)$, $i_0, \ldots, i_k \in \Lambda_0$ such that $i_0 < \cdots < i_k$.

\section{Review of the classical theory}
%
\label{sec:lagrange}


\subsection{Topological \texorpdfstring{$nBF$}{nBF} theories}
\label{sec:3BF}

For a given Lie group $G$ whose Lie algebra $\mathfrak{g}$ is equipped with the $G$-invariant symmetric nondegenerate bilinear form $\langle \_\,,\_\rangle{}_{\mathfrak{g}}$, and for a given $4$-dimensional spacetime manifold $\cM_4$, one can introduce the $BF$ action as
\begin{equation}\label{eq:bf}
    S_{BF} =\int_{\cM_4} \langle B \wedge F \rangle_\mathfrak{g}\,,
\end{equation}
where $2$-form $F\equiv \D \alpha+ \alpha\wedge \alpha$ is the curvature for the $\mathfrak{g}$-valued connection $1$-form $\alpha \in \cA^1(\cM_4\,, \mathfrak{g})$ and $2$-form $B \in \cA^2(\cM_4\,, \mathfrak{g})$ is an $\mathfrak{g}$-valued Lagrange multiplier. Varying the action (\ref{eq:bf}) with respect to the Lagrange multiplier $B$ and the connection $\alpha$, one obtains the equations of motion of the theory,
\begin{equation}
    F=0\,,\quad \quad \nabla B \equiv \D B + \alpha \wedge B  =0\,.
\end{equation}
From the first equation of motion, one sees that $\alpha$ is a flat connection, which then, together with the second equation of motion, implies that $B$ is constant. Therefore, the theory given by the $BF$ action has no local propagating degrees of freedom, i.e., the theory is topological. For more details about the $BF$ theory see \cite{BFgravity2016, plebanski1977, baez2000}.

Within the framework of Higher Gauge Theory, by passing from the notion of a gauge group to the notion of a gauge $2$-group, one defines the categorical generalization of the $BF$ action, called the $2BF$ action. A $2$-group has a naturally associated notion of a $2$-connection $(\alpha\,,\beta)$, described by the usual $\mathfrak{g}$-valued $1$-form $\alpha \in \cA^1(\cM_4\,,\mathfrak{g})$ and an $\mathfrak{h}$-valued $2$-form $\beta \in \cA^2(\cM_4\,,\mathfrak{h})$, where $\mathfrak{h}$ is a Lie algebra of the Lie group $H$. The $2$-connection gives rise to the so-called fake $2$-curvature $(\cF,\cG)$, where $\cF$ is a $\mathfrak{g}$-valued fake curvature $2$-form $\cF\in \cA^2(\cM_4\,,\mathfrak{g})$ and $\cG$ is a $\mathfrak{h}$-valued curvature $3$-form $\cG\in \cA^3(\cM_4\,,\mathfrak{h})$, defined as
\begin{equation}\label{eq:krivine}
    \cF=\D \alpha+ \alpha \wedge \alpha - \partial\beta\,, \quad \quad \cG= \D\beta+\alpha\wedge^\rhd \beta\,.
\end{equation}
Representing the $2$-group as a crossed-module $(H \stackrel{\del}{\to}G \,, \rhd)$, see the next section for the definition and notation, one introduces a $2BF$ action using the fake $2$-curvature (\ref{eq:krivine}) as
\begin{equation}\label{eq:bfcg}
S_{2BF} =\int_{\cM_4} \langle B \wedge \cF \rangle_{\mathfrak{ g}} +  \langle C \wedge \cG \rangle_{\mathfrak{h}} \,,
\end{equation}
where the $2$-form $B \in \cA^2(\cM_4\,, \mathfrak{g})$ and the $1$-form $C\in \cA^1(\cM_4\,,\mathfrak{h})$ are Lagrange multipliers, and $\langle \_\,,\_\rangle{}_{\mathfrak{g}}$ and $\langle \_\,,\_\rangle{}_{\mathfrak{h}}$ denote the $G$-invariant symmetric nondegenerate bilinear forms for the algebras $\mathfrak{g}$ and $\mathfrak{h}$, respectively. Similarly as in the case of the $BF$ theory, varying the $2BF$ action (\ref{eq:bfcg}) with respect to the Lagrange multipliers $B$ and $C$ one obtains the equations of motion,
\begin{equation}\label{eq:jedn2bf}
    \cF=0\,, \quad \quad \cG=0\,,
\end{equation}
\ie\,, the conditions that the curvature $2$-form $\cF$ and the curvature 3-form $\cG$ vanish, while varying with respect to the connections $\alpha$ and $\beta$ one obtains
\begin{equation}
\nabla B + C\wedge^\mathcal{T} \beta =0\,, \quad \quad
\nabla C - \partial(B) = 0\,.
\end{equation}
Similar to the case of the $BF$ action, the $2BF$ action defines a topological theory, i.e., a theory with no propagating degrees of freedom, see \cite{MikovicOliveira2014,MOV2016,GirelliPfeifferPopescu2008,FariaMartinsMikovic2011} for review and references.

Continuing the categorical ladder one step further, one can generalize the $2BF$ action to the $3BF$ action, by passing from the notion of a $2$-group to the notion of a $3$-group. Representing the $3$-group with a $2$-crossed module $ (L \stackrel {\delta} { \to} H \stackrel {\partial} {\to} G \,, \rhd \,, \{\_ \,, \_ \}{}_{\mathrm{pf}})$, see next section for definition and notation, one can define a $3$-connection as an ordered triple $ (\alpha, \beta, \gamma) $, where $\alpha$, $\beta$, and $\gamma$ are appropriate algebra-valued differential forms, $ \alpha \in \cA ^ 1 (\cM_4, \mathfrak {g}) $, $ \beta \in \cA ^ 2 (\cM_4, \mathfrak {h}) $, and $ \gamma \in \cA ^ 3 (\cM_4, \mathfrak {l}) $. The corresponding fake $ 3 $-curvature $ (\cal F, G, H) $ is defined as:
\begin{equation}\label{eq:3krivine}
\begin{array}{c}
\ds  \cF = \D \alpha+\alpha \wedge \alpha - \partial \beta \,, \quad \quad \cG = \D \beta + \alpha \wedge^\rhd \beta - \delta \gamma\,, \vphantom{\ds\int} \\
\ds  \cH = \D \gamma + \alpha\wedge^\rhd \gamma + \{\beta \wedge \beta\}{}_{\mathrm{pf}} \,.
\end{array}
\end{equation}
Then, similarly to the construction of $ BF $ and $ 2BF $ actions, one defines the $3BF$ action as
\begin{equation}\label{eq:bfcgdh}
S_{3BF} =\int_{\mathcal{M}_4} \langle B \wedge  {\cal F} \rangle_{\mathfrak{ g}} +  \langle C \wedge  {\cal G} \rangle_{\mathfrak{h}} + \langle D \wedge {\cal H} \rangle_{\mathfrak{l}} \,, 
\end{equation}
where $ \mathfrak {g} $, $ \mathfrak{h}$, and $ \mathfrak{l} $ denote the Lie algebras corresponding to the Lie groups $ G $, $ H $, and $ L $ and the forms $\killing{\_}{\_}_{\mathfrak{g}}$, $\killing{\_}{\_}_{\mathfrak{h}}$, and $\killing{\_}{\_}_{\mathfrak{l}}$ are $G$-invariant symmetric nondegenerate bilinear forms on $\mathfrak{g}$,  $\mathfrak{h}$, and $\mathfrak{l}$, respectively. The variables $B \in \cA^2(\cM_4,\mathfrak{g})$, $C \in \cA^1(\cM_4,\mathfrak{h})$, and $D \in \cA^0(\cM_4,\mathfrak{l})$ are Lagrange multipliers, and
their associated equations of motion are the conditions that the $3$-curvature $(\cF, \cG, \cH)$ vanishes,
\begin{equation}\label{eq:jed3bf}
\mathcal{F} = 0 \,, \quad \quad \mathcal{G} = 0\,, \quad \quad \mathcal{H} = 0 \,.
\end{equation}
Additionally, varying with respect to the $3$-connection variables $\alpha$, $\beta$, and $\gamma$ one gets:
\begin{gather}
\nabla B + C\wedge^\mathcal{T} \beta - D\wedge^\mathcal{S}\gamma=0\,, \\
\nabla C - \partial(B) - D\wedge^{(\chi_1+\chi_2)}\beta=0\,,\\
\nabla D +\delta (C)=0\,.
\end{gather}
For further details see \cite{martins2011,Wolf2014,Wang2014} for the definition of the $3$-group, and \cite{Radenkovic2019} for the definition of the pure $3BF$ action. 

All the above actions are topological, in the sense that they do not contain any local propagating degrees of freedom \cite{Radenkovic2020, RVojinovic2021}. In this sense, they are not very interesting for the description of realistic physics, which should feature nontrivial dynamics. Nevertheless, choosing the convenient underlying $2$-crossed module structure and imposing the appropriate simplicity constraints onto the degrees of freedom present in the $3BF$ action, one can obtain the nontrivial classical dynamics of the gravitational and matter fields, as we will see in the following subsection.

\subsection{Models with relevant dynamics}
Let us review how one can employ $n$-group structure to introduce the topological $nBF$ actions corresponding to gravity and matter fields, as well as the form of the appropriate simplicity constraints to be imposed on these fields to obtain the classical dynamics.

First we review the most important constrained $2BF$ actions. We begin by rewriting general relativity as a constrained $2BF$ action based on the underlying Poincar\'e $2$-group. The Poincar\'e $2$-group is equivalent to a crossed module $(G\overset{\partial}{\rightarrow} H, \rhd)$, where the groups are choosen as $G=SO(3,1)$ and $H=\realni^4$,
and the map $\partial$ is trivial. The action $\rhd$ is a natural action of $SO(3,1)$ on $\realni^4$, defined as
\begin{equation} \label{VektorskaRjaLorentza}
M_{ab}\rhd P_c = \eta_{[bc} P_{a]}\,,
\end{equation}
where $M_{ab}$ and $P_a$ are the generators of groups $SO(3,1)$ and $\realni^4$, respectively. The action $\rhd$ of $SO(3,1)$ on itself is given via conjugation, by definition of a crossed module. Then, Poincar\'e $2$-group gives rise to the $2$-connection $(\alpha, \beta)$, given by the algebra-valued differential forms
\begin{equation}
\alpha=\omega^{ab}M_{ab}\,, \qquad \beta = \beta^a P_a\,,
\end{equation}
where we have interpreted the connection $1$-form $\alpha^{ab}$ as the ordinary spin connection $\omega^{ab}$. Also, the corresponding $2$-curvature $(\mathcal{F},\mathcal{G})$ is given as
\begin{equation}  \label{eq:krivinezapoenc}
  \begin{array}{lclcl}
 {\cal F} & = & (\mathrm{d} \omega^{ab} + {\omega^a}_c\wedge\omega^{cb} )M_{ab} & \equiv & R^{ab}M_{ab} \,, \vphantom{\ds\int} \\
 {\cal G} & = & (\mathrm{d}\beta^a + {\omega^a}_b \wedge \beta^b)P_a & \equiv & \nabla\beta^a P_a \equiv G^a P_a \vphantom{\ds\int}\,,
\end{array}
\end{equation}
where we can recognize the standard Riemann curvature $2$-form $R^{ab}$ in $\mathcal{F}$. Having these variables in hand, one defines $2BF$ action (\ref{eq:bfcg}) for the Poincar\'e $2$-group as
\begin{equation}\label{eq:GravityTopoloski}
S_{2BF} = \int_{\cM_4} B^{ab}\wedge R_{ab} + e_a \wedge \nabla\beta^a \,. 
\end{equation}
Here, the crucial insight is that the Lagrange multiplier fields $C^a$ can be identified with the tetrads \cite{MikovicVojinovic2012}, since one can show that $1$-forms $C^a$ transform in the same way as the tetrad $1$-forms $e^a$ under the Lorentz transformations and diffeomorphisms. One can now construct the action for general relativity by simply adding the additional simplicity constraint term to the action (\ref{eq:GravityTopoloski}):
\begin{equation}\label{eq:GravityVeza}
  S = \int_{\cM_4} B^{ab}\wedge R_{ab} + e_a \wedge \nabla\beta^a - \lambda_{ab} \wedge \Big( B^{ab}-\frac{1}{16\pi l_p^2}\varepsilon^{abcd} e_c \wedge e_d \Big) \,.
\end{equation}
Here $\lambda_{ab}$ is a Lagrange multiplier $2$-form associated to the simplicity constraint term, and $l_p$ is the Planck length. It is straightforward to show that the corresponding equations of motion reduce to vacuum Einstein field equations. Thus the action \eqref{eq:GravityVeza} is classically equivalent to general relativity. The construction of the action \eqref{eq:GravityVeza} is analogous to the Plebanski model, where general relativity is constructed by adding a simplicity constraint to the $BF$ theory based on the Lorentz group. However, one clear advantage of this model over the Plebanski model is that the tetrads are explicitly present in the topological sector of the action. Upon the covariant quantization, tetrads are therefore fundamental, off-shell quantities, in contrast to the Plebanski model where they appear only on-shell, as solutions of the classical equations of motion. The off-shell presence of the tetrads facilitates the straightforward coupling of the matter fields to gravity, and thus overcomes the problems present in the spinfoam models \cite{RovelliSpinfoamFermions}.

The Poincar\' e $2$-group can be easily extended to include the coupling of the $SU(N)$ Yang-Mills fields to gravity \cite{Radenkovic2019}. To achieve this, one constructs the crossed module $(G\overset{\partial}{\rightarrow} H, \rhd)$, where the groups are chosen as $G=SO(3,1)\times SU(N)$ and $H=\mathbb{R}^4$, while the map $\partial$ remains trivial, as before. The action $\rhd$ of the group $G$ on $H$ is such that the $SO(3,1)$ subgroup acts on $\realni^4$ via the vector representation (\ref{VektorskaRjaLorentza}), while the action of the $SU(N)$ subgroup is trivial,
\begin{equation} \label{DelovanjeTauNaPeove}
\tau_I \rhd P_a = 0\,,
\end{equation}
where $\tau_I$ are the $SU(N)$ generators. This crossed module yields the $2$-connection $(\alpha,\beta)$, where algebra-valued $1$-form $\alpha$ and algebra valued $2$-form $\beta$ are defined as follows,
\begin{equation}
\alpha = \omega^{ab}M_{ab} + A^I \tau_I\,, \qquad \beta = \beta^a P_a\,,
\end{equation}
where we can identify the gauge connection $1$-form $A^I$. This connection gives rise to the $2$-curvature $(\cF,\cG)$, where $\cF$ as defined as
\begin{equation}
{\cal F} = R^{ab}M_{ab} + F^I \tau_I\,, \qquad F^I \equiv \rmd A^I + f_{JK}{}^I A^J \wedge A^K\,,
\end{equation}
while the curvature $\cG$ for $\beta$ remains the same as before. Given these variables, the Lagrange multiplier $B$ in the first term of the topological action (\ref{eq:bfcg}) also splits into two pieces corresponding to the direct product of the group $G$, giving
\begin{equation} \label{eq:bfcggauge}
    S_{2BF}=\int_{\cM_4} B^{ab}\wedge R_{ab} + B^I \wedge F_I + e_a \wedge \nabla \beta^a\,,
\end{equation}
where $2$-form $B^I \in \cA^2(\cM_4\,,\mathfrak{su}(N))$ is the second piece of the Lagrange multiplier. To obtain the non-trivial classical dynamics for gravity and the Yang-Mills field, we add the appropriate simplicity constraint terms to the action (\ref{eq:bfcggauge}), and construct the constrained $2BF$ action:
\begin{equation} \label{eq:YMplusGravity}
\begin{aligned}
S=&\int_{\cM_4} B^{ab}\wedge R_{ab} + B^I \wedge F_I + e_a \wedge \nabla \beta^a- \lambda_{ab} \wedge \Big(B^{ab}-\frac{1}{16\pi l_p^2}\varepsilon^{abcd} e_c \wedge e_d\Big) \\
  & + \lambda^I\wedge \Big(B_I-\frac{12}{g}{M_{ab}}_Ie^a\wedge e^b\Big) + {\zeta^{ab}}{}^I \Big( {M_{ab}}{}_I\varepsilon_{cdef}e^c\wedge e^d \wedge e^e \wedge e^f- g_{IJ}F^J \wedge e_a \wedge e_b \Big) \,. 
\end{aligned}
\end{equation}
Here, the first row is the topological sector and the familiar simplicity constraint for gravity from (\ref{eq:GravityVeza}), while the second row contains the appropriate simplicity constraints for Yang Mills field, featuring the Lagrange multipliers $\lambda^I$ and ${\zeta^{ab}}{}^I$. The action \eqref{eq:YMplusGravity} provides two dynamical equations -- the equation for $A^I$,
\begin{equation}
   \nabla_\rho F^{I \rho \mu}\equiv  \partial_\rho F^{I\rho\mu} + \itGamma^{\rho}{}_{\lambda\rho} F^{I\lambda\mu} + f_{JK}{}^I A^J{}_\rho F^{K\rho\mu} = 0\,,\label{eq:020}
\end{equation}
where $\itGamma^{\lambda}{}_{\mu\nu}$ is the standard Levi-Civita connection, and an equation for $e^a$ which is the Einstein field equation with the $SU(N)$ gauge field source term,
\begin{equation}
  R^{\mu \nu}-\frac{1}{2}g^{\mu \nu}R=8\pi l_p^2 \; T^{\mu \nu}\,, \quad 
  T^{\mu \nu} \equiv -\frac{1}{4g}\left(F_{\rho \sigma}{}^IF^{\rho \sigma}{}_Ig^{\mu \nu}+4F^{\mu \rho}{}_I{F_\rho}^{\nu}{}^I \right)\,.\label{eq:019}
\end{equation}
In this way, we see that both gravity and gauge fields can be successfully represented within a unified framework of higher gauge theory, based on a $2$-group structure. A generalization from $SU(N)$ Yang-Mills case to the more complicated cases, such as $SU(3)\times SU(2) \times U(1)$, is straightforward.

Let us now review how one can use the $3$-group structure and the corresponding constrained $3BF$ theory to describe general relativity coupled to Klein-Gordon and Dirac fields. To describe a single real Klein-Gordon field coupled to gravity, one begins by specifying a $2$-crossed module $(L\stackrel{\delta}{\to} H \stackrel{\partial}{\to}G\,, \rhd\,, \{\_\,,\_\})$, as follows. The groups are given as $G=SO(3,1)$, $H=\mathbb{R}^4$, and $L=\mathbb{R}$. The group $G$ acts on $H$ via the vector representation, and on $L$ via the trivial representation. The maps $\partial$ and  $\delta$ are chosen to be trivial, as well as the Peiffer lifting. Given this choice of a $2$-crossed module, the $3$-connection $(\alpha\,,\beta\,,\gamma)$ takes the form
\begin{equation}
 \alpha = \omega^{ab}M_{ab}\,, \quad \quad \beta=\beta^a P_a\,, \quad \quad \gamma = \gamma \mathbb{I}\,,
\end{equation}
where $\mathbb{I}$ is the sole generator of the Lie group $L$. This $2$-connection gives rise to the fake $3$-curvature $(\cF, \cG, \cH)$,
\begin{equation}
\cF = R^{ab} M_{ab}\,, \quad \quad \cG= \nabla \beta^a P_a\,, \quad \quad \cH= \D \gamma\,.
\end{equation}
The importance of the $3BF$ theory for this choice of the $2$-crossed module lies in the fact that the Lagrange multiplier $D$ can transform as a scalar with respect to Lorentz symmetry, $M_{ab} \rhd \mathbb{I}=0$, and it transforms as a scalar with respect to diffeomorphisms since $D$ is also a $0$-form. In other words, one can interpret the Lagrange multiplier $D$ to be a real scalar field, $D \equiv \phi$, and write the topological $3BF$ action (\ref{eq:bfcgdh}) as:
\begin{equation}\label{eq:Scalartopoloski}
    S_{3BF}=\int_{\cM_4} B^{ab}\wedge R_{ab} + e_a\wedge \nabla \beta^a + \phi \, \D \gamma\,.
\end{equation}
In order to obtain the Klein-Gordon field $\phi$ of mass $m$ coupled to gravity in the standard way, the appropriate simplicity constraints are imposed, and the constrained $3BF$ action takes the form:
\begin{equation}\label{eq:scalar}
\begin{aligned}
 S =\int_{\cM_4} & B^{ab}\wedge R_{ab} + e_a\wedge \nabla \beta^a + \phi \, \D \gamma \vphantom{\ds\int} 
 - \lambda_{ab} \wedge \Big(B^{ab}-\frac{1}{16\pi l_p^2}\varepsilon^{abcd} e_c \wedge e_d\Big)\vphantom{\ds\int}\\& + {\lambda}\wedge \Big(\gamma - \frac{1}{2} H_{abc} e^a \wedge e^b \wedge e^c\Big) 
 +\Lambda^{ab}\wedge \Big( H_{abc}\varepsilon^{cdef}e_d\wedge e_e \wedge e_f- \D \phi \wedge e_a \wedge e_b\Big) \vphantom{\ds\int} \\&-\frac{1}{2\cdot 4!} m^2\phi^2 \varepsilon_{abcd}e^a\wedge e^b \wedge e^c \wedge e^d\,.\vphantom{\ds\int}
\end{aligned}
\end{equation}
The first row is the topological sector (\ref{eq:Scalartopoloski}) and the simplicity constraint for gravity from the action (\ref{eq:GravityVeza}), the second row contains two new simplicity constraints featuring the Lagrange multiplier $1$-forms $\lambda$ and $\Lambda^{ab}$ and the $0$-form $H_{abc}$, and the third row features the mass term for the scalar field. The action \eqref{eq:scalar} has two dynamical equations of motion --- the equation for the scalar field $\phi$ is the covariant Klein-Gordon equation,
\begin{equation}
\left(\nabla_\mu\nabla^\mu -m^2\right)\phi=0\,,
\end{equation}
while the equation for the tetrads $e^a$ is the Einstein field equation with the scalar field source term,
\begin{equation}\label{eq:scalareomfore}
  {R}^{\mu\nu}-\frac{1}{2}g^{\mu\nu} R=8\pi l_p^2\; T^{\mu\nu}\,,\quad 
T^{\mu\nu}\equiv\partial^\mu \phi \partial^\nu \phi -\frac{1}{2}g^{\mu\nu} \left(\partial_\rho \phi \partial^\rho \phi+m^2\phi^2 \right)\,.
\end{equation}
We see that the obtained theory is classically equivalent to general relativity coupled to a scalar field. Most importantly, one sees that the choice of the group $L$ dictates the matter content of the theory, while the action $\rhd$ of $G$ on $L$ specifies the transformation properties of the matter fields.

Finally, in order to describe the Dirac field coupled to Einstein-Cartan gravity, the $2$-crossed module $(L\stackrel{\delta}{\to} H \stackrel{\partial}{\to}G\,, \rhd\,, \{\_\,,\_\})$ has to be chosen as follows. The groups are $G=SO(3,1)$. $H=\mathbb{R}^4$, and $L=\mathbb{R}^8(\mathbb{G})$, where $\mathbb{G}$ is the algebra of complex Grassmann numbers. The maps $\partial$,  $\delta$ and the Peiffer lifting are trivial, as before. The action of the group $G$ on $H$ is via vector representation, and on $L$ via spinor representation, in the following way. Denoting the eight generators of the Lie group $\mathbb{R}^8(\mathbb{G})$ as $P_{\alpha}$ and $P^{\alpha}$, where the bispinor index $\alpha$ takes the values $1,\dots,4$, the action $\rhd$ of $G$ on $L$ is given explicitly as
\begin{equation} \label{eq:actionOfGonLdirac}
M_{ab}\rhd P_{\alpha}=\frac{1}{2}(\sigma_{ab}){}^{\beta}{}_{\alpha} P_{\beta}\,, \qquad M_{ab} \rhd P^{\alpha}=-\frac{1}{2}(\sigma_{ab}){}^{\alpha}{}_{\beta} P^{\beta}\,,
\end{equation}
where $\sigma_{ab}=\frac{1}{4}[\gamma_a,\gamma_b]$, and $\gamma_a$ are the usual Dirac matrices. This choice of the $2$-crossed module gives rise to the $3$-connection $(\alpha\,,\beta\,,\gamma)$, defined as
\begin{equation}
\alpha = \omega^{ab}M_{ab}\,, \quad \quad \beta=\beta^a P_a\,, \quad \quad \gamma = \gamma^{\alpha} P_{\alpha}+\bar{\gamma}{}_{\alpha} P^{\alpha}\,,
\end{equation}
where the $3$-connection $3$-forms $\gamma^\alpha$ and $\bar{\gamma}_\alpha$ should not be confused with the Dirac matrices $\gamma_a$ due to different types of indices. The $3$-curvature $(\cF\,, \cG\,, \cH)$ is given as:
\begin{equation}
\begin{aligned}
    \cF = R^{ab} M_{ab}\,, \quad & \quad \cG= \nabla \beta^a P_a\,,\\
    \cH= \Big(\D \gamma^{\alpha}+\frac{1}{2}\omega^{ab}(\sigma_{ab}){}^{\alpha}{}_{\beta}\gamma^{\beta}\Big) P_{\alpha} + & \Big(\D \bar{\gamma}{}_{\alpha}-\frac{1}{2}\omega^{ab}\bar{\gamma}_{\beta}(\sigma_{ab}){}^{\beta}{}_{\alpha}\Big)P^{\alpha} 
    \equiv (\overset{\rightarrow}{\nabla} \gamma){}^{\alpha} P_{\alpha}+ (\bar{\gamma}\overset{\leftarrow}{\nabla}){}_{\alpha}P^{\alpha}\,. \hphantom{mmm} &
\end{aligned} 
\end{equation}
As in the case of the scalar field, the choice of the group $L$ and action $\rhd$ of $G$ on $L$ dictate the matter content of the theory and its transformation properties. The group $L$ prescribes that $D$ contains eight independent real anticommuting matter fields as its components. Then, since $D$ is a $0$-form and it transforms according to the spinorial representation of $SO(3,1)$,  these eight real Grassmann-valued fields can be identified with the four complex Dirac bispinor fields, and one can write the corresponding topological $3BF$ action as:
\begin{equation}\label{eq:DiracTopoloski}
    S_{3BF}=\int_{\cM_4} B^{ab}\wedge R_{ab} + e_a\wedge \nabla \beta^a + (\bar{\gamma}{\overset{\leftarrow}{\nabla}) {}_{\alpha} \psi^{\alpha} +\bar{\psi}_{\alpha}{({\overset{\rightarrow}{\nabla}}\gamma)}{}^{\alpha}}\,.
\end{equation}
In order to obtain the action that gives us the dynamics of Einstein-Cartan theory of gravity coupled to a Dirac field, we add the following simplicity constraints:
\begin{equation}\label{eq:Dirac}
\begin{aligned}
S= & \int_{\cM_4}  B^{ab}\wedge R_{ab} + e_a\wedge \nabla \beta^a + ( \bar{\gamma} {\overset{\leftarrow}{\nabla}} ) {}_{\alpha} \psi^{\alpha} +\bar{\psi}_{\alpha}{({\overset{\rightarrow}{\nabla}}\gamma)}{}^{\alpha} - \lambda_{ab} \wedge \Big(B^{ab}\vphantom{\ds\int}-\frac{1}{16\pi l_p^2}\varepsilon^{abcd} e_c \wedge e_d\Big)\vphantom{\ds\int}\\
&-\lambda^{\alpha}\wedge \Big({\bar{\gamma}}_{\alpha}-\frac{i}{6}\varepsilon_{abcd}e^a\wedge e^b \wedge e^c (\bar{\psi}\gamma^d)_{\alpha}\Big)\vphantom{\ds\int}+\bar{\lambda}_{\alpha}\wedge \Big({\gamma}^{\alpha}+\frac{i}{6}\varepsilon_{abcd}e^a\wedge e^b \wedge e^c (\gamma^d\psi){}^{\alpha}\Big)\vphantom{\ds\int}\\
 & -\frac{1}{12} m \, \bar{\psi}\psi\, \varepsilon_{abcd}e^a\wedge e^b \wedge e^c \wedge e^d +2 \pi i l_p^2 \, \bar{\psi}\gamma_5\gamma^a \psi \, \varepsilon_{abcd}e^b\wedge e^c \wedge \beta^d .\vphantom{\ds\int}
\end{aligned}
\end{equation}
The topological sector is in the first row, as well as the gravitational simplicity constraint, the second row contains the new simplicity constraints for the Dirac field, while the third row contains the mass term for the Dirac field and a term that ensures the correct coupling between the torsion and the spin of the Dirac field. Varying the action \eqref{eq:Dirac}, one obtains the following dynamical equations of motion --- the equations for $\psi$ and $\bar{\psi}$ which are the standard covariant Dirac equation and its conjugate,
\begin{equation} \label{eq:DiracEquation}
    (i\gamma^a e^{\mu}{}_a \overset{\rightarrow}{\nabla}_\mu -m)\psi=0\,, \quad
    \bar{\psi}(i\overset{\leftarrow}{\nabla}_\mu e^{\mu}{}_a \gamma^a+m)=0\,,
\end{equation}
and the differential equation of motion for $e^a$ which is the Einstein field equation with a Dirac field source term,
\begin{equation}\label{eq:diraceomfore}
R^{\mu\nu}-\frac{1}{2}g^{\mu\nu}R=8\pi l_p^2\; T^{\mu\nu}\,,\quad T^{\mu\nu} \equiv \frac{i}{2}\bar{\psi} \gamma^a{\overset{\leftrightarrow}{\nabla}}{}^{\nu} e^\mu{}_a \psi-\frac{1}{2}g^{\mu\nu}\bar{\psi} \Big(i\gamma^a\overset{\leftrightarrow}{\nabla}_\rho e^\rho{}_a-2m \Big)\psi\,,
\end{equation}
where ${\overset{\leftrightarrow}{\nabla}}={\overset{\rightarrow}{\nabla}}-{\overset{\leftarrow}{\nabla}}$. Moreover, one obtains the desired equation of motion for the torsion,
\begin{equation} \label{eq:DiracTorzijaJosJednom}
T_a \equiv \nabla e_a = 2\pi l_p^2 s_a\,,\quad s_a = i\varepsilon_{abcd} e^b \wedge e^c \bar\psi \gamma_5 \gamma^d \psi\,,
\end{equation}
where $s_a$ is the Dirac spin $2$-form. The equations of motion (\ref{eq:DiracEquation}), (\ref{eq:diraceomfore}), and (\ref{eq:DiracTorzijaJosJednom}) are precisely the equations of motion of the Einstein-Cartan-Dirac theory. 

The natural presence of a scalar and Dirac field in the $3BF$ action is an essential property of the specific choices of the $3$-group structures in a $4$-dimensional spacetime, just like the existence of the tetrad field $e^a$ in the topological $2BF$ action is an essential property of the $2BF$ action and the Poincar\'e $2$-group. In this way, both the scalar field and the Dirac field appear in the topological sector of the action, making the quantization procedure feasible. Similarly, one can introduce Weyl and Majorana fields as well, see \cite{Radenkovic2019}.

\section{A review of \texorpdfstring{$2$}{2}-groups and \texorpdfstring{$3$}{3}-groups}
\label{sec:preliminaries}

As we have seen in the previous section, the gauge symmetry of $3$-gauge theory is described by an algebraic structure known as a \emph{$3$-group}. In this section, we present the relevant definition of the $3$-group, and we briefly explain how this structure is used to equip curves, surfaces, and volumes with holonomies. The results obtained in this section are necessary for the construction of the topological invariant, which will be studied in section IV.

\subsection{3-Groups}
\label{sec:3groups}

In the category theory, a $2$-group is defined as a $2$-category consisting of only one object, where all the morphisms and $2$-morphisms are invertible. It has been shown that every strict $2$-group is equivalent to a crossed module $(H \stackrel{\del}{\to}G \,, \rhd)$.

A \emph{pre-crossed module} $(H \stackrel{\del}{\to}G \,, \rhd)$ of groups $G$ and $H$, is given by a group map $\partial : H \to G$,
together with a left action $\rhd$ of $G$ on both groups, by automorphisms, such that the group $G$ acts on itself via conjugation, \ie\,, for each $g_1, g_2 \in G$,
$$g_1\rhd g_2 = g_1 g_2 g_1^{-1}\,, $$
and for each $h_1\,,h_2 \in H$ and $g \in G$ the following identity holds:
$$g \partial h g^{-1} = \partial (g \rhd h)\,.$$
In a pre-crossed module the \emph{Peiffer commutator} is defined as:
\begin{equation}\label{Peiffer_comm}
    \langle h_1\,,h_2 \rangle{}_{\mathrm{p}}=h_1h_2h_1^{-1} \partial(h_1) \rhd h_2^{-1}\,.
\end{equation}
A pre-crossed module is said to be a \emph{crossed module} if all of its Peiffer commutators are trivial, which is to say that the \emph{Peiffer identity} is satisfied:
\begin{equation}
  (\partial h_1) \rhd h_2 = h_1 h_2 h_1^{-1}\,.  
\end{equation}

Continuing the categorical generalization one step further, one can generalize the notion of a $2$-group to the notion of a $3$-group. Similar to the definition of a group and a $2$-group within the category theory formalism, a $3$-group is defined as a $3$-category with only one object, where all morphisms, $2$-morphisms, and $3$-morphisms are invertible. Moreover, in analogy with how a crossed module encodes a strict 2-group, it has been proved that a semistrict $3$-group --- Gray group is equivalent to a $2$-crossed module \cite{Con, martins2011}. 

A \emph{$2$-crossed module} $(L\stackrel{\delta}{\to} H \stackrel{\partial}{\to}G,\,\rhd,\,\{\_,\,\_\}_\mathrm{p})$ is a chain complex of groups, given by three groups $G$, $H$, and $L$, together with maps $\partial$ and $\delta$,
\begin{equation*}
    L\stackrel{\delta}{\to} H \stackrel{\partial}{\to}G\,,
\end{equation*}
such that $\partial\delta=1_G$, an action $\rhd$ of the group $G$ on all three groups, and a map $\{ \_\,,\_ \}_\mathrm{p}$ called the \emph{Peiffer lifting}:
\begin{equation*}
\{ \_\,,\_ \}_\mathrm{p} : H\times H \to L\,.
\end{equation*}
The maps $\partial$ and $\delta$, and the Peiffer lifting are $G$-equivariant, \ie\,, for each $ g \in G$ and $h \in H $
\begin{equation*}\label{eq:ekvivarijantnost_partial}
    g\rhd \partial (h)=\partial(g\rhd h)\,, \quad \quad \quad g\rhd \delta (l) = \delta(g \rhd l)\,,
\end{equation*}
and for each $h_1,\, h_2 \in H$ and $g \in G$:
$$ g\rhd\{h_1\,,h_2\}_\mathrm{p}=\{ g\rhd h_1, \,g \rhd h_2\}_\mathrm{p}\,.$$
The action of the group $G$ on the groups $H$ and $L$ is a smooth left action by automorphisms, \ie\,, for each $g,g_1,g_2 \in G$, $\;h_1,h_2 \in H$, $\;l_1,l_2 \in L$ and $k \in H, L$,
\begin{equation*}\label{eq:ekvivarijantnost_delta}
    g_1 \rhd (g_2 \rhd k) =(g_1 g_2) \rhd k\,, \quad \quad g \rhd (h_1 h_2) =(g \rhd h_1)(g \rhd h_2)\,, \quad \quad g \rhd (l_1 l_2) =(g \rhd l_1)(g \rhd l_2)\,. 
\end{equation*}
The action of the group $G$ on itself is again via conjugation. Further, the following identities are satisfied:
\begin{subequations}
\begin{align}
 \;\; & \delta(\{h_1,h_2 \}_\mathrm{p})= \langle h_1\,,h_2\rangle{}_{\mathrm{p}}\,, \qquad\qquad \forall h_1,h_2 \in H\,;\;\label{prop1}\\
 \;\; & [l_1,l_2]=\{\delta(l_1)\,,\delta(l_2) \}_\mathrm{p}\,, \qquad \qquad\forall l_1\,, l_2 \in L\,. \quad\quad \quad\text{Here, the notation } [l,k]=lkl^{-1}k^{-1} \text{is used;}\label{prop2}\\
  \;\; & \{h_1h_2,h_3 \}_\mathrm{p}=\{h_1,h_2h_3h_2^{-1} \}_\mathrm{p}\partial(h_1)\rhd \{h_2,h_3 \}_\mathrm{p}\,, \qquad \qquad \forall h_1,h_2,h_3 \in H\,;\label{prop3}\\
 \;\;& \{h_1,h_2h_3 \}_\mathrm{p}= \{h_1,h_2\}_\mathrm{p} \{h_1,h_3\}_\mathrm{p}\{ \langle h_1,h_3 \rangle{}_{\mathrm{p}}^{-1}\,, \partial(h_1) \rhd h_2\}_\mathrm{p}\,, \quad\quad {\qquad\forall h_1,h_2,h_3 \in H}\,;\label{prop4}\\
 \;\;&\{\delta(l),h\}_\mathrm{p}\{h, \delta(l) \}_\mathrm{p}=l(\partial(h) \rhd l^{-1})\,,\qquad\qquad \forall h\in H\,, \quad \forall l \in L\,.\label{prop5}
 \end{align}
\end{subequations}
In a $2$-crossed module the structure $(L\stackrel{\delta}{\to}H,\,\rhd')$ is a crossed module, with action of the group $H$ on the group $L$ defined for each $h \in H$ and $l \in L$ as:
$$h \rhd' l = l \, \{\delta(l){}^{-1},\,h\}_\mathrm{p}\,,$$
and it follows that the Peiffer identity is satisfied for each $l_1,l_2 \in L$:
$$\delta(l_1)\rhd' \,l_2= l_1 \,l_2\, l_1^{-1}\,. $$
However, the structure $(H \stackrel{\del}{\to}G \,, \rhd)$ in the general case does not form a crossed module, but a pre-crossed module, and for each $h,h'\in H$ the Peiffer commutator does not necessarily vanish.

The following identities hold, for each $h_1,h_2,h_3 \in H$ \cite{martins2011}:
\begin{equation}\label{eq:id01}
    \{h_1h_2,h_3\}_{\mathrm{p}}= (h_1 \rhd' \{h_2,h_3\}_{\mathrm{p}})\{h_1,\partial(h_2)\rhd h_3\}_{\mathrm{p}}\,,
\end{equation}
\begin{equation}\label{eq:id02}
    \begin{aligned}
    \{h_1,h_2h_3\}_{\mathrm{p}}=\{h_1,h_2\}_{\mathrm{p}}(\partial(h_1)\rhd h_2)\rhd'\{h_1,h_3\}_{\mathrm{p}}\,,
    \end{aligned}
\end{equation}
and are of prime importance for the proof of the Pachner moves invariance. By using the condition \eqref{prop5} of the definition of a 2-crossed module, it follows that for each $h\in H$ and $l \in L$ the following identity holds:
\begin{equation}\label{identitet}
    \{h, \delta(l)^{-1}\}_{\mathrm{p}}=(h \rhd' l^{-1}) (\partial(h)\rhd l)\,.
\end{equation}
Moreover, for each $h_1,h_2 \in H$,
\begin{equation}\label{eq:id03}
    \{h_1,h_2\}_{\mathrm{p}}^{-1}=h_1 \rhd'\{h_1^{-1},\partial(h_1)\rhd h_2\}_{\mathrm{p}}\,,
\end{equation}
\begin{equation}\label{eq:id04}
  \{h_1,h_2\}_{\mathrm{p}}^{-1}=\partial(h_1)\rhd\{h_1^{-1},h_1h_2h_1^{-1}\}_{\mathrm{p}}\,,
\end{equation}
\begin{equation}\label{eq:id05}
    \{h_1,h_2\}^{-1}_{\mathrm{p}}= (h_1h_2h_1^{-1})\rhd'\{h_1,h_2^{-1}\}_{\mathrm{p}}\,,
\end{equation}
\begin{equation}\label{eq:id06}
   \{h_1,h_2\}_{\mathrm{p}}^{-1}= (\partial(h_1)\rhd h_2)\rhd'\{h_1,h_2^{-1}\}_{\mathrm{p}}\,. 
\end{equation}
A reader interested in more details about $3$-groups is referred to \cite{Wang2014}.
%

\subsection{3-gauge theory}\label{sec:3gaugetheory}

In this subsection, we will describe how the language of $3$-gauge theory can be used in order to define compositions of labeled paths, surfaces, and volumes. In a $3$-gauge theory, one labels geometric objects at three levels. Curves are labeled by elements of $G$. Their composition and orientation reversal is defined as in conventional gauge theory. In addition, surfaces are labeled with elements of $H$, and volumes are labeled with the elements of $L$. The reader interested in the formulation of a $2$-gauge theory is referred to \cite{GirelliPfeiffer}.

Curves are labeled with the elements of $G$, and the elements are composed as in the ordinary gauge theory, \ie\,, for each $g_1, g_2 \in G$,
\[ \xymatrix{ \bullet &
\ar@/_/[l]_{g_1}  
\bullet 
& 
\bullet 
\ar@/_/[l]_{g_2} 
} = 
\xymatrix{ \bullet & \bullet \ar@/_/[l]_{g_1g_2} }\,,
\]
the composition of the elements results in the element $g_1g_2 \in G$. The orientation of a curve can be reversed if it is labeled by the inverse element $g^{-1}$ instead.

Surfaces are labeled with the elements $h \in H$. For each surface, we choose two reference points on the boundary, and split the boundary into two curves, the source curve labeled with $g_1\in G$ and the target curve labeled with $g_2\in G$, as demonstrated in the diagram
\[          
  \xymatrix{
   \bullet
&& \bullet
  \ar@/_2.5ex/[ll]_{g_1}="g1"\ar@/^2.5ex/[ll]^{g_2}="g2"
  \ar@{=>}^{h} "g1"+<0ex,-2.5ex>;"g2"+<0ex,2.5ex>
}\,.
\]
The $2$-arrow $h \in H$ maps the curve $g_1 \in G$ to the curve $\partial(h) g_1 \in G$,
\[          
  \xymatrix{
   \bullet
&& \bullet
  \ar@/_2.5ex/[ll]_{1_\bullet}="g1"\ar@/^2.5ex/[ll]^{\partial h}="g2"
  \ar@{=>}^{h} "g1"+<0ex,-2.5ex>;"g2"+<0ex,2.5ex>
&& \bullet
  \ar@/_2.5ex/[ll]_{g_1}="g1"\ar@/^2.5ex/[ll]^{g_1}="g2"
  \ar@{=>}^{1_g} "g1"+<0ex,-2.5ex>;"g2"+<0ex,2.5ex>
} \; = \; \xymatrix{
   \bullet
&& \bullet
  \ar@/_2.5ex/[ll]_{g_1}="g1"\ar@/^2.5ex/[ll]^{\partial(h) g_1}="g2"
  \ar@{=>}^{h} "g1"+<0ex,-2.5ex>;"g2"+<0ex,2.5ex>
}\,,
\]
so that the label $h\in H$ of the surface is required to satisfy the following condition:
\begin{equation}
    \partial(h) =g_2g_1^{-1}\,.
\end{equation}
The orientation of the surface can be reversed and labeled with the inverse element instead,
\[          
  \xymatrix{
   \bullet
&& \bullet
  \ar@/_2.5ex/[ll]_{g_1}="g1"\ar@/^2.5ex/[ll]^{g_2}="g2"
  \ar@{<=}^{h^{-1}} "g1"+<0ex,-2.5ex>;"g2"+<0ex,2.5ex>
}\,,
\]
while the orientation reversal of the curves leads to the surface element labeled with $\tilde{h}=g_1^{-1}\rhd h^{-1}$: 
\[
\xymatrix{
\ar@/^2ex/[rr]^{g_1^{-1}}="g1"\ar@/_2ex/[rr]_{g_2^{-1}}="g2"
 \bullet && \bullet 
  \ar@{=>}^{\tilde{h}} "g1"+<0ex,-2.5ex>;"g2"+<0ex,2.5ex>
}\,.
\]
 One can now compose $2$-morphisms vertically. The vertical composition of $2$-morphisms $(g_1,h_1)$ and $(g_2,h_2)$, when they are compatible, \ie\,, when $\partial_2^+(h_1)=\partial_2^-(h_2)$,
\[
\xymatrix{
    \bullet && \bullet 
  \ar@/_5ex/[ll]_{g}="g1"
  \ar[ll]_(0.65){g_2}
  \ar@{}[ll]|{}="g2"
  \ar@/^5ex/[ll]^{g_3}="g3"
  \ar@{=>}^{h_1} "g1"+<0ex,-2ex>;"g2"+<0ex,0.5ex>
  \ar@{=>}^{h_2} "g2"+<0ex,-0.5ex>;"g3"+<0ex,2ex>
}
\quad =\quad
\xymatrix{
\bullet
&&\bullet 
\ar@/_3ex/[ll]_{g_1}="g1"
\ar@/^3ex/[ll]^{g_3}="g3"
\ar@{=>}^{h_2 h_1} "g1"+<-1ex,-2ex>;"g3"+<-1ex,2ex>
}\,,
\]
results in a $2$-morphism $(g_1,h_2h_1)$, 
\begin{equation}
    (g_2,h_2)\#_2(g_1,h_1) = (g_1,h_2h_1) \,.
\end{equation}

An important operation is known as whiskering. One can whisker a $2$-morphism $h$ with a morphism $g_1$ by attaching the whisker $g_1$ to the surface $h$ from the left, \ie\,, such that $\partial_1^-(g_1)=\partial_1^+(h)$,
\[
\xymatrix{
   \bullet 
&  \bullet
  \ar[l]_{g_1}
&&  \bullet 
  \ar@/_2.5ex/[ll]_{g_2}="h1"
  \ar@/^2.5ex/[ll]^{g_2'}="h3"
  \ar@{=>}^{h} "h1"+<0ex,-2ex>;"h3"+<0ex,2ex>
} = \xymatrix{
  \bullet &&&
 \bullet \ar@/_2.5ex/[lll]_{g_1g_2}="h1"
  \ar@/^2.5ex/[lll]^{g_1g_2'}="h3"
  \ar@{=>}^{g_1\rhd h} "h1"+<0ex,-2ex>;"h3"+<0ex,2ex>} \,,
\]
which results in the $2$-morphism with the source curve $g_1 g_2$ and target curve $g_1 g_2'$, carrying the label $g_1\rhd h$. 
Similarly, by attaching whisker $g_2$ to a surface $h$ from the right, \ie\,, such that $\partial_1^-(h)=\partial_1^+(g_2)$, 
\[
\xymatrix{
   \bullet
&&  \bullet 
  \ar@/_2.5ex/[ll]_{g_1}="h1"
  \ar@/^2.5ex/[ll]^{g_1'}="h3"
  \ar@{=>}^{h} "h1"+<0ex,-2ex>;"h3"+<0ex,2ex>
  &  \bullet 
   \ar[l]_{g_2} }
= \xymatrix{
  \bullet &&&
 \bullet \ar@/_2.5ex/[lll]_{g_1g_2}="h1"
  \ar@/^2.5ex/[lll]^{g_1'g_2}="h3"
  \ar@{=>}^{h} "h1"+<0ex,-2ex>;"h3"+<0ex,2ex>
}\,,
\]
one obtains the $2$-morphism with the source curve $g_1 g_2$ and target curve $g_1' g_2$, carrying the label $h$.

The volumes are labeled with the elements $l \in L$. For each volume, we split the boundary into two surfaces, the source surface labeled with $\partial_3^{-}(l)=h_1$ and the target surface labeled with $\partial_3^{+}(l)=h_2$. On the common boundary of the source and target surface, we choose two reference points, and split the boundary into two curves, the source curve labeled with $\partial_2^{-}(l)=g_1$ and the target curve labeled with $\partial_2^{+}(l)=g_2$, as demonstrated in the diagram below
\[
\xymatrix{  \bullet
&  \bullet 
\ar@/_4ex/[l]_{g_1}="g1"
\ar@/^4ex/[l]^{g_2}="g3"
\ar@{=>}^{h_1} "g1"+<0ex,-2.5ex>;"g3"+<0ex,2.5ex> } \; \stackrel{l}{\Rrightarrow} \; \xymatrix{\bullet
&  \bullet 
\ar@/_4ex/[l]_{g_1}="g2"
\ar@/^4ex/[l]^{g_2}="g4"
\ar@{=>}^{h_2} "g2"+<0ex,-2.5ex>;"g4"+<0ex,2.5ex> 
}\,,
\]
so that the volume label $l \in L$ is required to satisfy the following condition:
\begin{eqnarray}\label{deltal}
\delta(l)= h_2 h_1^{-1}\,.
\end{eqnarray}
The orientation of the volume can be reversed if one labels it with the inverse element $l^{-1}$:
\[
\xymatrix{  \bullet
&  \bullet 
\ar@/_4ex/[l]_{g_1}="g1"
\ar@/^4ex/[l]^{g_2}="g3"
\ar@{=>}^{h_1} "g1"+<0ex,-2.5ex>;"g3"+<0ex,2.5ex> } \; \stackrel{l^{-1}}{\Lleftarrow} \; \xymatrix{\bullet
&  \bullet 
\ar@/_4ex/[l]_{g_1}="g2"
\ar@/^4ex/[l]^{g_2}="g4"
\ar@{=>}^{h_2} "g2"+<0ex,-2.5ex>;"g4"+<0ex,2.5ex> 
}\,,
\]
while the orientation reversal of the curves and surfaces leads to the surface element labeled with $\tilde{l}=g_1^{-1}\rhd l$,
\[
\xymatrix{\ar@/_4ex/[rr]_{g_2^{-1}}="g1"
\ar@/^4ex/[rr]^{g_1^{-1}}="g3"  \bullet
&&  \bullet 
\ar@{=>}_{g_1^{-1}\rhd h_1} "g1"+<-1ex,+2.5ex>;"g3"+<-1ex,-2.5ex> } \; \stackrel{\tilde{l}}{\Rrightarrow} \; \xymatrix{
\ar@/_4ex/[rr]_{g_2^{-1}}="g2"
\ar@/^4ex/[rr]^{g_1^{-1}}="g4"
\bullet
&&  \bullet 
\ar@{=>}_{g_1^{-1}\rhd h_2} "g2"+<-1ex,+2.5ex>;"g4"+<-1ex,-2.5ex> 
}\,.
\]

One can compose two $3$-morphisms via the \emph{upward composition} (visualizing a third axis, orthogonal to the plane of the paper, as the direction up). The upward composition of $3$-morphisms $(g_1, h_1, l_1)$ and $(g_1, h_2, l_2)$, when they are compatible, \ie\,, when $\partial_3^+(l_1)=\partial_3^-(l_2)$, 
\[
\xymatrix{  \bullet
&  \bullet 
\ar@/_4ex/[l]_{g_1}="g1"
\ar@/^4ex/[l]^{g_2}="g3"
\ar@{=>}^{h_1} "g1"+<0ex,-2.5ex>;"g3"+<0ex,2.5ex> } \; \stackrel{l_1}{\Rrightarrow} \; \xymatrix{ \bullet
&  \bullet 
\ar@/_4ex/[l]_{g_1}="g2"
\ar@/^4ex/[l]^{g_2}="g4"
\ar@{=>}^{h_2} "g2"+<0ex,-2.5ex>;"g4"+<0ex,2.5ex> 
}\; \stackrel{l_2}{\Rrightarrow} \;\xymatrix{ \bullet &  \bullet  
\ar@/_4ex/[l]_{g_1}="g5"
\ar@/^4ex/[l]^{g_2}="g6" 
\ar@{=>}^{h_3} "g5"+<0ex,-2.5ex>;"g6"+<0ex,2.5ex> }
\quad = \quad \xymatrix{  \bullet
& \bullet 
\ar@/_4ex/[l]_{g_1}="g1"
\ar@/^4ex/[l]^{g_2}="g3"
\ar@{=>}^{h_1} "g1"+<0ex,-2.5ex>;"g3"+<0ex,2.5ex> }  \; \stackrel{l_2l_1}{\Rrightarrow}\; \xymatrix{ \bullet
& \bullet 
\ar@/_4ex/[l]_{g_1}="g2"
\ar@/^4ex/[l]^{g_2}="g4"
\ar@{=>}^{h_3} "g2"+<0ex,-2.5ex>;"g4"+<0ex,2.5ex> 
}\,,
\]
results in a $3$-morphism $(g_1, h_1, l_2 l_1)$:
\begin{equation}
(g_1,h_2,l_2)\#_3(g_1,h_1,l_1)=(g_1,h_1,l_2l_1)\,.
\end{equation}
The upward composition of $3$-morphisms is  associative, and for every $h \in H$ there is a $3$-morphism that is an identity for the upward composition of $3$-morphisms
\[
\xymatrix{ \bullet
& \bullet
\ar@/_3ex/[l]_{g_1}="g1"
\ar@/^3ex/[l]^{g_2}="g3"
\ar@{=>}^{h} "g1"+<0ex,-2.5ex>;"g3"+<0ex,2.5ex> } \; \stackrel{1_h}{\Rrightarrow} \;
\xymatrix{ \bullet
& \bullet
\ar@/_3ex/[l]_{g_1}="g1"
\ar@/^3ex/[l]^{g_2}="g3" 
\ar@{=>}^{h} "g1"+<0ex,-2.5ex>;"g3"+<0ex,2.5ex> }\,.
\]
\emph{The vertical composition} of two $3$-morphisms $(g_1, h_1, l_1)$ and $(g_2, h_2, l_2)$, when they are compatible, \ie\,, when $\partial_2^+(l_1)=\partial_2^-(l_2)$, 
    \[
        \begin{aligned}
 \xymatrix{  \bullet
&& \bullet 
\ar@/_4ex/[ll]_{g_1}="g1"
\ar[ll]^{g_2}="g3"
\ar@{=>}^{h_1} "g1"+<0ex,-2.5ex>;"g3"+<0ex,2ex> }& \; \stackrel{l_1}{\Rrightarrow} & \; \xymatrix{ \bullet
&&  \bullet 
\ar@/_4ex/[ll]_{g_1}="g2"
\ar[ll]^{g_2}="g4"
\ar@{=>}^{h_1'} "g2"+<0ex,-2.5ex>;"g4"+<0ex,2ex> 
}\\ \xymatrix{  \bullet
&& \bullet 
\ar@/^4ex/[ll]^{g_3}="g1"
\ar[ll]_{g_2}="g3"
\ar@{=>}^{h_2} "g3"+<0ex,-2.5ex>;"g1"+<0ex,+2ex> } & \; \stackrel{l_2}{\Rrightarrow} \;& \xymatrix{ \bullet
&& \bullet 
\ar@/^4ex/[ll]^{g_3}="g2"
\ar[ll]_{g_2}="g4"
\ar@{=>}^{h_2'} "g4"+<0ex,-2.5ex>;"g2"+<0ex,+2ex> 
} &&&,
        \end{aligned}
        \]
results in a $3$-morphism $(g_1,h_2h_1,l_2(h_2\rhd' l_1))$,
\[ \xymatrix{  \bullet
&&&& \bullet 
\ar@/_4ex/[llll]_{g_1}="g1"
\ar@/^4ex/[llll]^{g_3}="g3"
\ar@{=>}^{h_2h_1} "g1"+<0ex,-2.5ex>;"g3"+<0ex,2ex> } \; \stackrel{l_2(h_2\rhd' l_1)}{\Rrightarrow} \; \xymatrix{ \bullet
&&&& \bullet 
\ar@/_4ex/[llll]_{g_1}="g2"
\ar@/^4ex/[llll]^{g_3}="g4"
\ar@{=>}^{\delta\big(l_2(h_2\rhd' l_1)\big)h_2h_1} "g2"+<-4ex,-3ex>;"g4"+<-4ex,3ex> 
}\,. \]
One can write, for $(g_1, h_1, l_1)$ and $(g_2, h_2, l_2)$,
\begin{equation}
(g_2,h_2,l_2)\#_2(g_1,h_1,l_1)=(g_1,h_2h_1,l_2(h_2\rhd' l_1))\,.
\end{equation}
The vertical composition of $3$-morphisms is an associative operation. Composition of $3$-morphisms is invariant under the change of order of upward composition and vertical composition of $3$-morphisms, \ie\,,
\begin{equation}\label{eq:inv1}
\big((g_2,h_2',l_2') \#_3(g_2,h_2,l_2)\big)\#_2\big((g_1,h_1',l_1')\#_3(g_1,h_1,l_1)\big)=\big((g_2,h_2',l_2')\#_2(g_1,h_1',l_1')\big)\#_3\big((g_2,h_2,l_2)\#_2(g_1,h_1,l_1)\big)\,,
\end{equation}
which is demonstrated in the diagram notation, where the diagram
\[
\begin{aligned}
 \xymatrix{  \bullet
& \bullet 
\ar@/_4ex/[l]_{g_1}="g1"
\ar[l]^{g_2}="g3"
\ar@{=>}^{h_1} "g1"+<0ex,-2.5ex>;"g3"+<0ex,2ex> }& \; \stackrel{l_1}{\Rrightarrow} & \; \xymatrix{ \bullet
&  \bullet 
\ar@/_4ex/[l]_{g_1}="g2"
\ar[l]^{g_2}="g4"
\ar@{=>}^{h_1'} "g2"+<0ex,-2.5ex>;"g4"+<0ex,2ex> 
}&\; \stackrel{l_1'}{\Rrightarrow} &\;\xymatrix{  \bullet
& \bullet 
\ar@/_4ex/[l]_{g_1}="g1"
\ar[l]^{g_2}="g3"
\ar@{=>}^{h_1''} "g1"+<0ex,-2.5ex>;"g3"+<0ex,2ex> } \\ \xymatrix{  \bullet
& \bullet 
\ar@/^4ex/[l]^{g_3}="g1"
\ar[l]_{g_2}="g3"
\ar@{=>}^{h_2} "g3"+<0ex,-2ex>;"g1"+<0ex,+2.5ex> } & \; \stackrel{l_2}{\Rrightarrow} \;& \xymatrix{ \bullet
& \bullet 
\ar@/^4ex/[l]^{g_3}="g2"
\ar[l]_{g_2}="g4"
\ar@{=>}^{h_2'} "g4"+<0ex,-2ex>;"g2"+<0ex,+2.5ex> 
}& \; \stackrel{l_2'}{\Rrightarrow} &\xymatrix{ \bullet
& \bullet 
\ar@/^4ex/[l]^{g_3}="g2"
\ar[l]_{g_2}="g4"
\ar@{=>}^{h_2''} "g4"+<0ex,-2ex>;"g2"+<0ex,+2.5ex> 
}
\end{aligned}
\]
uniquely determines the $3$-morphism.

One can whisker the $3$-morphisms with morphisms and $2$-morphisms. Whiskering of a $3$-morphism by a morphism from the left is the composition of a volume $l \in L$ and curve $g_1 \in G$ from the left, when they are compatible, \ie\,, when $\partial_1^+(l)=\partial_1^-(g_1)$,
\[
\xymatrix{
  \bullet 
&  \bullet
  \ar[l]_{g_1}
& \bullet
\ar@/_4ex/[l]_{g_2}="g1"
\ar@/^4ex/[l]^{g_2'}="g3"
\ar@{=>}^{h_1} "g1"+<0ex,-2.5ex>;"g3"+<0ex,2.5ex> }\; \stackrel{l}{\Rrightarrow} \; \xymatrix{\bullet 
&  \bullet
  \ar[l]_{g_1}
& \bullet 
\ar@/_4ex/[l]_{g_2}="g2"
\ar@/^4ex/[l]^{g_2'}="g4"
\ar@{=>}^{h_2} "g2"+<0ex,-2.5ex>;"g4"+<0ex,2.5ex> 
} = \xymatrix{
   \bullet 
&& \bullet 
\ar@/_4ex/[ll]_{g_1g_2}="g1"
\ar@/^4ex/[ll]^{g_1g_2'}="g3"
\ar@{=>}^{g_1\rhd h_1} "g1"+<0ex,-2.5ex>;"g3"+<0ex,2.5ex> }\; \stackrel{g_1\rhd l}{\Rrightarrow} \; \xymatrix{   \bullet 
 && \bullet 
\ar@/_4ex/[ll]_{g_1g_2}="g2"
\ar@/^4ex/[ll]^{g_1g_2'}="g4"
\ar@{=>}^{g_1\rhd h_2} "g2"+<0ex,-2.5ex>;"g4"+<0ex,2.5ex>}\,.
\]
The composition results in a $3$-morphism:
\begin{equation}
    g_1\#_1(g_2,h_1,l)=(g_1g_2, g_1\rhd h,g_1\rhd l)\,.
\end{equation}
Similarly, one can whisker a $3$-morphism by a morphism from the right, when they are compatible, \ie\,, $\partial_1^-(l)=\partial_1^+(g_2)$,
\[
\xymatrix{
 \bullet  & \bullet 
\ar@/_4ex/[l]_{g_1}="g1"
\ar@/^4ex/[l]^{g_1'}="g3"
\ar@{=>}^{h_1} "g1"+<0ex,-2.5ex>;"g3"+<0ex,2.5ex>  & \bullet 
  \ar[l]_{g_2}}\; \stackrel{l}{\Rrightarrow} \; \xymatrix{
 \bullet  & \bullet 
\ar@/_4ex/[l]_{g_1}="g1"
\ar@/^4ex/[l]^{g_1'}="g3"
\ar@{=>}^{h_2} "g1"+<0ex,-2.5ex>;"g3"+<0ex,2.5ex>  & \bullet 
  \ar[l]_{g_2}}=\xymatrix{
   \bullet 
&& \bullet 
\ar@/_4ex/[ll]_{g_1g_2}="g1"
\ar@/^4ex/[ll]^{g_1'g_2}="g3"
\ar@{=>}^{h_1} "g1"+<0ex,-2.5ex>;"g3"+<0ex,2.5ex> }\; \stackrel{l}{\Rrightarrow} \; \xymatrix{  \bullet 
 && \bullet 
\ar@/_4ex/[ll]_{g_1g_2}="g2"
\ar@/^4ex/[ll]^{g_1'g_2}="g4"
\ar@{=>}^{h_2} "g2"+<0ex,-2.5ex>;"g4"+<0ex,2.5ex>}\,,
\]
which results in the $3$-morphism:
\begin{equation}
(g_1,h_1,l)\#_1g_2=(g_1g_2, h_1,l)\,.
\end{equation}
\emph{Whiskering of a $3$-morphism with a $2$-morphisms from below}, when they are compatible, \ie\,, $\partial_2^+(l)=\partial_2^-(h_2)$, is formed as a vertical composition of $3$-morphisms $(g_1,h
    _1,l)$ and $(g_2,h_2,1_{h_2})$,
\[
\begin{aligned}
 \xymatrix{  \bullet
&& \bullet 
\ar@/_4ex/[ll]_{g_1}="g1"
\ar[ll]^{g_2}="g3"
\ar@{=>}^{h_1} "g1"+<0ex,-2.5ex>;"g3"+<0ex,2ex> }&\; \stackrel{l}{\Rrightarrow} \;&  \xymatrix{ \bullet
&&  \bullet 
\ar@/_4ex/[ll]_{g_1}="g2"
\ar[ll]^{g_2}="g4"
\ar@{=>}^{h_1'} "g2"+<0ex,-2.5ex>;"g4"+<0ex,2ex> 
}\,\,\,\\ \xymatrix{  \bullet
&& \bullet 
\ar@/^4ex/[ll]^{g_3}="g1"
\ar[ll]_{g_2}="g3"
\ar@{=>}^{h_2} "g3"+<0ex,-2.5ex>;"g1"+<0ex,+2ex> } & \; \stackrel{1_{h_2}}{\Rrightarrow} \;& \xymatrix{ \bullet
&& \bullet 
\ar@/^4ex/[ll]^{g_3}="g2"
\ar[ll]_{g_2}="g4"
\ar@{=>}^{h_2} "g4"+<0ex,-2.5ex>;"g2"+<0ex,+2ex> 
} \,,
        \end{aligned}
    \]
which results in a $3$-morphism
\[ \xymatrix{ \bullet
&& \bullet 
\ar@/_4ex/[ll]_{g_1}="g1"
\ar@/^4ex/[ll]^{g_3}="g3"
\ar@{=>}^{h_2h_1} "g1"+<0ex,-2.5ex>;"g3"+<0ex,2ex> } \; \stackrel{h_2\rhd' l}{\Rrightarrow} \; \xymatrix{ \bullet
&&& \bullet 
\ar@/_4ex/[lll]_{g_1}="g2"
\ar@/^4ex/[lll]^{g_3}="g4"
\ar@{=>}^{\delta(h_2\rhd' l)h_2h_1} "g2"+<-4ex,-3ex>;"g4"+<-4ex,3ex> 
}\,. \]
One writes,
\begin{equation}
(g_1,h_1,l)\#_2(g_2,h_2)=(g_1,h_2h_1,h_2\rhd' l)\,.
\end{equation}
\emph{Whiskering a $3$-morphism by $2$-morphism from above}, when they are compatible, \ie\,, when $\partial_2^-(l)=\partial_2^+(h_1)$, is formed as a vertical composition of $3$-morphisms $(g_1,h_1,1_{h_1})$ and $(g_2,h_2,l)$,
\[
        \begin{aligned}
 \xymatrix{  \bullet
&& \bullet 
\ar@/_4ex/[ll]_{g_1}="g1"
\ar[ll]^{g_2}="g3"
\ar@{=>}^{h_1} "g1"+<0ex,-2.5ex>;"g3"+<0ex,2ex> }& \; \stackrel{1_{h_1}}{\Rrightarrow} & \; \xymatrix{\bullet
&&  \bullet 
\ar@/_4ex/[ll]_{g_1}="g2"
\ar[ll]^{g_2}="g4"
\ar@{=>}^{h_1} "g2"+<0ex,-2.5ex>;"g4"+<0ex,2ex> 
}\,\,\,\\ \xymatrix{  \bullet
&& \bullet 
\ar@/^4ex/[ll]^{g_3}="g1"
\ar[ll]_{g_2}="g3"
\ar@{=>}^{h_2} "g3"+<0ex,-2.5ex>;"g1"+<0ex,+2ex> } & \; \stackrel{l}{\Rrightarrow} \;& \xymatrix{\bullet
&& \bullet 
\ar@/^4ex/[ll]^{g_3}="g2"
\ar[ll]_{g_2}="g4"
\ar@{=>}^{h_2'} "g4"+<0ex,-2.5ex>;"g2"+<0ex,+2ex> 
}\,,
        \end{aligned}
    \]
which results in a $3$-morphism,
\[ \xymatrix{ \bullet
&& \bullet 
\ar@/_4ex/[ll]_{g_1}="g1"
\ar@/^4ex/[ll]^{g_3}="g3"
\ar@{=>}^{h_2h_1} "g1"+<0ex,-2.5ex>;"g3"+<0ex,2ex> } \; \stackrel{l}{\Rrightarrow} \; \xymatrix{\bullet
&& \bullet 
\ar@/_4ex/[ll]_{g_1}="g2"
\ar@/^4ex/[ll]^{g_3}="g4"
\ar@{=>}^{\delta(l)h_2h_1} "g2"+<-4ex,-3ex>;"g4"+<-4ex,3ex> 
}\,.\]
One obtains
\begin{equation}
(g_1,h_1)\#_2(g_2,h_2,l)=(g_1,h_2h_1,l)\,.
\end{equation}
\textit{The interchanging $3$-arrow} is the horizontal composition of two $2$-morphisms $h_1$ and $h_2$, when they are compatible, \ie\,, when $\partial_1^-(h_1)=\partial_1^+(h_2)$, 
\[
  \xymatrix{
  \bullet
&& \bullet
  \ar@/_2.5ex/[ll]_{g_1}="g1"\ar@/^2.5ex/[ll]^{g'_1}="g2"
  \ar@{=>}^{h_1} "g1"+<0ex,-2.5ex>;"g2"+<0ex,2.5ex>
&& \bullet 
  \ar@/_2.5ex/[ll]_{g_2}="g1"\ar@/^2.5ex/[ll]^{g'_2}="g2"
  \ar@{=>}^{h_2} "g1"+<0ex,-2.5ex>;"g2"+<0ex,2.5ex>
}\,,
\]
that results in a $3$-morphism $l$, with source surface
\[\partial_{3}^-(l)=\big((g_1,h_1)\#_1 g_2'\big)\#_2\big(g_1\#_1 (g_2,h_2)\big)\,,\] 
and target surface 
\[\partial_3^+(l)=\big(g_1'\#_1 (g_2,h_2)\big)\#_2\big((g_1,h_1)\#_1 g_2\big)\,,\]
\[
  \xymatrix{
   \bullet
&& \bullet
  \ar@/_2.5ex/[ll]_{g_1}="g1"\ar@/^2.5ex/[ll]^{g'_1}="g2"
  \ar@{=>}^{h_1} "g1"+<0ex,-2.5ex>;"g2"+<0ex,2.5ex>
&& \bullet 
  \ar@/_2.5ex/[ll]_{g_2}="g1"\ar@/^2.5ex/[ll]^{g'_2}="g2"
  \ar@{=>}^{h_2} "g1"+<0ex,-2.5ex>;"g2"+<0ex,2.5ex>
}
\quad =\quad
\xymatrix{
  \bullet
&&& \bullet 
\ar@/_3ex/[lll]_{g_1 g_2}="g1"
\ar@/^3ex/[lll]^{g'_1 g_2'}="g3"
\ar@{=>}^{h_1 g_1\rhd h_2} "g1"+<-1ex,-2ex>;"g3"+<-1ex,2ex>
}\stackrel{l}{\Rrightarrow}\xymatrix{
  \bullet
&&& \bullet 
\ar@/_3ex/[lll]_{g_1 g_2}="g1"
\ar@/^3ex/[lll]^{g'_1 g_2'}="g3"
\ar@{=>}^{g_1'\rhd h_2 h_1} "g1"+<-1ex,-2ex>;"g3"+<-1ex,2ex>
}\,.
\]
One obtains,
\begin{equation}
    (g_1,h_1)\#_1(g_2,h_2)=(g_1g_2,h_1 g_1\rhd h_2,l)\,,
\end{equation}
where the $3$-morphism $l$ is Peiffer lifting $\{h_1,g_1\rhd h_2\}_{\mathrm{p}}^{-1}$. Using the condition (\ref{deltal}), one obtains 
\begin{equation}
    \big((\partial(h_1)g_1)\rhd h_2\big) h_1=\delta(l)h_1\big(g_1\rhd h_2\big) \,,
\end{equation}
and from the definition of the Peiffer commutator, the identity (\ref{Peiffer_comm}), and the property \eqref{prop1} of the $2$-crossed module, \ie\,, $\delta(\{h_1,h_2 \}_\mathrm{p})= \langle h_1\,,h_2\rangle{}_{\mathrm{p}}$, one obtains
\begin{eqnarray}
\delta(l)^{-1}= h_1 g_1 \rhd h_2 h_1^{-1}(\partial(h_1)g_1)\rhd h_2{}^{-1}=\langle h_1, g_1\rhd h_2 \rangle_{\mathrm{p}}=\delta(\{h_1,g_1\rhd h_2 \}_\mathrm{p})\,.
\end{eqnarray}

Given any collection of curves, surfaces, and volumes, a configuration of $3$-gauge theory is an assignment of elements of $G$ to the curves, elements of $H$ to the surfaces, and elements of $L$ to volumes so that the following conditions hold: 
\begin{enumerate}
    \item For each surface labeled by $h\in H$, one has that $\partial(h) =g_2 g_1^{-1}$ where $g_1$ and $g_2$ are the source and target curve, respectively;
    \item For each volume labeled by $l \in L$, one has that $\delta(l) =h_2 h_1^{-1}$, where $h_1$ and $h_2$ are the source and target surface, respectively;
    \item For each $4$-simplex labeled by
$(jk\ell m n)\in\Lambda_4$, the volume holonomy around it is trivial.
\end{enumerate}
The configurations thus defined can be viewed as the classical configurations of $3$-gauge theory or, in a path integral quantum theory, these are the configurations over which we integrate in the path integral.
\subsection{Gauge invariant quantities}

In subsection \ref{sec:3gaugetheory}, we have introduced a number of operations by which we can combine labeled paths, surfaces, and volumes, in order to calculate the composition of elementary paths, surfaces, and volumes, to arbitrarily large ones. In this subsection, we will make use of these compositions in order to construct gauge invariant quantities that are associated with closed paths, surfaces, and volumes. In lemmas \ref{Th01}, \ref{Th02}, and \ref{Th:03}, this procedure is used for the boundary path of a triangle, the boundary surface of a tetrahedron, and the boundary volume of the $4$-simplex. The result of the lemma \ref{Th01} is already derived for the case of $2$-groups and remains unchanged in the $3$-gauge theory, see \cite{GirelliPfeifferPopescu2008}. The higher flatness condition for the boundary surface of a tetrahedron derived in \cite{GirelliPfeifferPopescu2008}, is generalized for the case of $3$-groups in the lemma \ref{Th02}. One of the main results of the paper is the lemma \ref{Th:03} where we derived the higher flatness condition for the boundary volume of the $4$-simplex.
\begin{lemma}\label{Th01}
Let us consider a triangle, $(jk\ell)$. The edges $(jk)\,, j < k$, are labeled by group elements $g_{jk}\in G$ and the triangle $(jk\ell)\,, j < k < \ell$, by element $h_{jk\ell}\in H$. Consider the diagram (\ref{diag1}). 
\begin{equation}\label{diag1}
  \xymatrix{l
\bullet & k\bullet
\ar@/_2ex/[l]_{g_{kl}}="g1" & \bullet j
\ar@/_2ex/[l]_{g_{jk}}="g2"
\ar@/^6ex/[ll]^{g_{jl}}="g3"
\ar@{=>}^{h_{jkl}} "g1"+<4.7ex,-4.8ex>;"g3"+<0ex,2.5ex>
}\quad = \quad \xymatrix{l \bullet && l \bullet 
\ar@/_2ex/[ll]_{1_\bullet}="g1"
\ar@/^2ex/[ll]^{\partial(h_{jkl})}="g2"
\ar@{=>}^{h_{jkl}} "g1"+<0ex,-2ex>;"g2"+<0ex,2ex>
& k\bullet
\ar@/_2ex/[l]_{g_{kl}}="g4" & \bullet j
\ar@/_2ex/[l]_{g_{jk}}="g5"
\ar@/^7ex/[ll]^{g_{kl}g_{jk}}="g3"
\ar@{=>}^(0.25){1_{g_{kl}g_{jk}}} "g4"+<4.7ex,-4.8ex>;"g3"+<0ex,2.5ex>
}\quad = \quad \xymatrix{l
\bullet & k\bullet
\ar@/_2ex/[l]_{g_{kl}}="g1" & \bullet j
\ar@/_2ex/[l]_{g_{jk}}="g2"
\ar@/^6ex/[ll]^{\partial(h_{jkl}) \,g_{kl}g_{jk}}="g3"
\ar@{=>}^{h_{jkl}} "g1"+<4.7ex,-4.8ex>;"g3"+<0ex,2.5ex>
}\,.
\end{equation}
The curve $\gamma_1=g_{ k\ell}g_{j k}$ is the source and the curve $\gamma_2=g_{j \ell}$ is the target of the surface morphism $\Sigma:\gamma_1\to\gamma_2$, labeled by the group element $h_{j k \ell}$, \ie\,,
\begin{equation}
g_{j\ell}=\partial(h_{jk\ell})g_{k\ell}g_{jk}\,.
\end{equation}

\end{lemma}
\begin{lemma}\label{Th02}
Let us consider a tetrahedron, $(jk\ell m)$. The edges $(jk)\,, j < k$, are labeled by group elements $g_{jk}\in G$ and the triangles $(jk\ell)\,, j < k < \ell$, by elements $h_{jk\ell}\in H$, and the tetrahedron $(j k \ell m)\,, j<k<\ell <m$ by the group element $l_{j k \ell m}\in L$.  We have oriented the triangles $(jk\ell)$ so that they have the source is $g_{k\ell}g_{jk}$ and the target is $g_{j\ell}$, \ie\ $g_{j\ell}=\partial(h_{jk\ell})g_{k\ell}g_{jk}$\,.

Let us first cut the tetrahedron surface along the edge $(jm)$. This determines the ordering of the vertical composition of the constituent surfaces. We just have to make sure that all surfaces are composable, \ie\,, they have the suitable reference points and the correct orientation in order to compose them vertically.

Consider the diagram (\ref{diag4}). We first move the curve from $g_{k\ell} g_{jk}$ to the curve $g_{j\ell}$. At this stage, one cannot compose the result with the triangle $(j\ell m)$, and one first has to whisker it from the left by $g_{\ell m}$. Now the two morphisms are vertically composable, and this moves the curve to $g_{j m}$. The following $2$-morphism is obtained
\begin{equation}\label{diag4}
\xymatrix{m\bullet & \bullet \ell
\ar[l]_{g_{\ell m}}="g1"
& \bullet k
\ar@/_2ex/[l]_{g_{k\ell}}="g2" 
& \bullet j
\ar@/_2ex/[l]_{g_{jk}}="g3"
\ar@/^6ex/[ll]^{g_{j\ell}}="g4"
\ar@{=>}^{h_{jk\ell}} "g2"+<4.7ex,-5ex>;"g4"+<0ex,2.5ex>
\ar@/^9ex/[lll]^{g_{jm}}="g5"
\ar@{=>}^{h_{j\ell m}} "g1"+<4.1ex,-3.7ex>;"g5"+<-6ex,5.3ex>
} \quad = (g_{\ell m} g_{j\ell}, h_{j\ell m})\#_2\big(g_{\ell m}\#_1 (g_{k \ell}g_{jk}, h_{jk\ell})\big)=\big(g_{\ell m} g_{k\ell}g_{jk}, h_{j\ell m} (g_{\ell m}\rhd h_{jk\ell})\big)\,.
\end{equation}
Let us then consider the diagram (\ref{diag3}). We first move the curve from $g_{\ell m} g_{k\ell}$ to the curve $g_{k m}$. At this stage, one cannot compose the result with the triangle $(jkm)$, and one first has to whisker it from the right by $g_{jk}$. Now the two morphisms are vertically composable, and this moves the curve to $g_{j m}$. The following $2$-morphism is obtained
\begin{equation}\label{diag3}
\xymatrix{m\bullet & \bullet \ell
\ar@/_2ex/[l]_{g_{\ell m}}="g4" & \bullet k
\ar@/_2ex/[l]_{g_{k\ell}}="g3"
\ar@/^6ex/[ll]^{g_{km}}="g2"
\ar@{=>}^{h_{k\ell m}} "g4"+<4.7ex,-5ex>;"g2"+<0ex,2.5ex>
& \bullet j
\ar[l]_{g_{jk}}="g1"
\ar@/^9ex/[lll]^{g_{jm}}="g5"
\ar@{=>}^{h_{jkm}} "g3"+<4.6ex,-5.6ex>;"g5"+<6ex,5ex>
} \quad = (g_{km} g_{jk},h_{jkm})\#_2\big((g_{\ell m} g_{k \ell},h_{k \ell m})\#_1 g_{j k}\big)=(g_{\ell m} g_{k\ell} g_{jk},h_{jkm}h_{k\ell m}) \,.
\end{equation}

The two surfaces have the same source and target, $\Sigma_1 : g_{\ell m} g_{k\ell}g_{jk} \to g_{jm}$ and $\Sigma_2 :g_{\ell m} g_{k\ell}g_{jk} \to g_{jm}$. Now, moving from surface shown on the diagram \eqref{diag4} to the surface shown on the diagram \eqref{diag4} is given by the volume morphism $\mathcal{V}: \Sigma_1 \to \Sigma_2$ determined by the group element $l_{jk \ell m}$, \ie\,,
\begin{equation}
    (g_{\ell m} g_{k\ell} g_{jk},h_{jkm}h_{k\ell m}) = \big(g_{\ell m} g_{k\ell}g_{jk}, \delta(l_{jk\ell m}) h_{j\ell m} (g_{\ell m}\rhd h_{jk\ell})\big)\,,
\end{equation}
that gives the relation,
\begin{equation}
    h_{jkm}h_{k\ell m}  = \delta(l_{jk\ell m}) h_{j\ell m} (g_{\ell m}\rhd h_{jk\ell})\,.
\end{equation}
\end{lemma}

\begin{lemma}\label{Th:03}
Let us consider a $4$-simplex, $(jk\ell m n)$. The edges $(jk)\,, j < k $, are labeled by group elements $g_{jk}\in G$, the triangles $(jk\ell)\,, j < k  < \ell$, by elements $h_{jk\ell}\in H$, and the tetrahedrons $(jk\ell m)\,, j<k <\ell <m$, by the group element $l_{jk\ell m}\in L$.  We have oriented the triangles $(j k \ell)$ so that the source curve is $g_{k \ell}g_{jk}$ and the target curve is $g_{j \ell}$, \ie\,, $g_{j \ell}=\partial(h_{j k \ell})g_{k\ell}g_{jk}$\,, and the tetrahedrons $(j k \ell m)$ so that the source surface is $h_{j\ell m} (g_{\ell m}\rhd h_{jk\ell})$ and the target surface is $ h_{jkm}h_{k\ell m} $, \ie\,, $ h_{jkm}h_{k\ell m}  = \delta(l_{jk\ell m}) h_{j\ell m} (g_{\ell m}\rhd h_{jk\ell})$.

Let us first cut the $4$-simplex volume along the surface $h_{jm n}g_{m n}\rhd(h_{j\ell m }g_{\ell m }\rhd h_{jk\ell })$. This surface determines the ordering of the vertical composition of the constituent volumes. We have to make sure that all volumes are composable, \ie\,, they have the suitable reference points and the correct orientation in order to compose them vertically. First, let us consider the diagram (\ref{diag5}). 
We first move the surface from $h_{j\ell m}g_{\ell m}\rhd h_{jk\ell }$ to surface $h_{jkm}h_{k\ell m}$ with the $3$-arrow $l_{jk\ell  m}$. To compose the resulting $3$-morphism with the surface $h_{jm n}$ one must first whisker it from the left with $g_{m n}$. The obtained $3$-morphism $(g_{m n }g_{\ell  m}g_{k \ell }g_{jk}, g_{m n}\rhd(h_{j\ell m}g_{\ell m}\rhd h_{jk\ell }), g_{m n }\rhd l_{jk\ell m})$ can be whiskered from below with the $2$-morphism $(g_{m n}g_{j m}, h_{j m n})$, and the resulting $3$-morphism is $(g_{m n}g_{\ell m}g_{k\ell }g_{jk}, h_{jm n}g_{m n}\rhd(h_{j\ell m}g_{\ell m}\rhd h_{jk\ell }), h_{jm n}\rhd'(g_{m n}\rhd l_{jk\ell m}))$, with the source surface $h_{jm n}g_{m n}\rhd(h_{j\ell m}g_{\ell m}\rhd h_{jk\ell })$ and the target surface $h_{jm n}g_{m n}\rhd(h_{jkm }h_{k\ell m})$,
\begin{equation}\label{diag5}
\begin{aligned}
\xymatrix{n \bullet & \bullet m
\ar@/_2ex/[l]_{g_{m n}}="g6"& \bullet \ell 
\ar@/_2ex/[l]_{g_{\ell m}}="g1"
& \bullet k
\ar@/_2ex/[l]_{g_{k\ell }}="g2" 
& \bullet j
\ar@/_2ex/[l]_{g_{jk}}="g3"
\ar@/^6ex/[ll]^{g_{j\ell }}="g4"
\ar@{=>}^{h_{jk\ell }} "g2"+<4.7ex,-5ex>;"g4"+<0ex,2.5ex>
\ar@/^9ex/[lll]^{g_{jm}}="g5"
\ar@{=>}^{h_{j\ell m}} "g1"+<4.1ex,-4ex>;"g5"+<-6ex,5.3ex>
\ar@/^13ex/[llll]^{g_{jn}}="g7"
\ar@{=>}^{h_{jm n}} "g6"+<4.6ex,-5.6ex>;"g7"+<-10ex,7ex>
} \; \stackrel{h_{jm n}\rhd'(g_{m n}\rhd l_{jk\ell m})}{\Rrightarrow} \; \xymatrix{n \bullet 
 & \bullet m 
 \ar@/_2ex/[l]_{g_{m n}}="g6" & \bullet \ell 
\ar@/_2ex/[l]_{g_{\ell m }}="g4" & \bullet k
\ar@/_2ex/[l]_{g_{k\ell }}="g3"
\ar@/^6ex/[ll]^{g_{km }}="g2"
\ar@{=>}^{h_{k\ell m }} "g4"+<4.7ex,-5ex>;"g2"+<0ex,2.5ex>
& \bullet j
\ar@/_2ex/[l]_{g_{jk}}="g1"
\ar@/^9ex/[lll]^{g_{jm }}="g5"
\ar@/^13ex/[llll]^{g_{jn}}="g7"
\ar@{=>}^{h_{jkm }} "g3"+<4.6ex,-5.6ex>;"g5"+<6ex,5ex>
\ar@{=>}^{h_{jm n}} "g6"+<4.6ex,-5.6ex>;"g7"+<-10ex,7ex>
}\,.
\end{aligned}
\end{equation}
Let us move the surface to $h_{jkn}h_{km n}g_{m \ell }\rhd h_{k\ell m }$, see diagram (\ref{diag6}). To do that, we consider the $3$-morphism $(g_{mn}g_{km}g_{jk},h_{jmn}g_{mn}\rhd h_{jkm},l_{jk m n})$ with the source surface $ h_{j m n}g_{m n}\rhd h_{jkm}$ and target surface $h_{j k n} h_{k m n}$. This $3$-morphism can be whiskered from above with the $2$-morphism $(g_{m n}g_{\ell l}g_{k\ell }g_{jk},g_{m n}\rhd h_{k\ell m})$, and the obtained $3$-morphism is $(g_{m n}g_{\ell m }g_{k\ell }g_{jk},h_{jm n}g_{m n}\rhd(h_{jkm }h_{k\ell m }),l_{jkm n})$, with the source surface $h_{jm n}g_{m n}\rhd(h_{jkm }h_{k\ell m})$ and target surface $h_{jkn}h_{km n}g_{m n}\rhd h_{k\ell m }$,

\begin{equation}\label{diag6}
    \begin{aligned}
  \xymatrix{n \bullet 
 & \bullet m 
 \ar@/_2ex/[l]_{g_{m n}}="g6" & \bullet \ell 
\ar@/_2ex/[l]_{g_{\ell m }}="g4" & \bullet k
\ar@/_2ex/[l]_{g_{k\ell }}="g3"
\ar@/^6ex/[ll]^{g_{km }}="g2"
\ar@{=>}^{h_{k\ell m }} "g4"+<4.7ex,-5ex>;"g2"+<0ex,2.5ex>
& \bullet j
\ar@/_2ex/[l]_{g_{jk}}="g1"
\ar@/^9ex/[lll]^{g_{jm }}="g5"
\ar@/^13ex/[llll]^{g_{jn}}="g7"
\ar@{=>}^{h_{jkm }} "g3"+<4.6ex,-5.6ex>;"g5"+<6ex,5ex>
\ar@{=>}^{h_{jm n}} "g6"+<4.6ex,-5.6ex>;"g7"+<-10ex,7ex>
}\;\stackrel{l_{jkm n}}{\Rrightarrow} \; \xymatrix{n \bullet 
 & \bullet m  
 \ar@/_2ex/[l]_{g_{m n}}="g6" & \bullet \ell 
\ar@/_2ex/[l]_{g_{\ell m }}="g4" & \bullet k
\ar@/_2ex/[l]_{g_{k\ell }}="g3"
\ar@/^6ex/[ll]^{g_{km }}="g2"
\ar@/^9ex/[lll]^{g_{kn}}="g5"
\ar@{=>}^{h_{k\ell m }} "g4"+<4.6ex,-5.6ex>;"g2"+<0ex,2.5ex>
& \bullet j
\ar@/_2ex/[l]_{g_{jk}}="g1"
\ar@/^13ex/[llll]^{g_{jn}}="g7"
\ar@{=>}^{h_{km n}} "g6"+<4.6ex,-5.6ex>;"g7"+<-10ex,7ex>
\ar@{=>}^{h_{jkn}} "g3"+<4.7ex,-5ex>;"g7"+<13ex,10ex>}\,.
\end{aligned}
\end{equation}
Next, we want to move the surface $h_{jkn}h_{km n}g_{m n}\rhd h_{k\ell m }$ to surface $h_{jkn}h_{k\ell n}h_{\ell m n}$, as shown on the diagram (\ref{diag7}). We whisker the $3$-morphism $(g_{m n}g_{\ell  m}g_{k\ell }, h_{km n}g_{m n}\rhd h_{k\ell m}, l_{k\ell m n})$, with the source surface $ h_{km n}g_{m n}\rhd h_{k\ell m}$ and target surface $h_{k\ell n}h_{\ell m n}$, with the morphism $g_{jk}$ from the right, obtaining the $3$-morphism $(g_{m n}g_{\ell  m}g_{k\ell }g_{jk}, h_{km n}g_{m n}\rhd h_{k\ell m}, l_{k\ell m n})$. Now, we whisker this $3$-morphism with the $2$-morphism $(g_{kn}g_{jk}, h_{jkn})$ from below, and we obtain the $3$-morphism $(g_{m n}g_{\ell  m}g_{k\ell }g_{jk}, h_{jkn}h_{km n}g_{m n}\rhd h_{k\ell m}, h_{jkn}\rhd' l_{k\ell m n})$, 
\begin{equation}\label{diag7}
    \begin{aligned}
\xymatrix{n \bullet 
 & \bullet m  
 \ar@/_2ex/[l]_{g_{m n}}="g6" & \bullet \ell 
\ar@/_2ex/[l]_{g_{\ell m }}="g4" & \bullet k
\ar@/_2ex/[l]_{g_{k\ell }}="g3"
\ar@/^6ex/[ll]^{g_{km }}="g2"
\ar@/^9ex/[lll]^{g_{kn}}="g5"
\ar@{=>}^{h_{k\ell m }} "g4"+<4.6ex,-5.6ex>;"g2"+<0ex,2.5ex>
& \bullet j
\ar@/_2ex/[l]_{g_{jk}}="g1"
\ar@/^13ex/[llll]^{g_{jn}}="g7"
\ar@{=>}^{h_{km n}} "g6"+<4.6ex,-5.6ex>;"g7"+<-10ex,7ex>
\ar@{=>}^{h_{jkn}} "g3"+<4.7ex,-5ex>;"g7"+<13ex,10ex>}\;\stackrel{h_{jkn}\rhd' l_{k\ell m n}}{\Rrightarrow} \; \xymatrix{n \bullet 
 & \bullet m  
 \ar@/_2ex/[l]_{g_{m n}}="g6" & \bullet \ell 
\ar@/_2ex/[l]_{g_{\ell m }}="g4"
\ar@/^6ex/[ll]^{g_{\ell n}}="g2"& \bullet k
\ar@/_2ex/[l]_{g_{k\ell }}="g3"
\ar@/^9ex/[lll]^{g_{kn}}="g5"
\ar@{=>}^{h_{k\ell n}} "g4"+<5ex,-5.6ex>;"g5"+<8ex,5ex>
& \bullet j
\ar@/_2ex/[l]_{g_{jk}}="g1"
\ar@/^13ex/[llll]^{g_{jn}}="g7"
\ar@{=>}^{h_{\ell m n}} "g6"+<4.6ex,-5.6ex>;"g2"+<0ex,2ex>
\ar@{=>}^{h_{jkn}} "g3"+<4.7ex,-5ex>;"g7"+<13ex,10ex>}\,.
\end{aligned}
\end{equation}
The mapping of the surface $h_{jkn}h_{k\ell n}h_{\ell m n}$ to the surface $h_{j\ell n}g_{\ell n}\rhd h_{jk\ell } h_{\ell m n}$ in shown on the diagram (\ref{diag8}). The $3$-morphism with the appropriate source and target is constructed by whiskering the $3$-morphism $(g_{\ell n}g_{k\ell }g_{jk},h_{jkn}h_{k\ell n},l_{jk\ell n}^{-1})$ with $2$-morphism $(g_{m n}g_{\ell m }g_{k\ell }g_{jk},h_{\ell m n})$ from above.
The obtained $3$-morphism is $(g_{m n}g_{\ell m }g_{k\ell }g_{jk},h_{jkn}h_{k\ell n}h_{\ell m n},l_{jk\ell n}^{-1})$, 
\begin{equation}\label{diag8}
    \begin{aligned}
 \xymatrix{n \bullet 
 & \bullet m  
 \ar@/_2ex/[l]_{g_{m n}}="g6" & \bullet \ell 
\ar@/_2ex/[l]_{g_{\ell m }}="g4"
\ar@/^6ex/[ll]^{g_{\ell n}}="g2"& \bullet k
\ar@/_2ex/[l]_{g_{k\ell }}="g3"
\ar@/^9ex/[lll]^{g_{kn}}="g5"
\ar@{=>}^{h_{k\ell n}} "g4"+<5ex,-5.6ex>;"g5"+<8ex,5ex>
& \bullet j
\ar@/_2ex/[l]_{g_{jk}}="g1"
\ar@/^13ex/[llll]^{g_{jn}}="g7"
\ar@{=>}^{h_{\ell m n}} "g6"+<4.6ex,-5.6ex>;"g2"+<0ex,2ex>
\ar@{=>}^{h_{jkn}} "g3"+<4.7ex,-5ex>;"g7"+<13ex,10ex>}\;\stackrel{l_{jk\ell n}^{-1}}{\Rrightarrow} \; \xymatrix{n \bullet 
 & \bullet m  
 \ar@/_2ex/[l]_{g_{m n}}="g6" & \bullet \ell 
\ar@/_2ex/[l]_{g_{\ell m }}="g4"
\ar@/^6ex/[ll]^{g_{\ell n}}="g2"& \bullet k
\ar@/_2ex/[l]_{g_{k\ell }}="g3"
& \bullet j
\ar@/_2ex/[l]_{g_{jk}}="g1"
\ar@/^6ex/[ll]^{g_{j\ell }}="g5"
\ar@/^13ex/[llll]^{g_{jn}}="g7"
\ar@{=>}^{h_{jk\ell }} "g3"+<5ex,-5.6ex>;"g5"+<0.4ex,2ex>
\ar@{=>}^{h_{\ell m n}} "g6"+<4.6ex,-5.6ex>;"g2"+<0ex,2ex>
    \ar@{=>}^{h_{j\ell n}} "g4"+<4.7ex,-5ex>;"g7"+<0ex,2ex>}\,.
\end{aligned}
\end{equation}
Next we map the surface $h_{j\ell n}g_{\ell n}\rhd h_{jk\ell } h_{\ell m n}$ to the surface $h_{j\ell n}h_{\ell m n} (g_{m n}g_{\ell  m})\rhd h_{jk\ell }$, see the diagram (\ref{diag9}). We use the inverse interchanging $2$-arrow composition to map the surface $g_{\ell n}\rhd h_{jk\ell } h_{\ell m n}$ to the surface $h_{\ell m n} (g_{m n}g_{\ell  m})\rhd h_{jk\ell }$, resulting in the $3$-morphism $(g_{m n}g_{\ell m }g_{k\ell }g_{jk},g_{\ell n}\rhd h_{jk\ell } h_{\ell m n},\{h_{\ell m n}, (g_{m n}g_{\ell m})\rhd h_{jk\ell }\}_{\mathrm{p}})$. Next, we whisker the obtained $3$-morphism with the $2$-morphism $(g_{\ell n}g_{j\ell }, h_{j\ell n})$ from below. The obtained $3$-morphism with the appropriate source and target surfaces is $
(g_{m n}g_{\ell m }g_{k\ell }g_{jk},h_{j\ell n}g_{\ell n}\rhd h_{jk\ell } h_{\ell m n},h_{j\ell n}\rhd'\{h_{\ell m n}, (g_{m n}g_{\ell m})\rhd h_{jk\ell }\}_{\mathrm{p}})$,
\begin{equation}
    \begin{aligned}\label{diag9}
\xymatrix{n \bullet 
 & \bullet m  
 \ar@/_2ex/[l]_{g_{m n}}="g6" & \bullet \ell 
\ar@/_2ex/[l]_{g_{\ell m }}="g4"
\ar@/^6ex/[ll]^{g_{\ell n}}="g2"& \bullet k
\ar@/_2ex/[l]_{g_{k\ell }}="g3"
& \bullet j
\ar@/_2ex/[l]_{g_{jk}}="g1"
\ar@/^6ex/[ll]^{g_{j\ell }}="g5"
\ar@/^13ex/[llll]^{g_{jn}}="g7"
\ar@{=>}^{h_{jk\ell }} "g3"+<5ex,-5.6ex>;"g5"+<0.4ex,2ex>
\ar@{=>}^{h_{\ell m n}} "g6"+<4.6ex,-5.6ex>;"g2"+<0ex,2ex>
    \ar@{=>}^{h_{j\ell n}} "g4"+<4.7ex,-5ex>;"g7"+<0ex,2ex>}\;\stackrel{h_{j\ell n}\rhd'\{h_{\ell m n}, (g_{m n}g_{\ell m})\rhd h_{jk\ell }\}_{\mathrm{p}}}{\Rrightarrow} \; \xymatrix{n \bullet 
 & \bullet m  
 \ar@/_2ex/[l]_{g_{m n}}="g6" & \bullet \ell 
\ar@/_2ex/[l]_{g_{\ell m }}="g4"
\ar@/^6ex/[ll]^{g_{\ell n}}="g2"& \bullet k
\ar@/_2ex/[l]_{g_{k\ell }}="g3"
& \bullet j
\ar@/_2ex/[l]_{g_{jk}}="g1"
\ar@/^6ex/[ll]^{g_{j\ell }}="g5"
\ar@/^13ex/[llll]^{g_{jn}}="g7"
\ar@{=>}^{h_{jk\ell }} "g3"+<5ex,-5.6ex>;"g5"+<0.4ex,2ex>
\ar@{=>}^{h_{\ell m n}} "g6"+<4.6ex,-5.6ex>;"g2"+<0ex,2ex>
\ar@{=>}^{h_{j\ell n}} "g4"+<4.7ex,-5ex>;"g7"+<0ex,2ex>}\,.
\end{aligned}
\end{equation}
Finally, we construct the $3$-morphism that maps the surface $h_{j\ell n}h_{\ell m n} (g_{m n}g_{\ell  m})\rhd h_{jk\ell }$ to the starting surface $h_{jm n}g_{m n}\rhd(h_{j\ell m}g_{\ell m}\rhd h_{jk\ell })$. To obtain the $3$-morphism with the appropriate source and target surfaces we first move the surface $h_{j\ell n}h_{\ell m n}$ to the surface $h_{jm n}g_{m n} \rhd h_{j\ell  m}$ with the $3$-arrow $(g_{mn}g_{\ell m}g_{j\ell},h_{j\ell n}h_{\ell m n},l_{j\ell m n}^{-1})$. Next, we whisker the $3$-morphism $(g_{m n}g_{\ell m }g_{j\ell },h_{j\ell n}h_{\ell m n} ,l_{j\ell m n}^{-1})$ with the $2$-morphism $(g_{m n}g_{\ell m }g_{k\ell }g_{jk},(g_{m n}g_{\ell  m})\rhd h_{jk\ell })$ from above. The obtained $3$-morphism $(g_{m n}g_{\ell m }g_{k\ell }g_{jk},h_{j\ell n}h_{\ell m n} (g_{m n}g_{\ell  m})\rhd h_{jk\ell },l_{j\ell m n}^{-1})$ moves the surface to the starting surface, as shown on the diagram (\ref{diag10}),
\begin{equation}
    \begin{aligned}\label{diag10}
\xymatrix{n \bullet 
 & \bullet m  
 \ar@/_2ex/[l]_{g_{m n}}="g6" & \bullet \ell 
\ar@/_2ex/[l]_{g_{\ell m }}="g4"
\ar@/^6ex/[ll]^{g_{\ell n}}="g2"& \bullet k
\ar@/_2ex/[l]_{g_{k\ell }}="g3"
& \bullet j
\ar@/_2ex/[l]_{g_{jk}}="g1"
\ar@/^6ex/[ll]^{g_{j\ell }}="g5"
\ar@/^13ex/[llll]^{g_{jn}}="g7"
\ar@{=>}^{h_{jk\ell }} "g3"+<5ex,-5.6ex>;"g5"+<0.4ex,2ex>
\ar@{=>}^{h_{\ell m n}} "g6"+<4.6ex,-5.6ex>;"g2"+<0ex,2ex>
\ar@{=>}^{h_{j\ell n}} "g4"+<4.7ex,-5ex>;"g7"+<0ex,2ex>}\; \stackrel{ l_{j\ell m n}^{-1}}{\Rrightarrow} \;\xymatrix{n \bullet & \bullet m 
\ar@/_2ex/[l]_{g_{m n}}="g6"& \bullet \ell 
\ar@/_2ex/[l]_{g_{\ell m }}="g1"
& \bullet k
\ar@/_2ex/[l]_{g_{k\ell }}="g2" 
& \bullet j
\ar@/_2ex/[l]_{g_{jk}}="g3"
\ar@/^6ex/[ll]^{g_{j\ell }}="g4"
\ar@{=>}^{h_{jk\ell }} "g2"+<4.7ex,-5ex>;"g4"+<0ex,2.5ex>
\ar@/^9ex/[lll]^{g_{jm }}="g5"
\ar@{=>}^{h_{j\ell m }} "g1"+<4.1ex,-4ex>;"g5"+<-6ex,5.3ex>
\ar@/^13ex/[llll]^{g_{jn}}="g7"
\ar@{=>}^{h_{jm n}} "g6"+<4.6ex,-5.6ex>;"g7"+<-10ex,7ex>
} \,.
\end{aligned}
\end{equation}
After the upward composition of the $3$-morphisms given by the diagrams (\ref{diag5})-(\ref{diag10}), the obtained $3$-morphism is:
\begin{equation}
\begin{aligned}
(g_{m n}g_{\ell m }g_{k\ell }g_{jk},h_{j\ell n}h_{\ell m n} (g_{m n}g_{\ell  m})\rhd h_{jk\ell },l_{j\ell m n}^{-1})\#_3\\
(g_{m n}g_{\ell m }g_{k\ell }g_{jk},g_{\ell n}\rhd h_{jk\ell } h_{\ell m n},h_{j\ell n}\rhd'\{h_{\ell m n}, (g_{m n}g_{\ell m})\rhd h_{jk\ell }\}_{\mathrm{p}})\#_3\\
(g_{m n}g_{\ell m }g_{k\ell }g_{jk},h_{jkn}h_{k\ell n}h_{\ell m n},l_{jk\ell n}^{-1}) \#_3\\
(g_{m n}g_{\ell m }g_{k\ell }g_{jk},h_{jkn}h_{km n}g_{m \ell }\rhd h_{k\ell m },h_{jkn}\rhd'l_{jkm n}) \#_3\\
(g_{m n}g_{\ell m }g_{k\ell }g_{jk},h_{jm n}g_{m n}\rhd(h_{jkm }h_{k\ell m }),l_{jkm n}{}) \#_3\\(g_{m n}g_{\ell m }g_{k\ell }g_{jk}, h_{jm n}g_{m n}\rhd(h_{j\ell m }g_{\ell m }\rhd h_{jk\ell }), h_{jm n}\rhd'(g_{m n}\rhd l_{jk\ell m }))\\=(g_{m n}g_{\ell m }g_{k\ell }g_{jk},h_{jm n}g_{m n}\rhd(h_{j\ell m }g_{\ell m }\rhd h_{jk\ell }),  l_{j\ell m n}^{-1}\,h_{j\ell n}\rhd'\{h_{\ell m n}, (g_{m n}g_{\ell m})\rhd h_{jk\ell }\}_{\mathrm{p}}\,\\ l_{jk\ell n}^{-1}(h_{jkn}\rhd'l_{k\ell m n}) l_{jkm n} h_{jm n}\rhd'(g_{m n}\rhd l_{jk\ell m }))\,.
\end{aligned}
\end{equation}
The obtained $3$-morphism is the identity morphism with source and target surface $\mathcal{V}_1=\mathcal{V}_2=h_{jm n}g_{m n}\rhd(h_{j\ell m }g_{\ell m }\rhd h_{jk\ell })$, \ie\,,
\begin{equation}
     l_{j\ell m n}^{-1}\,h_{j\ell n}\rhd'\{h_{\ell m n}, (g_{m n}g_{\ell m})\rhd h_{jk\ell }\}_{\mathrm{p}}\,l_{jk\ell n}^{-1}(h_{jkn}\rhd'l_{k\ell m n}) l_{jkm n} h_{jm n}\rhd'(g_{m n}\rhd l_{jk\ell m })=e\,.
\end{equation}

\end{lemma}

\section{Quantization of the topological \texorpdfstring{$3BF$}{3BF} theory}
%
\label{sec:discrete}
In conventional $BF$ theory, one chooses the action in such a way
that the theory does not depend on any background field, but only the spacetime
manifold. The classical field equations of the theory require the gauge connection to be flat, \ie\,, in terms of the holonomy variables, that any null-homotopic closed curve corresponds to the identity of the gauge group. In the framework of higher gauge theory, specifically $2$-gauge theory, one generalizes this idea by imposing the \emph{higher flatness condition} requiring that the surface holonomy around the boundary $2$-sphere of any $3$-ball be trivial instead. One can continue further categorical generalization by choosing a $3$-group structure to describe the gauge symmetry of the theory, and formulate a $3BF$ theory whose equations of motion impose a higher flatness condition for a $3$-curvature $(\cF,\cG, \cH)$. In this section, we present a combinatorial description of such a model for any triangulation of any smooth manifold of dimension $d=4$. This model coincides with Porter's abstract definition of a TQFT \cite{Porter98} for $d=4$ and $n=3$, which is itself a generalization of Yetter's work \cite{Y1, Y2}.

Let us show how to construct a state sum model from the classical action~\eqref{eq:bfcgdh} by the usual heuristic spinfoam quantization procedure. We consider the path integral for the action $S_{3BF}$,

\begin{equation}\label{eq:partition1}
\begin{aligned}
Z &= \int \mathcal{D} \alpha \,\mathcal{D} \beta\, \mathcal{D}\gamma \,\mathcal{D} B \,\mathcal{D} C\, \mathcal{D}D \, \exp\left( i\int_{M_4} \langle B\wedge \mathcal{F} \rangle{}_{\mathfrak{g}}  + \langle C\wedge \mathcal{G}\rangle_{\mathfrak{h}} + \langle D\wedge \mathcal{H}\rangle_{\mathfrak{l}} \right)\,.
\end{aligned}
\end{equation}
The formal integration over the Lagrange multipliers $B$, $C$, and $D$ leads to:
\begin{equation}\label{eq:partition111}
\begin{aligned}
Z &= \mathcal{N}\int \mathcal{D} \alpha \,\mathcal{D} \beta\, \mathcal{D}\gamma \ \delta(\cF)\delta(\cG)\delta(\cH)\,.
\end{aligned}
\end{equation}
Similarly to conventional gauge theory, the connection $1$-form $\alpha\in \cA^1(\cM_4,\mathfrak{g})$ is discretized by colouring the edges $\epsilon=(jk)\in \Lambda_1$ of the triangulation with group elements $g_\epsilon\in G$. The connection $2$-form $\beta \in \cA^2(\cM_4\,,\mathfrak{h})$ is represented by group elements $h_\Delta\in H$ coloring the triangles $\Delta=(jk\ell)\in\Lambda_2$. The connection $3$-form $\gamma \in \cA^3(\cM_4\,,\mathfrak{l})$ is represented by group elements $l_\tau\in L$ coloring the tetrahedrons $\tau=(jk\ell m)\in\Lambda_3$. 

The path integral measures of~\eqref{eq:partition1} are discretized by replacing
\begin{eqnarray}
\label{measures}
  \int\dd \alpha\quad&\mapsto&\quad \prod_{(jk)\in\Lambda_1}\int_G dg_{jk}\,,\\
\label{measures2}
  \int\dd\beta\quad &\mapsto&\quad \prod_{(jk\ell)\in\Lambda_2}\int_H dh_{jk\ell},\\
  \label{measures3}
  \int\dd\gamma\quad &\mapsto&\quad \prod_{(jk\ell m)\in\Lambda_3}\int_L dl_{jk\ell m}\,,
\end{eqnarray}
where $dg_{jk}$, $dh_{jk\ell}$, and $dl_{j k \ell m}$ denote integration with respect to the Haar measures of $G$, $H$, and $L$, respectively.
The vanishing fake curvature condition is discretized on each triangle $(jkl)\in \Lambda_2$ by discretizing $\delta(\mathcal{F})$. When passing from a smooth manifold to its triangulation, the $\delta$ distribution is defined over the appropriate set of simplices as follows,
\begin{equation}\label{id:g1}
    \delta(\mathcal{F}) = \prod_{(jk\ell)\in \Lambda_2} \delta_G( g_{jk\ell} )\,,
\end{equation}
where for each $(jkl) \in \Lambda_2$ the $\delta$-function $\delta_G( g_{jkl} )$ is given by:
\begin{equation}\label{id:g}
    \delta_G(g_{jk\ell})=\delta_G\big(\partial(h_{jk\ell})\,g_{k\ell}\,g_{jk}\,g_{j\ell}^{-1}\big)\,.
\end{equation}
Similarly, on the triangulated manifold the condition $\delta(\mathcal{G})$ on the fake curvature $3$-form reads
\begin{equation}\label{id:h1}
    \delta(\mathcal{G}) = \prod_{(jk\ell m)\in \Lambda_3} \delta_H( h_{jk\ell m} )\,,
\end{equation}
where for every tetrahedron $(jk\ell m)\in\Lambda_3$ one has:
\begin{equation}\label{id:h}
    \delta_H(h_{jk \ell m})=\delta_H\big(\delta(l_{jk\ell m})h_{j\ell m}\,(g_{\ell m }\rhd h_{jk\ell})\,h_{k\ell m}^{-1}\,h_{jkm}^{-1}\big)\,.
\end{equation}
Finally, the condition $\delta(\mathcal{H})$ is disretized as
\begin{equation}\label{id:l1}
    \delta(\mathcal{H}) = \prod_{(jk\ell mn)\in \Lambda_4} \delta_L( l_{jk\ell m n} )\,,
\end{equation}
where for each $4$-simplex $(jk\ell m n)\in\Lambda_4$ one has:
\begin{equation}\label{id:l}
\delta_L(l_{jk\ell m n})=\delta_L\big(l_{j\ell m n}^{-1}\,h_{j\ell n}\rhd'\{h_{\ell m n}, (g_{m n}g_{\ell m)}\rhd h_{jk\ell }\}_{\mathrm{p}}\,l_{jk\ell n}^{-1}(h_{jkn}\rhd'l_{k\ell m n}) l_{jkm n} h_{jm n}\rhd'(g_{m n}\rhd l_{jk\ell m })\big)\,.
\end{equation}
The identities \eqref{id:g}, \eqref{id:h}, and \eqref{id:l} are the results of Lemmas \ref{Th01}, \ref{Th02}, and \ref{Th:03}, respectively.

After substituting the expressions for discretized measures \eqref{measures}-\eqref{measures3} and $\delta$-functions \eqref{id:g1}, \eqref{id:h1}, and \eqref{id:l1} into the equation \eqref{eq:partition111} one obtains:
\begin{equation}
\label{eq:bfdisc}
\resizebox{0.99\hsize}{!}{$
\begin{aligned} 
Z&=& \mathcal{N}\prod_{(jk)\in\Lambda_1}\int\limits_G dg_{jk}\,
        \prod_{(jk\ell)\in\Lambda_2}\int\limits_H dh_{jk\ell}\,\prod_{(jk\ell m)\in\Lambda_3}\int\limits_L dl_{jk\ell m}\biggl(\prod_{(jk\ell)\in\Lambda_2}\delta_G\bigl(g_{j k \ell}\bigr)\biggr) \biggl(\prod_{(jk\ell m)\in\Lambda_3}\delta_H\bigl(h_{j k \ell m}\bigr)\biggr) \biggl(\prod_{(jk\ell mn)\in\Lambda_4}\delta_L\bigl(l_{j k \ell m n}\bigr)\biggr).
\end{aligned}$}
\end{equation}
By inserting (\ref{id:g}), (\ref{id:h}), and (\ref{id:l}) into (\ref{eq:bfdisc}), we obtain an explicit expression for the state sum over a given triangulation of the manifold $\mathcal{M}_4$. This expression can be made independent of the triangulation if one appropriately chooses the constant factor $\mathcal{N}$, obtained after the integration over the Lagrange multipliers $B$, $C$, and $D$. This is done by requiring that the state sum is invariant under the Pachner moves, which leads us to the appropriate form of the constant factor $\mathcal{N}$, as given by the definition \ref{def_statesum}.
\begin{definition}
\label{def_statesum}
Let $\cM_4$ be a compact and oriented combinatorial $d$-manifold,
$d=4$, and $(L\stackrel{\delta}{\to} H \stackrel{\partial}{\to}G\,, \rhd\,, \{\_\,,\_\}{}_{\mathrm{pf}})$ be a $2$-crossed module. The state sum of \emph{topological higher gauge theory} is defined by
\begin{equation}
\resizebox{0.97\hsize}{!}{$\begin{array}{lcl}
\label{eq_partition}
  Z &=& {|G|}^{-|\Lambda_0|+|\Lambda_1|-|\Lambda_2|}{|H|}^{|\Lambda_0|-|\Lambda_1|+|\Lambda_2|-|\Lambda_3|}\,{|L|}^{-|\Lambda_0|+|\Lambda_1|-|\Lambda_2|+|\Lambda_3|-|\Lambda_4|} \vphantom{\ds\int}\\&& \times\biggl(\prod_{(jk)\in\Lambda_1}\int\limits_G dg_{jk}\biggr)\,
        \biggl(\prod_{(jk\ell)\in\Lambda_2}\int\limits_H dh_{jk\ell}\biggr) \biggl(\prod_{(jk\ell m)\in\Lambda_3}\int\limits_L dl_{jk\ell m}\biggr)
  \vphantom{\ds\int} \\&&\times \biggl(\prod_{(jk\ell)\in\Lambda_2}\delta_G\bigl(\partial(h_{jk\ell})\,g_{k\ell}\,g_{jk}\,g_{j\ell}^{-1}\bigr)\biggr) \biggl(\prod_{(jk\ell m)\in\Lambda_3}\delta_H\bigl(\delta(l_{jk\ell m})h_{j\ell m}\,(g_{\ell m }\rhd h_{jk\ell})\,h_{k\ell m}^{-1}\,h_{jkm}^{-1}\bigr)\biggr)\vphantom{\ds\int}\\&&\times\biggl(\prod_{(jk\ell mn)\in\Lambda_4}\delta_L\bigl(l_{j\ell m n}^{-1}\,h_{j\ell n}\rhd'\{h_{\ell m n}, (g_{m n}g_{\ell m)}\rhd h_{jk\ell }\}_{\mathrm{p}}\,l_{jk\ell n}^{-1}(h_{jkn}\rhd'l_{k\ell m n}) l_{jkm n} h_{jm n}\rhd'(g_{m n}\rhd l_{jk\ell m })\bigr)\biggr)\,.\vphantom{\ds\int}\end{array}$}
\end{equation}
\end{definition}

Here we integrate over $g_{jk}\in G$ for every edge $(jk)\in\Lambda_1$, over $h_{jk\ell}\in H$ for every triangle $(jk\ell)\in\Lambda_2$ and over $l_{jklm}$ for every tetrahedron $(jk\ell m)\in\Lambda_3$ . The $\delta$-distributions under the integral
impose the following conditions. First, the condition that $\partial(h_{jk\ell})\,g_{k\ell}\,g_{jk}=g_{j\ell}$ for each triangle $(jk\ell)\in\Lambda_2$, \ie\,, that each surface label $h_{jk\ell}$ has got the appropriate source and target, see Lemma \ref{Th01}. Second, the condition that $h_{jkm}\,h_{k\ell m}=\delta(l_{jk\ell m})h_{j\ell m}\,(g_{\ell m }\rhd h_{jk\ell})$ for each tetrahedron $(jk\ell m)\in\Lambda_3$, \ie\,, that each volume label $l_{jk \ell m}$ has got the appropriate source and target, see Lemma \ref{Th02}. Finally, the condition that the volume holonomy around every $4$-simplex
$(jk\ell m n)\in\Lambda_4$ is trivial, \ie\,, that $l_{j\ell m n}^{-1}\,h_{j\ell n}\rhd'\{h_{\ell m n}, (g_{m n}g_{\ell m)}\rhd h_{jk\ell }\}_{\mathrm{p}}\,l_{jk\ell n}^{-1}(h_{jkn}\rhd'l_{k\ell m n}) l_{jkm n} h_{jm n}\rhd'(g_{m n}\rhd l_{jk\ell m })$ is equal to the neutral element of the group $L$ for each $4$-simplex $(jk\ell m n)\in\Lambda_4$, see Lemma \ref{Th:03}.

\begin{theorem}
Let ${\cal{M}}_4$ be a closed and oriented combinatorial $4$-manifold and $(L\stackrel{\delta}{\to} H \stackrel{\partial}{\to}G\,, \rhd\,, \{\_\,,\_\}{}_{\mathrm{pf}})$ be a $2$-crossed module. The state sum \eqref{eq_partition} is invariant under Pachner moves.
\end{theorem}

The statements of Pachner move invariance are formulated in the following subsections, while corresponding proofs are given in the Appendix \ref{app:pachner}.
\subsection{Pachner move \texorpdfstring{$1\leftrightarrow 5$}{1-5}}
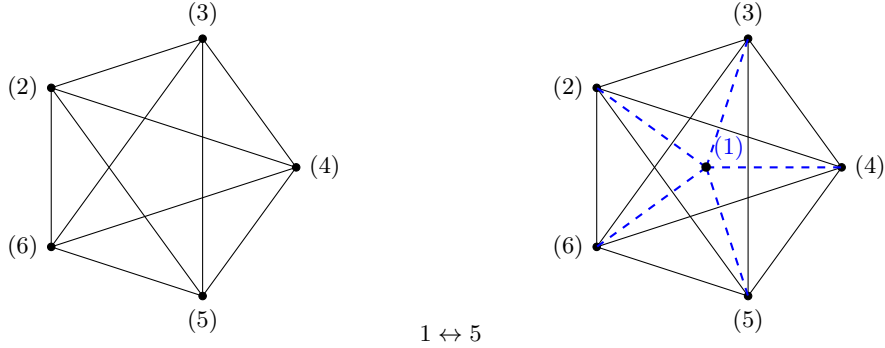
\begin{figure}[h]
\begingroup
\setlength{\tabcolsep}{8pt} 
\renewcommand{\arraystretch}{1.5}
\begin{tabular}{ccc}
\begin{tikzpicture}
   \newdimen\R
   \R=1.8cm
   \draw (0:\R) \foreach \x in {72,144,...,360} {  -- (\x:\R) };
   \foreach \x/\l/\p in
     { 72/{(3)}/above,
      144/{(2)}/left,
      216/{(6)}/left,
      288/{(5)}/below,
      360/{(4)}/right
    }
     \node[inner sep=1pt,circle,draw,fill,label={\p:\l}] at (\x:\R) {};
     \draw (0:\R) -- (144:\R);
     \draw (0:\R) -- (216:\R);
     \draw (72:\R) -- (216:\R);
     \draw (72:\R) -- (288:\R);
     \draw (144:\R) -- (288:\R);
\end{tikzpicture} &\;\;\;
$1\leftrightarrow 5$\;\;\; &
\begin{tikzpicture}
   \newdimen\R
   \R=1.8cm
   \draw (0:\R) \foreach \x in {72,144,...,360} {  -- (\x:\R) };
   \foreach \x/\l/\p in
     { 72/{(3)}/above,
      144/{(2)}/left,
      216/{(6)}/left,
      288/{(5)}/below,
      360/{(4)}/right
    }
     \node[inner sep=1pt,circle,draw,fill,label={\p:\l}] at (\x:\R) {};
     \draw (0:\R) -- (144:\R);
     \draw (0:\R) -- (216:\R);
     \draw (72:\R) -- (216:\R);
     \draw (72:\R) -- (288:\R);
     \draw (144:\R) -- (288:\R);
     \draw[blue,thick,dashed] (0,0) -- (144:\R);
     \draw[blue,thick,dashed] (0,0) -- (216:\R);
     \draw[blue,thick,dashed] (0,0) -- (72:\R);
     \draw[blue,thick,dashed] (0,0) -- (288:\R);
     \draw[blue,thick,dashed] (0,0) -- (0:\R);
     \node[label={[shift={(0.3,-0.2)},text = blue](1)}] {$\bullet$};
\end{tikzpicture} 
\end{tabular}
\endgroup
\end{figure}

Since the partition function (\ref{eq_partition}) is independent of the total order of vertices, we need to verify the move only in one case. Let us denote the vertices of the $4$-simplex on the left hand side of the $1\to 5$ Pachner move as $(23456)$. Adding a vertex $1$ on the right hand side of the Pachner move there are now five $4$-simplices $M_4= \{(13456),(12456),(12356),(12346),(12345)\}$. On the r.h.s. there are tetrahedrons $M_3=\{(1234),(1235),(1236),(1245),(1246),(1256),(1345),(1346),(1356),(1456)\}$, triangles $(jk\ell)\in
M_2=\{(123),(124),(125),(126),(134),(135),(136),(145),(146),(156)\}$, edges $(jk)\in M_1=\{(12),(13),(14),(15),(16)\}$ and vertices $(j)\in M_0=\{(1)\}$. All other simplices are present on both sides of the move.

If the $1\leftrightarrow 5$ Pachner move does not change the state sum (\ref{eq_partition}), then the state sum of the right hand side,
\begin{equation}\label{eq:51move}
\begin{aligned}
  Z_{\text{right}}^{1\leftrightarrow 5}=|G|^{-11}|H|^{-4}|L|^{-1}\int_{G^5}\prod_{(jk)\in M_1}dg_{jk}\,
  \int_{H^{10}}\prod_{(jk\ell)\in M_2}dh_{jk\ell}\int_{L^{10}} \prod_{(jklm)\in M_3} dl_{jklm}\\ \cdot
  \Biggl(\prod_{(jk\ell)\in M_2}\delta_G(g_{jk\ell})\Biggr)\,
  \Biggl(\prod_{(jk\ell m)\in M_3}\delta_H(h_{jk\ell m})\Biggr)\,\Biggl(\prod_{(jk\ell m n)\in M_4}\delta_L(l_{jk\ell mn})\Biggr)Z_\text{remainder}\,,
  \end{aligned}
\end{equation}
should be equal to the state sum of the left hand side, 
\begin{equation}
    Z_{\text{left}}^{1\leftrightarrow 5}=|G|^{-5} |H|^0 |L|^{-1} \delta_L(l_{23456})Z_\text{remainder}\,.
\end{equation}
Here, the prefactors $|G|^{-|\Lambda_0|+|\Lambda_1|-|\Lambda_2|}$, $ |H|^{|\Lambda_0|-|\Lambda_1|+|\Lambda_2|-|\Lambda_3|}$, and $|L|^{-|\Lambda_0|+|\Lambda_1|-|\Lambda_2|+|\Lambda_3|-|\Lambda_4|}$ are $|G|^{-11}|H|^{-4}|L|^{-1}$ on the r.h.s. and $|G|^{-5} |H|^0 |L|^{-1}$ on the l.h.s., as obtained by counting the numbers of the $k$-simplices on both sides of the $1\to 5$ move, shown in the Table \ref{Tabela15}. The $Z_\text{remainder}$ denotes the part of the state sum that is the same on both sides of the move, and thus irrelevant for the proof of invariance. The proof that $Z_{\rm left} = Z_{\rm right}$ is given in the Appendix A.

\begin{table}
    \centering
    \begin{tabular}{|c|c|c|c|c|c|}
    & $\quad|\Lambda_0|\quad$ & $\quad|\Lambda_1|\quad$ & $\quad|\Lambda_2|\quad$ & $\quad|\Lambda_3|\quad$ & $\quad|\Lambda_4|\quad$\\
         \hline
         l.h.s. & 5& 10& 10 & 5& 1\\
         r.h.s. &6 & 15& 20 & 15 & 5
    \end{tabular}
      \caption{Number of vertices $|\Lambda_0|$, edges $|\Lambda_1|$, triangles $|\Lambda_2|$, tetrahedrons $|\Lambda_3|$, and 4-simplices $|\Lambda_4|$ on both sides of the $1\leftrightarrow5$ move.\label{Tabela15}}
\end{table}

\subsection{Pachner move \texorpdfstring{$2\leftrightarrow 4$}{2-4}}

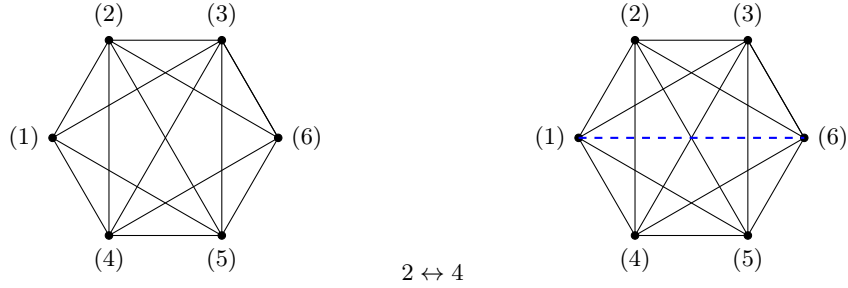
\begin{figure}[ht]
\begingroup
\setlength{\tabcolsep}{8pt} 
\renewcommand{\arraystretch}{1.5}
\begin{tabular}{ccc}
\begin{tikzpicture}
   \newdimen\R
   \R=1.5cm
   \draw (0:\R) \foreach \x in {60,120,...,360} {  -- (\x:\R) };
   \foreach \x/\l/\p in
     { 60/{(3)}/above,
      120/{(2)}/above,
      180/{(1)}/left,
      240/{(4)}/below,
      300/{(5)}/below,
      360/{(6)}/right
     }
     \node[inner sep=1pt,circle,draw,fill,label={\p:\l}] at (\x:\R) {};
     \draw (120:\R) -- (240:\R);
     \draw (120:\R) -- (300:\R);
     \draw (60:\R) -- (240:\R);
     \draw (60:\R) -- (300:\R);
     \draw (0:\R) -- (240:\R);
     \draw (0:\R) -- (120:\R);
     \draw (0:\R) -- (60:\R);
     \draw (180:\R) -- (60:\R);
     \draw (180:\R) -- (300:\R);
\end{tikzpicture} &\;\;\;
$2\leftrightarrow 4$\;\;\; &
\begin{tikzpicture}
   \newdimen\R
   \R=1.5cm
   \draw (0:\R) \foreach \x in {60,120,...,360} {  -- (\x:\R) };
   \foreach \x/\l/\p in
     { 60/{(3)}/above,
      120/{(2)}/above,
      180/{(1)}/left,
      240/{(4)}/below,
      300/{(5)}/below,
      360/{(6)}/right
     }
     \node[inner sep=1pt,circle,draw,fill,label={\p:\l}] at (\x:\R) {};
     \draw (120:\R) -- (240:\R);
     \draw (120:\R) -- (300:\R);
     \draw (60:\R) -- (240:\R);
     \draw (60:\R) -- (300:\R);
     \draw (0:\R) -- (240:\R);
     \draw (0:\R) -- (120:\R);
     \draw (0:\R) -- (60:\R);
     \draw (180:\R) -- (60:\R);
     \draw (180:\R) -- (300:\R);
     \draw[blue,thick,dashed] (180:\R) -- (0:\R);
\end{tikzpicture}
\end{tabular}
\endgroup
\end{figure}
We order the vertices in such a way that on the l.h.s. of the move we have the $4$-simplices $(23456)$ and $(12345)$, while on the r.h.s. we have the $4$-simplices
$(12346)$, $(12356)$, $(12456)$,  and $(13456)$. On the l.h.s. we have one tetrahedron $(2345)$, whereas on the r.h.s. there are six tetrahedrons $(1236)$, $(1246)$, $(1256)$, $(1346)$, $(1356)$, and $(1456)$. All other tetrahedrons appear on both sides of the move. On the r.h.s. there are triangles $(126)$, $(136)$, $(146)$, and $(156)$ and edge $(16)$, while the rest of the triangles and edges appear on both sides of the move.

On the l.h.s. there is the state sum,
\begin{equation}
  Z_{\mathrm{left}}^{2\leftrightarrow 4}=|G|^{-8}|H|^{-1}|L|^{-1}\int_{L} dl_{2345}\delta_H(h_{2345})\Biggl(\prod_{(jk\ell m n)\in M_4}\delta_L(l_{jk\ell mn})\Biggr)Z_{\text{remainder}}\,,
\end{equation}
where $M_4=\{(23456),(12345)\}$,
whereas on the r.h.s. the state sum reads 
\begin{equation}\label{eq:42rhs}
\begin{aligned}
   Z_{\mathrm{right}}^{2\leftrightarrow 4}=|G|^{-11}|H|^{-3}|L|^{-1}\int_G dg_{16}\,\int_{H^4}dh_{126}dh_{136}dh_{146}dh_{156}\int_{L} dl_{1236}dl_{1246}dl_{1256}dl_{1346}dl_{1356}dl_{1456}\\
  \Biggl(\prod_{(jk\ell)\in M_2}\delta_G(g_{jk\ell})\Biggr)\,
  \Biggl(\prod_{(jk\ell m)\in M_3}\delta_H(h_{jk\ell m})\Biggr)\Biggl(\prod_{(jk\ell m n)\in M_4}\delta_L(l_{jk\ell mn})\Biggr)Z_{\text{remainder}},
\end{aligned}
\end{equation}
where $M_2=\{(126),(136),(146),(156)\}$,
$M_3=\{(1236),(1246),(1256),(1346),(1356),(1456)\}$, and $M_4=\{(12346),(12356),(12456),(13456)\}$.

Here the prefactors $  |G|^{-8}|H|^{-1}|L|^{-1}$ on the l.h.s. and $  |G|^{-11}|H|^{-3}|L|^{-1}$ on the r.h.s. are obtained by counting the numbers of $k$-simplices on both sides of the $2\leftrightarrow 4$ move, as shown in the Table \ref{Tabela24}. The term $Z_{\text{remainder}}$ denotes the part of the state sum that is identical on both sides of the move, as before. The proof that $Z_{\mathrm{left}} = Z_{\mathrm{right}}$ is given in the Appendix A.

\begin{table}[h!]
    \centering
    \begin{tabular}{|c|c|c|c|c|c|}
         & $\quad|\Lambda_0|\quad$ & $\quad|\Lambda_1|\quad$ & $\quad|\Lambda_2|\quad$ & $\quad|\Lambda_3|\quad$ & $\quad|\Lambda_4|\quad$\\
         \hline
         l.h.s. & 6& 14& 16 & 9& 2\\
         r.h.s. &6 & 15& 20 & 14 & 4\\
    \end{tabular}
    \caption{\label{Tabela24}Number of vertices $|\Lambda_0|$, edges $|\Lambda_1|$, triangles $|\Lambda_2|$, tetrahedrons $|\Lambda_3|$, and 4-simplices $|\Lambda_4|$ on both sides of the $2\leftrightarrow4$ move.}
\end{table}

\subsection{Pachner move \texorpdfstring{$3 \leftrightarrow 3$}{3-3}}

\begin{figure}[h]
\begingroup
\setlength{\tabcolsep}{8pt} 
\renewcommand{\arraystretch}{1.5}
\begin{tabular}{ccc}
\begin{tikzpicture}
   \newdimen\R
   \R=1.5cm
   \draw (0:\R) \foreach \x in {60,120,...,360} {  -- (\x:\R) };
   \foreach \x/\l/\p in
     { 60/{(2)}/above,
      120/{(4)}/above,
      180/{(1)}/left,
      240/{(6)}/below,
      300/{(3)}/below,
      360/{(5)}/right
     }
     \node[inner sep=1pt,circle,draw,fill,label={\p:\l}] at (\x:\R) {};
     \fill[fill=gray!20] (120:\R)--(0:\R)--(240:\R);
     \draw (120:\R) -- (240:\R);
     \draw (120:\R) -- (300:\R);
     \draw (60:\R) -- (240:\R);
     \draw (60:\R) -- (300:\R);
     \draw (0:\R) -- (240:\R);
     \draw (0:\R) -- (120:\R);
     \draw (180:\R) -- (60:\R);
     \draw (180:\R) -- (300:\R);
     \draw (180:\R) -- (0:\R);
\end{tikzpicture} &\;\;\;
$3\leftrightarrow 3$\;\;\; &
\begin{tikzpicture}
   \newdimen\R
   \R=1.5cm
   \draw (0:\R) \foreach \x in {60,120,...,360} {  -- (\x:\R) };
   \foreach \x/\l/\p in
     { 60/{(2)}/above,
      120/{(4)}/above,
      180/{(1)}/left,
      240/{(6)}/below,
      300/{(3)}/below,
      360/{(5)}/right
     }
     \node[inner sep=1pt,circle,draw,fill,label={\p:\l}] at (\x:\R) {};
     \fill[fill=gray!20] (60:\R)--(180:\R)--(300:\R);
     \draw (120:\R) -- (240:\R);
     \draw (120:\R) -- (300:\R);
     \draw (60:\R) -- (240:\R);
     \draw (60:\R) -- (300:\R);
     \draw (0:\R) -- (240:\R);
     \draw (0:\R) -- (120:\R);
     \draw (180:\R) -- (60:\R);
     \draw (180:\R) -- (300:\R);
     \draw (180:\R) -- (0:\R);
\end{tikzpicture}
\end{tabular}
\endgroup
\end{figure}
We order the vertices in such a way that on the l.h.s.\ of the $3\leftrightarrow 3$ move, we have the $4$-simplices $(23456)$, $(13456)$, and
$(12456)$, whereas on the r.h.s. we have the $4$-simplices $(12356)$, $(12346)$, and $(12345)$.
Six tetrahedrons therefore form the common boundary of both sides of the move
whereas on each side there are three tetrahedrons shared by two $4$-simplices.
On the l.h.s.\ these are $(1456)$, $(2456)$, and $(3456)$ and on the r.h.s.\
$(1234)$, $(1235)$, and $(1236)$. On the l.h.s\ we therefore have the triangle
$(456)$ and on the r.h.s\ $(123)$. All other triangles appear on both sides of
the move.

Therefore on the l.h.s. there is the state sum,
\begin{equation}
  Z_{\mathrm{left}}^{3\leftrightarrow 3}= \int_Hdh_{456}\int_{L^3}dl_{1456}dl_{2456}dl_{3456}\delta_G(g_{456})\,\delta_H(h_{3456})\delta_H(h_{2456})\delta_H(h_{1456})\delta_L(l_{23456})\delta_L(l_{13456})\delta_L(l_{12456})Z_{\text{remainder}}\,,
\end{equation}
whereas on the r.h.s. the state sum reads
\begin{equation}
   Z_{\mathrm{right}}^{3\leftrightarrow 3}=\int_H dh_{123}\int_{L^3}dl_{1234}dl_{1235}dl_{1236}\delta_G(g_{123})\,\delta_H(h_{1234})\delta_H(h_{1235})\delta_H(h_{1236})\delta_L(l_{12356})\delta_L(l_{12346})\delta_L(l_{12345})Z_{\text{remainder}}\,.
\end{equation}
The numbers of $k$-simplices agree on both sides of the $3\leftrightarrow 3$ move for all
$k$, and the prefactors play no role in this case, so they are part of the $Z_{\mathrm{remainder}}$. The proof that $Z_{\mathrm{left}} = Z_{\mathrm{right}}$ is given in the Appendix A.

We conclude that the state sum given by the definition \ref{def_statesum} is invariant under all three Pachner moves, and thus independent of triangulation of the underlying $4$-dimensional manifold.

%
\section{Conclusions}
Let us summarize the results of the paper. In Section \ref{sec:lagrange} we reviewed the pure the constrained $2BF$ actions describing the Yang-Mills field and Einstein-Cartan gravity, and constrained $3BF$ actions describing the Klein-Gordon and Dirac fields coupled to Yang-Mills fields and gravity in the standard way. In Section~\ref{sec:preliminaries}, we reviewed the relevant algebraic tools involved in the description of higher gauge theory,  $2$-crossed modules, and $3$-gauge theory and generalized the integral picture of an ordinary gauge theory to a $3$-gauge theory that involves curves, surfaces, and volumes labeled with elements of non-Abelian groups. We have also proved three key results, stated in Lemmas III.1, III.2, and III.3, which are crucial for the construction of the invariant state sum. In Section~\ref{sec:discrete}, we have presented the two main results of the paper. First, we constructed a triangulation independent state sum $Z$ of a topological higher gauge theory for a general $3$-group and a $4$-dimensional spacetime manifold $\cM_4$. Second, we proved the theorem that the constructed state sum is indeed independent of the choice of triangulation, i.e., that it is a genuine topological invariant.

The constructed state sum coincides with the Porter’s TQFT \cite{Porter98, Porter96} for $d = 4$ and $n = 3$. The proof that the state sum is invariant under the local changes of triangulation called the Pachner moves and thus independent of the chosen triangulation is presented in Appendix~\ref{app:pachner}. It is obtained that the state sum is invariant under all five different Pachner moves: the $3-3$ move, $4-2$ move, and $5-1$ move, and their inverses. The state sum constructed this way can be thought of as a combinatorial construction of a topological quantum field theory (TQFT) in the sense of Atiyah’s axioms, a topic that is beyond the scope of this paper and will be studied in a future work.

In order to finish the second step of the spinfoam quantization procedure, however, the generalizations of the Peter-Weyl and Plancharel theorems to $2$-groups and $3$-groups are required, which so far represent open problems. Namely, these theorems should provide a decomposition of a function on a $3$-group into a sum over the corresponding irreducible representations of a $3$-group. In this way, the spectrum of labels for the simplices, \ie\,, the domain of values of the fields living on the simplices of the triangulation, would be specified. Nonetheless, one can still try to guess the irreducible representations of $3$-groups, as was done for example in the case of $2$-groups in the spincube model of quantum gravity \cite{MikovicVojinovic2012}, or obtain the state sum using other techniques, see for example \cite{Quantum, BaratinFreidel, Tsimiklis}).

However, if one wants to describe a real physical theory, \ie\,, the theory which contains local propagating degrees of freedom, one needs to construct the nontopological state sum, with the non-trivial dynamics. To do so, once the topological state sum is constructed, the final third step of the spinfoam quantization procedure is to impose the constraints that deform the topological theory into a realistic theory of gravity coupled to matter fields (as defined in \cite{Radenkovic2019}) at the quantum level. We leave the construction of the constrained state sum model for future work.

In addition to the above topics, there are also many other possible applications of the invariant state sum, both in physics and mathematics.

\acknowledgments
This research was supported by the Ministry of Education, Science and Technological Development of the Republic of Serbia (MPNTR), and by the Science Fund of the Republic of Serbia, Program DIASPORA, No. 6427195, SQ2020. The contents of this publication are the sole responsibility of the authors and can in no way be taken to reflect the views of the Science Fund of the Republic of Serbia.

\appendix
%
\section{Proof of Pachner move invariance}
%
\label{app:pachner}

Here we give a self contained proof in terms of Pachner
moves that the partition function~\eqref{eq_partition} is independent
of the chosen triangulation.

\subsection{Pachner move \texorpdfstring{$1 \leftrightarrow 5$}{1-5}}
On the {\emph{left hand side of the move}} there is the integrand $\delta_L(l_{23456})$:
\begin{equation}
\delta_L(l_{23456})=\delta_L\big( l_{2346}{}^{-1}(h_{236}\rhd'l_{3456}) l_{2356} h_{256}\rhd'(g_{56}\rhd l_{2345})l_{2456}{}^{-1}h_{246}\rhd'\{h_{456}, (g_{56}g_{45})\rhd h_{234}\}_{\mathrm{p}}\big).
\end{equation}
Let us examine the {\emph{right hand side of the move}}, given by the equation (\ref{eq:51move}). First, one integrates out $g_{12}$ using $\delta_G{(g_{123})}$, $g_{13}$ using $\delta_G{(g_{134})}$, $g_{14}$ using $\delta_G{(g_{145})}$, and $g_{15}$ using $\delta_G{(g_{156})}$, and obtains:
\begin{equation}
\begin{aligned}
g_{12}&=& \,g_{23}^{-1}\,\partial(h_{123})^{-1}\,g_{13}\,,\\
g_{13}&=& \,g_{34}^{-1}\,\partial(h_{134})^{-1}\,g_{14}\,,\\g_{14}&=& \,g_{45}^{-1}\,\partial(h_{145})^{-1}\,g_{15}\,,\\
g_{15}&=& \,g_{56}^{-1}\,\partial(h_{156})^{-1}\,g_{16}\,.
\end{aligned}
\end{equation}
One integrates out $h_{123}$ using $\delta_H{(h_{1234})}$, $h_{124}$ using $\delta_H{(h_{1245})}$, $h_{125}$ using $\delta_H{(h_{1256})}$, $h_{134}$ using $\delta_H{(h_{1345})}$, $h_{135}$ using $\delta_H{(h_{1356})}$, and $h_{145}$ using $\delta_H{(h_{1456})}$, and obtains:
\begin{equation}
\begin{aligned}
h_{123}&=& g_{34}^{-1}\rhd h_{134}^{-1}\,g_{34}^{-1}\rhd \delta(l_{1234})^{-1}\,g_{34}^{-1}\rhd h_{124}\,g_{34}^{-1}\rhd h_{234}\,,\\
h_{124}&=& g_{45}^{-1}\rhd h_{145}^{-1}\,g_{45}^{-1}\rhd \delta(l_{1245})^{-1}\,g_{45}^{-1}\rhd h_{125}\,g_{45}^{-1}\rhd h_{245}\,,\\
 h_{125}&=& g_{56}^{-1}\rhd h_{156}^{-1}\,g_{56}^{-1}\rhd \delta(l_{1256})^{-1}\,g_{56}^{-1}\rhd h_{126}\,g_{56}^{-1}\rhd h_{256}\,,\\
  h_{134}&=& g_{45}^{-1}\rhd h_{145}^{-1}\,g_{45}^{-1}\rhd \delta(l_{1345})^{-1}\,g_{45}^{-1}\rhd h_{135}\,g_{45}^{-1}\rhd h_{345}\,,\\
   h_{135}&=& g_{56}^{-1}\rhd h_{156}^{-1}\,g_{56}^{-1}\rhd \delta(l_{1356})^{-1}\,g_{56}^{-1}\rhd h_{136}\,g_{56}^{-1}\rhd h_{356}\,,\\
    h_{145}&=& g_{56}^{-1}\rhd h_{156}^{-1}\,g_{56}^{-1}\rhd \delta(l_{1456})^{-1}\,g_{56}^{-1}\rhd h_{146}\,g_{56}^{-1}\rhd h_{456}\,.
\end{aligned}
\end{equation}
The $\delta$-functions on the group $G$ now read $\delta_G(e)^6$. Let us show this. First, for $\delta_G(g_{124})$ one obtains
\begin{equation}
\begin{aligned}
    \delta_G(g_{124})&=&\delta_G\big(\partial(h_{124})\,g_{24}\,g_{12}\,g_{14}^{-1}\big)\\&=&\delta_G\big(\partial(h_{124})\,g_{24}\,g_{23}^{-1}\,\partial(h_{123})^{-1}\,g_{13}\,g_{14}^{-1}\big)\\&=&\delta_G\big(\partial(h_{124})\,g_{24}\,g_{23}^{-1}\,g_{34}^{-1}\,\partial(h_{234})^{-1} \partial(h_{124})^{-1} \partial(h_{134})\,g_{34}\,g_{13}\,g_{14}^{-1}\big)\,\\&=&\delta_G\big(\partial(h_{124})\,g_{24}\,g_{23}^{-1}\,g_{34}^{-1}\,(g_{34}\,g_{23}^{-1}\,g_{24}^{-1})\, \partial(h_{124})^{-1}\, e\big)\\&=&\delta_G(e)\,,
    \end{aligned}
\end{equation}
Next, for $\delta$-function $\delta_G(g_{125})$ one obtains,
\begin{equation}
\begin{aligned}
    \delta_G(g_{125})&=&\delta_G\big(\partial(h_{125})\,g_{25}\,g_{12}\,g_{15}^{-1}\big)\,,\\&=&\delta_G\big(\partial(h_{125})\,g_{25}\,g_{23}^{-1}\,\partial(h_{123})^{-1}\,g_{13}\,g_{15}^{-1}\big)\\&=&\delta_G\big(\partial(h_{125})\,g_{25}\,g_{23}^{-1}\,g_{34}^{-1}\,\partial(h_{234})^{-1} \partial(h_{124})^{-1} \partial(h_{134})\,g_{34}\,g_{13}\,g_{15}^{-1}\big)\\&=&\delta_G\big(\partial(h_{125})\,g_{25}\,g_{23}^{-1}\,g_{34}^{-1}\,\partial(h_{234})^{-1} g_{45}^{-1}(\partial(h_{245})^{-1}\partial(h_{125})^{-1}\partial(h_{145}))\,g_{45} g_{14}\,g_{15}^{-1}\big)\,\\&=&\delta_G\big(\partial(h_{125})\,g_{25}\,g_{23}^{-1}\,g_{34}^{-1}\,(g_{34}\,g_{23}^{-1}\,g_{24}^{-1}) g_{45}^{-1}(g_{45}\,g_{24}^{-1}\,g_{25}^{-1})\partial(h_{125})^{-1} e\big)\,\\&=&\delta_H(e)\,.
    \end{aligned}
\end{equation}
Similarly, $ \delta_G(g_{126})$ becomes
\begin{equation}
\thinmuskip=0mu
\begin{aligned}
    \delta_G(g_{126})&=&\delta_G\big(\partial(h_{126})\,g_{26}\,g_{12}\,g_{16}^{-1}\big)\,\\&=&\delta_G\big(\partial(h_{126})\,g_{26}\,g_{23}^{-1}\,\partial(h_{123})^{-1}\,g_{13}\,g_{16}^{-1}\big)\\&=&\delta_G\big(\partial(h_{126})\,g_{26}\,g_{23}^{-1}\,g_{34}^{-1}\,\partial(h_{234})^{-1} \partial(h_{124})^{-1} \partial(h_{134})\,g_{34}\,g_{13}\,g_{16}^{-1}\big)\,\\&=&\delta_G\big(\partial(h_{126})\,g_{26}\,g_{23}^{-1}\,g_{34}^{-1}\,\partial(h_{234})^{-1} g_{45}^{-1}(\partial(h_{245})^{-1}\partial(h_{125})^{-1}\partial(h_{145}))\,g_{45} \partial(h_{134})\,g_{34}\,g_{13}\,g_{16}^{-1}\big)\,\\&=&\delta_G\big(\partial(h_{126})\,g_{26}\,g_{23}^{-1}\,g_{34}^{-1}\,\partial(h_{234})^{-1} g_{45}^{-1}(\partial(h_{245})^{-1}g_{56}^{-1}\partial(h_{256})^{-1}\partial(h_{126})^{-1}\partial(h_{156})g_{56}\\&&\,\partial(h_{145}))\,g_{45}g_{14}\,g_{16}^{-1}\big)\,\\&=&\delta_G\big(\partial(h_{126})\,g_{26}\,g_{23}^{-1}\,g_{34}^{-1}\,(g_{34}\,g_{23}^{-1}\,g_{24}^{-1}) g_{45}^{-1}(g_{45}\,g_{24}^{-1}\,g_{25}^{-1})g_{56}^{-1}(g_{56}\,g_{25}^{-1}\,g_{26}^{-1})\partial(h_{126})^{-1}\\&&(g_{16}\,g_{15}^{-1}\,g_{56}^{-1})g_{56}\,g_{15}\,g_{16}^{-1}\big)\,\\&=&\delta_G(e)\,,
    \end{aligned}
\end{equation}
and $\delta_G(g_{135})$ now reads,
\begin{equation}
\begin{aligned}
    \delta_G(g_{135})&=&\delta_G\big(\partial(h_{135})\,g_{35}\,g_{13}\,g_{15}^{-1}\big)\,,\\
    &=&\delta_G\big(\partial(h_{135})\,g_{35}\,g_{34}^{-1}\,\partial(h_{134})^{-1}\,g_{14}\,g_{15}^{-1}\big)\\
    &=&\delta_G\big(\partial(h_{135})\,g_{35}\,g_{34}^{-1}\,g_{45}^{-1}\,\partial(h_{345})^{-1} \partial(h_{135})^{-1} \partial(h_{145})\,g_{45}\,g_{14}\,g_{15}^{-1}\big)\\
    &=&\delta_G\big(\partial(h_{135})\,g_{35}\,g_{34}^{-1}\,g_{45}^{-1}\,\partial(h_{345})^{-1}\partial(h_{135})^{-1}\,\partial(h_{145})\,g_{45} \, g_{45}^{-1}\,\partial(h_{145})^{-1}\,g_{15}\,g_{15}^{-1}\big)\,\\
    &=&\delta_G\big(\partial(h_{135})\,g_{35}\,g_{34}^{-1}\,g_{45}^{-1}\,(g_{45}\,g_{34}^{-1}\,g_{35}^{-1})\partial(h_{135})^{-1}\big)\,\\&=&\delta_G(e)\,,
    \end{aligned}
\end{equation}
while $\delta_G(g_{136})$ reads:
\begin{equation}
\begin{aligned}
    \delta_G(g_{136})&=&\delta_G\big(\partial(h_{136})\,g_{36}\,g_{13}\,g_{16}^{-1}\big)\,\\&=&\delta_G\big(\partial(h_{136})\,g_{36}\,g_{34}^{-1}\,\partial(h_{134})^{-1}\,g_{14}\,g_{16}^{-1}\big)\\
    &=&\delta_G\big(\partial(h_{136})\,g_{36}\,g_{34}^{-1}\,g_{45}^{-1}\,\partial(h_{345})^{-1} \partial(h_{135})^{-1} \partial(h_{145})\,g_{45}\,g_{14}\,g_{16}^{-1}\big)\\
    &=&\delta_G\big(\partial(h_{136})\,g_{36}\,g_{34}^{-1}\,g_{45}^{-1}\,\partial(h_{345})^{-1} g_{56}^{-1}(\partial(h_{356})^{-1}\partial(h_{136})^{-1}\partial(h_{156}))\,g_{56} \partial(h_{145})\,g_{45}\,g_{14}\,g_{16}^{-1}\big)\,\\
    &=&\delta_G\big(\partial(h_{136})\,g_{36}\,g_{34}^{-1}\,g_{45}^{-1}\,(g_{45}\,g_{34}^{-1}\,g_{35}^{-1}) g_{56}^{-1}(g_{56}\,g_{35}^{-1}\,g_{36}^{-1})\partial(h_{136})^{-1}e\big)\,\\&=&\delta_G(e)\,.
    \end{aligned}
\end{equation}
Finally, the $\delta$-function $\delta_G(g_{146})$ reads: 
\begin{equation}
\begin{aligned}
    \delta_G(g_{146})&=&\delta_G\big(\partial(h_{146})\,g_{46}\,g_{14}\,g_{16}^{-1}\big)\,\\&=&\delta_G\big(\partial(h_{146})\,g_{46}\,(g_{45}^{-1}\,\partial(h_{145})^{-1}\,g_{15})\,g_{16}^{-1}\big)\\&=&\delta_G\big(\partial(h_{146})\,g_{46}\,g_{45}^{-1}\,\partial(h_{145})^{-1}\,(g_{56}^{-1}\,\partial(h_{156})^{-1}\,g_{16})\,g_{16}^{-1}\big)\\&=&\delta_G\big(\partial(h_{146})\,g_{46}\,g_{45}^{-1}\,g_{56}^{-1}\partial(h_{456})^{-1}\partial(h_{146})^{-1}\partial(h_{156})g_{56}\,(g_{56}^{-1}\,\partial(h_{156})^{-1}\,g_{16})\,g_{16}^{-1}\big)\\&=&\delta_G(e)\,.
\end{aligned}
\end{equation}
Next, one integrates out $l_{1235}$ using $\delta_L(l_{12345})$, $l_{1236}$ using $\delta_L(l_{12346})$, $l_{1246}$ using $\delta_L(l_{12456})$, and $l_{1346}$ using $\delta_L(l_{13456})$, and obtains
\begin{equation}\label{l:1235}
    l_{1235}= (h_{125}\rhd'l_{2345})l_{1245} h_{145}\rhd'(g_{45}\rhd l_{1234})l_{1345}^{-1}\,h_{135}\rhd'\{h_{345}, (g_{45}g_{34})\rhd h_{123}\}_{\mathrm{p}}\,,
\end{equation}
\begin{equation}\label{l:1236'}
     l_{1236} = (h_{126}\rhd'l_{2346}) l_{1246} h_{146}\rhd'(g_{46}\rhd l_{1234})l_{1346}^{-1}\,h_{136}\rhd'\{h_{346}, (g_{46}g_{34})\rhd h_{123}\}_{\mathrm{p}}\,,
\end{equation}
\begin{equation}\label{l:1246'}
    l_{1246}= (h_{126}\rhd'l_{2456}) l_{1256} h_{156}\rhd'(g_{56}\rhd l_{1245})l_{1456}{}^{-1}\,h_{146}\rhd'\{h_{456}, (g_{56}g_{45})\rhd h_{124}\}_{\mathrm{p}}\,,
\end{equation}
\begin{equation}\label{l:1346'}
    l_{1346}= (h_{136}\rhd'l_{3456}) l_{1356} h_{156}\rhd'(g_{56}\rhd l_{1345})l_{1456}{}^{-1}\,h_{146}\rhd'\{h_{456}, (g_{56}g_{45})\rhd h_{134}\}_{\mathrm{p}}\,.
\end{equation}
Let us now show that the remaining $\delta$-functions on the group $H$ equal $\delta_H(e)^4$.
First, $\delta_H(h_{1235})$ becomes:
\begin{equation}
\medmuskip=0mu
\thinmuskip=0mu
\thickmuskip=0mu
\resizebox{0.96\hsize}{!}{$
\begin{aligned}
    \delta_H(h_{1235})&=&\delta_H\big(\delta(l_{1235})h_{135}\,(g_{35}\rhd h_{123})\,h_{235}^{-1}\,h_{125}^{-1}\big)\\
    &=&\delta_H\Big(\delta\big(({h_{125}}\rhd'l_{2345})l_{1245} h_{145}\rhd'(g_{45}\rhd l_{1234})l_{1345}^{-1}\,h_{135}\rhd'\{h_{345}, (g_{45}g_{34})\rhd h_{123}\}_{\mathrm{p}}\big)h_{135}\,(g_{35}\rhd h_{123})\,h_{235}^{-1}\,h_{125}^{-1}\Big)\,\\
    &=&\delta_H\Big(\big(h_{125}\delta(l_{2345})h_{125}^{-1}\delta(l_{1245}) h_{145}(g_{45}\rhd \delta(l_{1234}))h_{145}^{-1}\delta(l_{1345})^{-1}\,h_{135}\delta(\{h_{345}, (g_{45}g_{34})\rhd h_{123}\}_{\mathrm{p}})h_{135}^{-1}\big)\\&&h_{135}\,(g_{35}\rhd h_{123})\,h_{235}^{-1}\,h_{125}^{-1}\Big)\,\\
    &=&\delta_H\Big( h_{235} \,h_{345} (g_{45}\rhd h_{234}^{-1})\, h_{245}^{-1}h_{125}^{-1} h_{125} \,h_{245} (g_{45}\rhd h_{124}^{-1})\, h_{145}^{-1} h_{145}(g_{45}\rhd ( h_{124} \,h_{234} (g_{34}\rhd h_{123}^{-1})\, h_{134}^{-1}))\\&&h_{145}^{-1}(h_{145}\,(g_{45}\rhd h_{134})\,h_{345}^{-1}\,h_{135}^{-1})h_{135}\delta(\{h_{345}, (g_{45}g_{34})\rhd h_{123}\}_{\mathrm{p}})h_{135}^{-1}h_{135}\,(g_{35}\rhd h_{123})\,h_{235}^{-1}\,\Big)\,\\&=&\delta_H(h_{345}\big((g_{45}g_{34})\rhd h_{123}^{-1}\big)h_{345}^{-1}\delta(\{h_{345}, (g_{45}g_{34})\rhd h_{123}\}_{\mathrm{p}})(g_{35}\rhd h_{123})\,.
    \end{aligned}$}
\end{equation}
Here, one uses the following identity
\begin{equation}\label{eq:Pfajferovkomutator2}
   \delta\{ h_1\,,h_2 \}_{\mathrm{p}}(\partial(h_1) \rhd h_2)h_1h_2^{-1}h_1^{-1}=e \,.
\end{equation}
Substituting $g_{35}=\partial(h_{345})g_{45}g_{34}$, and applying the (\ref{eq:Pfajferovkomutator2}) identity for $h_1=h_{345}$ and $h_2=(g_{45}g_{34})\rhd h_{123}$, one obtains
\begin{equation}
\begin{aligned}
    \delta_H(h_{1235})=\delta_H(e).
    \end{aligned}
\end{equation}
Similarly, one obtains for $\delta_H(h_{1236})$:
\begin{equation}
\medmuskip=0mu
\thinmuskip=0mu
\thickmuskip=0mu
\resizebox{0.96\hsize}{!}{$
\begin{aligned}
    \delta_H(h_{1236})&=&\delta_H\big(\delta(l_{1236})h_{136}\,(g_{36}\rhd h_{123})\,h_{236}^{-1}\,h_{126}^{-1}\big)\,\\
    &=&\delta_H\Big(\delta\big(({h_{126}}\rhd'l_{2346})l_{1246} h_{146}\rhd'(g_{46}\rhd l_{1236}l_{1346}^{-1}\,h_{136}\rhd'\{h_{346}, (g_{46}g_{34})\rhd h_{123}\}_{\mathrm{p}}\big)h_{136}\,(g_{36}\rhd h_{123})\,h_{236}^{-1}\,h_{126}^{-1}\Big)\,\\
    &=&\delta_H\Big(\big(h_{126}\delta(l_{2346})h_{126}^{-1}\delta(l_{1246}) h_{146}(g_{46}\rhd \delta(l_{1234}))h_{146}^{-1}\delta(l_{1346})^{-1}\,h_{136}\delta(\{h_{346}, (g_{46}g_{34})\rhd h_{123}\}_{\mathrm{p}})h_{136}^{-1}\big)\\&&h_{136}\,(g_{36}\rhd h_{123})\,h_{236}^{-1}\,h_{126}^{-1}\Big)\,\\
    &=&\delta_H\Big(h_{236} \,h_{346} (g_{46}\rhd h_{234}^{-1})\, h_{246}^{-1}h_{126}^{-1}{h_{126} \,h_{246} (g_{46}\rhd h_{124}^{-1})}\, h_{146}^{-1} h_{146}(g_{46}\rhd ({ h_{124}} \,h_{234} (g_{34}\rhd h_{123}^{-1})\, h_{134}^{-1}))\\&&h_{146}^{-1}({h_{146}\,(g_{46}\rhd h_{134}})\,h_{346}^{-1}\,h_{136}^{-1})h_{136}\delta(\{h_{346}, (g_{46}g_{34})\rhd h_{123}\}_{\mathrm{p}})h_{136}^{-1}h_{136}\,(g_{36}\rhd h_{123})\,h_{236}^{-1}\,\Big)\,\\&=&\delta_H(h_{346}\big((g_{46}g_{34})\rhd h_{123}^{-1}\big)h_{346}^{-1}\delta(\{h_{346}, (g_{46}g_{34})\rhd h_{123}\}_{\mathrm{p}})(g_{36}\rhd h_{123})\,.
    \end{aligned}$}
\end{equation}
Substituting $g_{36}=\partial(h_{346})g_{46}g_{34}$, and applying the (\ref{eq:Pfajferovkomutator2}) identity for $h_1=h_{346}$ and $h_2=(g_{46}g_{34})\rhd h_{123}$, one obtains
\begin{equation}
\begin{aligned}
    \delta_H(h_{1236})=\delta_H(e)\,.
    \end{aligned}
\end{equation}
Similarly, one obtains that $\delta_H(h_{1246})=\delta_H(h_{1346})=\delta_H(e)$. The remaining $\delta$-function on the group $L$ $\delta_L(l_{12356})$ reads:
\begin{equation}
\begin{aligned}
    \delta_L(l_{12356})&=&\delta_L\big( l_{1236}{}^{-1}(h_{126}\rhd'{l_{2356}}) l_{1256} h_{156}\rhd'(g_{56}\rhd l_{1235})l_{1356}{}^{-1}h_{136}\rhd'\{h_{356}, (g_{56}g_{35})\rhd h_{123}\}_{\mathrm{p}}\big)\,.
    \end{aligned}
    \end{equation}
After substituting the equations (\ref{l:1235}), (\ref{l:1236'}), (\ref{l:1246'}), and (\ref{l:1346'}), one obtains:     
    \begin{equation}
    \medmuskip=0mu
\thinmuskip=0mu
\thickmuskip=0mu
    \begin{aligned}
    \delta_L(l_{12356})&=&\delta_L\Big(h_{136}\rhd'\{h_{346}, (g_{46}g_{34})\rhd h_{123}\}^{-1}_{\mathrm{p}} (h_{136}\rhd'{ l_{3456}}) l_{1356} h_{156}\rhd'(g_{56}\rhd l_{1345})l_{1456}{}^{-1}\\&&h_{146}\rhd'\{h_{456}, (g_{56}g_{45})\rhd h_{134}\}_{\mathrm{p}}h_{146}\rhd'(g_{46}\rhd l_{1234}){}^{-1}h_{146}\rhd'\{h_{456}, (g_{56}g_{45})\rhd h_{124}\}^{-1}_{\mathrm{p}} l_{1456}\\&&h_{156}\rhd'(g_{56}\rhd l_{1245})^{-1}l_{1256}^{-1} (h_{126}\rhd'l_{2456})^{-1}(h_{126}\rhd'{ l_{2346}}^{-1}) (h_{126}\rhd'{ l_{2356}}) l_{1256}\\&&h_{156}\rhd'(g_{56}\rhd ({(h_{125}\rhd'{ l_{2345}})l_{1245} h_{145}\rhd'(g_{45}\rhd l_{1234})l_{1345}^{-1}}h_{135}\rhd'\{h_{345}, (g_{45}g_{34})\rhd h_{123}\}_{\mathrm{p}}))\\&&l_{1356}{}^{-1}h_{136}\rhd'\{h_{356}, (g_{56}g_{35})\rhd h_{123}\}_{\mathrm{p}}\Big)\,.
    \end{aligned}
    \end{equation}
    Using the identity (\ref{eq:id01}) the delta function $\delta_L(l_{12356})$ becomes:
\begin{equation}
\resizebox{0.91\hsize}{!}{$
    \medmuskip=0mu
\thinmuskip=0mu
\thickmuskip=0mu
        \begin{aligned}
        \delta_L(l_{12356})&=&\delta_L\Big((h_{136}\rhd'{ l_{3456}}) l_{1356} h_{156}\rhd'(g_{56}\rhd l_{1345})l_{1456}{}^{-1}\\&&h_{146}\rhd'\{h_{456}, (g_{56}g_{45})\rhd h_{134}\}_{\mathrm{p}}h_{146}\rhd'(g_{46}\rhd l_{1234}){}^{-1}h_{146}\rhd'\{h_{456}, (g_{56}g_{45})\rhd h_{124}\}^{-1}_{\mathrm{p}} l_{1456}\\&&\delta(h_{156}\rhd'(g_{56}\rhd l_{1245})^{-1})\rhd'\Big(\big(\delta(l_{1256})^{-1} h_{126}\big)\rhd'\big(l_{2456}^{-1}l_{2346}^{-1}l_{2356}\big)h_{156}\rhd'(g_{56}\rhd (h_{125}\rhd'{ l_{2345}}))\Big)\\&&h_{156}\rhd'(g_{56}\rhd( h_{145}\rhd'(g_{45}\rhd l_{1234})l_{1345}^{-1}))l_{1356}{}^{-1} (h_{136}h_{346})\rhd'\{h_{346}^{-1}h_{356}g_{56}\rhd h_{345}, (g_{56}g_{45}g_{34})\rhd h_{123}\}_{\mathrm{p}}\Big)\,.
        \end{aligned}$}
        \end{equation}
        Commuting the elements, one obtains
        \begin{equation}\label{eq:1235602}
        \medmuskip=0mu
\thinmuskip=0mu
\thickmuskip=0mu
            \begin{aligned}
         \delta_L(l_{12356})&=&\delta_L\Big( (h_{156}\rhd'(g_{56}\rhd \delta(l_{1245})^{-1})\delta(l_{1256})^{-1} h_{126})\rhd'\big(l_{2456}^{-1}l_{2346}^{-1}l_{2356}h_{256}\rhd'(g_{56}\rhd l_{2345})\big)\\&&h_{156}\rhd'\big(g_{56}\rhd( h_{145}\rhd'(g_{45}\rhd l_{1234})l_{1345}^{-1})\big)l_{1356}{}^{-1} (h_{136}h_{346})\rhd'\{h_{346}^{-1}h_{356}g_{56}\rhd h_{345}, (g_{56}g_{45}g_{34})\rhd h_{123}\}_{\mathrm{p}}\\&&h_{136}\rhd' l_{3456}l_{1356}h_{156}\rhd'(g_{56}\rhd l_{1345})(\delta(l_{1456}){}^{-1}h_{146})\rhd'\big(\{h_{456}, (g_{56}g_{45})\rhd h_{134}\}_{\mathrm{p}}\big)\\&&(\delta(l_{1456}){}^{-1}h_{146})\rhd'\big((g_{46}\rhd l_{1234}){}^{-1}\big)(\delta(l_{1456}){}^{-1}h_{146})\rhd'\{h_{456}, (g_{56}g_{45})\rhd h_{124}\}^{-1}_{\mathrm{p}}\Big)\,.
         \end{aligned}
         \end{equation}
The tetrahedron $(3456)$ is part of the integrand on both sides of the move, so using the condition (\ref{id:h}) for $\delta_H(h_{3456})$ one can write $h_{346}^{-1}h_{356}g_{56}\rhd h_{345}=h_{346}^{-1}\rhd' \delta(l_{3456})^{-1} h_{456}$. Then, using the identity (\ref{eq:id01}) one obtains that 
         \begin{equation}\label{eq:pf346}
         \begin{aligned}
         \{h_{346}^{-1}h_{356}g_{56}\rhd h_{345}, (g_{56}g_{45}g_{34})\rhd h_{123}\}_{\mathrm{p}}&= & \{h_{346}^{-1}\rhd'\delta(l_{3456})^{-1} h_{456}, (g_{56}g_{45}g_{34})\rhd h_{123}\}_{\mathrm{p}}\\&= & \big(h_{346}^{-1}\rhd'\delta(l_{3456})^{-1}\big)\rhd'\{ h_{456}, (g_{56}g_{45}g_{34})\rhd h_{123}\}_{\mathrm{p}}\\&& \{h_{346}^{-1}\rhd'\delta(l_{3456})^{-1}, (g_{46}g_{34})\rhd h_{123}\}_{\mathrm{p}}\\ &=&h_{346}^{-1}\rhd'l_{3456}^{-1}\{ h_{456}, (g_{56}g_{45}g_{34})\rhd h_{123}\}_{\mathrm{p}}\\&& \big((g_{46}g_{34})\rhd h_{123} h_{346}^{-1}\big)\rhd' l_{3456}\,,
         \end{aligned}
         \end{equation}
where in the last row the definition of the action $\rhd'$ is used. Substituting the equation (\ref{eq:pf346}) in the equation (\ref{eq:1235602}) one obtains
         \begin{equation}
         \medmuskip=0mu
\thinmuskip=0mu
\thickmuskip=0mu
             \begin{aligned}
           \delta_L(l_{12356})&=&\delta_L\Big( (h_{156}\rhd'(g_{56}\rhd \delta(l_{1245})^{-1})\delta(l_{1256})^{-1} h_{126}\delta(l_{2456})^{-1})\rhd'\Big(l_{2346}^{-1}l_{2356}h_{256}\rhd'(g_{56}\rhd l_{2345})l_{2456}^{-1}\Big)\\&&h_{156}\rhd'(g_{56}\rhd( h_{145}\rhd'(g_{45}\rhd l_{1234})))(h_{156}\rhd' (g_{56}\rhd \delta(l_{1345})^{-1})\delta(l_{1356}){}^{-1} h_{136}\delta(l_{3456})^{-1}h_{346})\rhd'\\&&\big(\{h_{456}, (g_{56}g_{45}g_{34})\rhd h_{123}\}_{\mathrm{p}}((g_{46}g_{34})\rhd h_{123})\rhd' l_{3456}\big)(\delta(l_{1456}){}^{-1}h_{146})\rhd'\big(\{h_{456}, (g_{56}g_{45})\rhd h_{134}\}_{\mathrm{p}}\big)\\&&(\delta(l_{1456}){}^{-1}h_{146})\rhd'\big((g_{46}\rhd l_{1234}){}^{-1}\big)(\delta(l_{1456}){}^{-1}h_{146})\rhd'\{h_{456}, (g_{56}g_{45})\rhd h_{124}\}^{-1}_{\mathrm{p}}\Big)\,.
           \end{aligned}
           \end{equation}
Commuting the element $l_{3456}$ to the end of the expression, one obtains
           \begin{equation}
           \medmuskip=0mu
\thinmuskip=0mu
\thickmuskip=0mu
               \begin{aligned}
              \delta_L(l_{12356}) &=&\delta_L\big( (h_{156}\rhd'(g_{56}\rhd \delta(l_{1245})^{-1})\delta(l_{1256})^{-1} h_{126}\delta(l_{2456})^{-1})\rhd'\big(l_{2346}^{-1}l_{2356}h_{256}\rhd'(g_{56}\rhd l_{2345})l_{2456}^{-1}\big)\\&&h_{156}\rhd'(g_{56}\rhd( h_{145}\rhd'(g_{45}\rhd l_{1234})))(h_{156}\rhd' (g_{56}\rhd \delta(l_{1345})^{-1})\delta(l_{1356}){}^{-1} h_{136}\delta(l_{3456})^{-1}h_{346})\rhd'\\&&\big(\{h_{456}, (g_{56}g_{45}g_{34})\rhd h_{123}\}_{\mathrm{p}}\big)(\delta(l_{1456}){}^{-1}h_{146})\rhd'\big(\{h_{456}, (g_{56}g_{45})\rhd h_{134}\}_{\mathrm{p}}\big)\\&&(\delta(l_{1456}){}^{-1}h_{146})\rhd'\big((g_{46}\rhd l_{1234}){}^{-1}\big)(\delta(l_{1456}){}^{-1}h_{146})\rhd'\{h_{456}, (g_{56}g_{45})\rhd h_{124}\}^{-1}_{\mathrm{p}}\\&&(h_{156}g_{56}\rhd h_{145}h_{246}g_{46}\rhd h_{234}h_{346}^{-1})\rhd' l_{3456}\big)\big)\,.
              \end{aligned}
              \end{equation}
Acting to the whole expression with $(h_{156}\rhd'(g_{56}\rhd \delta(l_{1245})^{-1})\delta(l_{1256})^{-1} h_{126}\delta(l_{2456})^{-1})^{-1}\rhd'$, one obtains,
              \begin{equation}
              \medmuskip=0mu
\thinmuskip=0mu
\thickmuskip=0mu
                  \begin{aligned}
               \delta_L(l_{12356})&=&\delta_L\big(l_{2346}^{-1}l_{2356}h_{256}\rhd'(g_{56}\rhd l_{2345})l_{2456}^{-1} \big(h_{246}h_{456}(g_{56}g_{45})\rhd h_{124}^{-1}\big)\rhd' \\&&\Big( (g_{56}g_{45})\rhd l_{1234} \big((g_{56}g_{45})\rhd h_{134} h_{456}^{-1}\big)\rhd' \{h_{456}, (g_{56}g_{45}g_{34})\rhd h_{123}\}_{\mathrm{p}}\\&&h_{456}^{-1}\rhd' \{h_{456}, (g_{56}g_{45})\rhd h_{134}\}_{\mathrm{p}} h_{456}^{-1}\rhd g_{46}\rhd l_{1234}^{-1} \big(h_{456}^{-1}g_{46}\rhd h_{124}\big)\rhd' \{h_{456}, (g_{56}g_{45})\rhd h_{124}^{-1}\}_{\mathrm{p}}\Big)\\&&(h_{246}g_{46}\rhd h_{234}h_{346}^{-1})\rhd' l_{3456}\,.
               \end{aligned}
               \end{equation}
Using the identity (\ref{eq:id02}) for $\{h_{456}, (g_{56}g_{45})\rhd(h_{134}g_{34}\rhd h_{123})\}_{\mathrm{p}}$,
\begin{equation}
\begin{aligned}
\{h_{456}, (g_{56}g_{45})\rhd(h_{134}g_{34}\rhd h_{123})\}_{\mathrm{p}}&=&\{h_{456}, (g_{56}g_{45})\rhd h_{134}\}_{\mathrm{p}} (g_{46}\rhd h_{134})\rhd'  \{h_{456}, (g_{56}g_{45}  g_{34})\rhd h_{123}\}_{\mathrm{p}}\,,
\end{aligned}
\end{equation}
one obtains:
\begin{equation}
\begin{aligned}
\delta_L(l_{12356})&=&\delta_L\big(l_{2346}^{-1}l_{2356}h_{256}\rhd'(g_{56}\rhd l_{2345})l_{2456}^{-1} \\&&h_{246}\rhd'\Big(\big(h_{456}(g_{56}g_{45})\rhd h_{124}^{-1}\big)\rhd' \Big( (g_{56}g_{45})\rhd l_{1234} h_{456}^{-1}\rhd' \{h_{456}, (g_{56}g_{45})\rhd(h_{134}g_{34}\rhd h_{123})\}_{\mathrm{p}}\\&& h_{456}^{-1}\rhd g_{46}\rhd l_{1234}^{-1}\Big) \{h_{456}, (g_{56}g_{45})\rhd h_{124}^{-1}\}_{\mathrm{p}}\Big)(h_{246}g_{46}\rhd h_{234}h_{346}^{-1})\rhd' l_{3456}\,.
\end{aligned}
\end{equation}
Using the identity (\ref{eq:id02}) for $\{h_{456}, (g_{56}g_{45})\rhd (h_{124}^{-1}\delta(l_{1234})h_{134}g_{34}\rhd h_{123})\}_{\mathrm{p}}$ one obtains the terms featuring $l_{1234}$ cancel, \ie\,,          
\begin{equation}\medmuskip=0mu
\thinmuskip=0mu
\thickmuskip=0mu\begin{aligned}\delta_L(l_{12356})&=&\delta_L\big(l_{2346}^{-1}l_{2356}h_{256}\rhd'(g_{56}\rhd l_{2345})l_{2456}^{-1} \\&&h_{246}\rhd'\{h_{456}, (g_{56}g_{45})\rhd (h_{124}^{-1}\delta(l_{1234})h_{134}g_{34}\rhd h_{123})\}_{\mathrm{p}}(h_{246}g_{46}\rhd h_{234}h_{346}^{-1})\rhd' l_{3456}\\&=&\delta_L\big( l_{2346}{}^{-1}l_{2356} h_{256}\rhd'(g_{56}\rhd l_{2345})l_{2456}{}^{-1}h_{246}\rhd'\{h_{456},(g_{56}g_{45})\rhd h_{234}\}_{\mathrm{p}}(\delta(l_{2346}){}^{-1}h_{236})\rhd'l_{3456})\big)\\&=&
    \delta_L(l_{23456})\,,\\
    \end{aligned}\end{equation}
the delta function $\delta_L(l_{12356})$ on the r.h.s. reduces to the delta function $\delta_L(l_{23456})$ of the l.h.s.
The integrations over $l_{1234}$, $l_{1245}$, $l_{1256}$, $l_{1345}$, $l_{1356}$, and $l_{1456}$ are trivial, and finally one obtains,
\begin{equation}
    r.h.s.=\delta_G(e)^6 \delta_H(e)^4 \delta_L(l_{23456})=|G|^6 |H|^4 \delta_L(l_{23456}) \,.
\end{equation}
The prefactors $|G|^{-11}|H|^{-4}|L|^{-1}$ on the r.h.s. and $|G|^{-5} |H|^0 |L|^{-1}$ on the l.h.s., compensate for left-over factors.
\subsection{Pachner move \texorpdfstring{$2\leftrightarrow 4$}{2-4}}

On the left hand side of the move one has the following integrals and the integrand,
\begin{equation}
  \int_{L} dl_{2345}\delta_H(h_{2345})\delta_L(l_{23456})\delta_L(l_{12345}).
\end{equation}
Integrating out $l_{2345}$ using $\delta_L(l_{12345})$, one obtains
\begin{equation}
    l_{2345}= h_{125}{}^{-1}\rhd'\big(l_{1235}h_{135}\rhd'\{h_{345}, (g_{45}g_{34})\rhd h_{123}\}_{\mathrm{p}}^{-1}l_{1345}h_{145}\rhd'(g_{45}\rhd l_{1234})^{-1}l_{1245}^{-1}\big)\,.
\end{equation}
The $\delta$-function $\delta_H(h_{2345})$ now reads,
\begin{equation}\label{eq:h3456}
\begin{aligned}
     \delta_H(h_{2345})&=&
     \delta_H\big(\delta(l_{2345})h_{245}\,(g_{45}\rhd h_{234})\,h_{345}^{-1}\,h_{235}^{-1}\Big)\\&=&\delta_H\Big(h_{125}{}^{-1}\delta(l_{1235})h_{135}\delta(\{h_{345}, (g_{45}g_{34})\rhd h_{123}\}_{\mathrm{p}}^{-1})h_{135}^{-1}\delta(l_{1345})h_{145}(g_{45}\rhd \delta(l_{1234}))^{-1}h_{145}^{-1}\\&&\delta(l_{1245})^{-1}h_{125}h_{245}\,(g_{45}\rhd h_{234})\,h_{345}^{-1}\,h_{235}^{-1}\big)\,.
\end{aligned}
\end{equation}
Using the identity (\ref{id:h}) for the tetrahedras $(1235)$, $(1345)$, $(1234)$, and $(1245)$, the equation (\ref{eq:h3456}) reduces to:
\begin{equation}
\resizebox{0.93\hsize}{!}{$
\begin{aligned}
     \delta_H(h_{2345})&=&\delta_H\Big(h_{125}{}^{-1}h_{125}\,h_{235}\,(g_{35}\rhd h_{123}^{-1}) \,h_{135}^{-1}h_{135}\delta(\{h_{345}, (g_{45}g_{34})\rhd h_{123}\}_{\mathrm{p}}^{-1})h_{135}^{-1}h_{135}\,h_{345}\,(g_{45}\rhd h_{134}^{-1})\\&& \,h_{145}^{-1}h_{145}g_{45}\rhd (h_{134}(g_{34}\rhd h_{123})h_{234}^{-1}h_{124}^{-1})h_{145}^{-1}h_{145}(g_{45}\rhd h_{124})h_{245}^{-1}h_{125}^{-1}h_{125}h_{245}\,(g_{45}\rhd h_{234})\,h_{345}^{-1}\,h_{235}^{-1}\big)\\
     &=&\delta_H\Big((g_{35}\rhd h_{123}^{-1}) \,\delta(\{h_{345}, (g_{45}g_{34})\rhd h_{123}\}_{\mathrm{p}}^{-1})\,h_{345} \,(g_{45}g_{34})\rhd  h_{123})\,\,h_{345}^{-1}\big)\,.
\end{aligned}$}
\end{equation}
Here, one uses the following identity
\begin{equation}\label{eq:Pfajferovkomutator1}
   \delta\{ h_1\,,h_2 \}_{\mathrm{p}}(\partial(h_1) \rhd h_2)h_1h_2^{-1}h_1^{-1}=e \,,
\end{equation}
for $h_1=h_{345}$ and $h_2=(g_{45}g_{34})\rhd h_{123}$, and the identity $g_{35}=\partial(h_{345}) g_{45}g_{34}$, and obtains 
\begin{equation}
\begin{aligned}
     \delta_H(h_{2345})&=&\delta_H(e)\,.
\end{aligned}
\end{equation}
The remaining $\delta$-function $\delta_L(l_{23456})$, reads
\begin{equation}
\begin{aligned}
\delta_L(l_{23456})&=&\delta_L\big( l_{2346}{}^{-1}(h_{236}\rhd'l_{3456}) l_{2356} h_{256}\rhd'(g_{56}\rhd l_{2345})l_{2456}{}^{-1}h_{246}\rhd'\{h_{456}, (g_{56}g_{45})\rhd h_{234}\}_{\mathrm{p}}\big)\,.
\end{aligned}
\end{equation}
Substituting the equation (\ref{eq:h3456}), one obtains
\begin{equation}
\begin{aligned}
\delta_L(l_{23456})&=&\delta_L\Big( l_{2346}{}^{-1}(h_{236}\rhd'l_{3456}) l_{2356} h_{256}\rhd'\Big(g_{56}\rhd \big(h_{125}{}^{-1}\rhd'\big(l_{1235}h_{135}\rhd'\{h_{345}, (g_{45}g_{34})\rhd h_{123}\}_{\mathrm{p}}^{-1}\\&&l_{1345}h_{145}\rhd'(g_{45}\rhd l_{1234})^{-1}l_{1245}^{-1}\big)\big)\Big)l_{2456}{}^{-1}h_{246}\rhd'\{h_{456}, (g_{56}g_{45})\rhd h_{234}\}_{\mathrm{p}}\Big)\,.
\end{aligned}
\end{equation}
Commuting the elements one obtains
\begin{equation}\label{l23456}
\begin{aligned}\delta_L(l_{23456})&=&\delta_L\Big(l_{2456}{}^{-1} l_{2346}{}^{-1} l_{2356} (h_{256}g_{56}\rhd h_{125}{}^{-1})\rhd'g_{56}\rhd l_{1235}\big(h_{256}g_{56}\rhd h_{125}{}^{-1}g_{56}\rhd h_{135}\big)\rhd'\\&&\Big((g_{35}\rhd h_{123}h_{356}^{-1})\rhd'l_{3456})\{g_{56}\rhd h_{345}, (g_{56}g_{45}g_{34})\rhd h_{123}\}_{\mathrm{p}}^{-1}(g_{56}\rhd h_{345}(g_{56}g_{45})\rhd (h_{123}h_{234}^{-1})h_{456}^{-1})\rhd'\\&&\{h_{456}, (g_{56}g_{45})\rhd h_{234}\}_{\mathrm{p}}\Big)(h_{256}g_{56}\rhd h_{125}{}^{-1})\rhd'g_{56}\rhd l_{1345}\\&&(h_{256}g_{56}\rhd h_{125}{}^{-1}g_{56}\rhd h_{145})\rhd'((g_{56}g_{45})\rhd l_{1234})^{-1}(h_{256}g_{56}\rhd h_{125}{}^{-1})\rhd'g_{56}\rhd l_{1245}^{-1}\Big) \,.
\end{aligned}
\end{equation}
Finally, the l.h.s. reads:
\begin{equation}
    l.h.s.=\delta_H(e)\delta_L(l_{23456})=|H|\delta_L(l_{23456})\,.
\end{equation}

Let us now examine the right hand side of the move, \ie\,, the integral (\ref{eq:42rhs}). First, one integrates out $g_{16}$ using $\delta_{G}(g_{126})$, and obtains
\begin{equation}\label{eq:g16}
    g_{16}= \partial(h_{126})\,g_{26}\,g_{12}\,.
\end{equation}
Next, one integrates out $h_{126}$ using $\delta_H(h_{1236})$, $h_{136}$ using $\delta_H(h_{1346})$, and $h_{146}$ using $\delta_H(h_{1456})$, and obtains
\begin{equation}\label{h24}
\begin{aligned}
    h_{126}&=& \delta(l_{1236})h_{136}\,(g_{36}\rhd h_{123})\,h_{236}^{-1}\,,\\
    h_{136}&=& \delta(l_{1346})h_{146}\,(g_{46}\rhd h_{134})\,h_{346}^{-1}\,,\\ h_{146}&=&\delta(l_{1456})h_{156}\,(g_{56}\rhd h_{145})\,h_{456}^{-1}\,.
\end{aligned}
\end{equation}
The remaining $\delta$-functions on the group $G$ reduces to $\delta_G(e)^3$. The $\delta$-function $\delta_G(g_{136})$ 
\begin{equation}
    \delta_G(g_{136})=\delta_G\big(\partial(h_{136})\,g_{36}\,g_{13}\,g_{16}^{-1}\big)\,,
\end{equation}
after substituting the equation (\ref{eq:g16}) reads:
\begin{equation}
\delta_G(g_{136})=\delta_G\big(\partial(h_{136})\,g_{36}\,g_{13}\,g_{12}^{-1}g_{26}^{-1}\partial(h_{126})^{-1}\big)\,.
    \end{equation}
Using the equations (\ref{h24}) for $h_{126}$, and $h_{136}$, and $h_{146}$, and the identity $\partial( \delta l) = 0$ for every element $l \in L$, the $\delta$-function $\delta_G(g_{136})$
reduces to $\delta_G(e)$ after implementing the identity (\ref{id:g}) for the triangles $(156)$, $(145)$, $(456)$ $(134)$, $(346)$, $(236)$, and $(123)$. Similarly, one obtains $\delta_G(g_{146})=\delta_G(g_{156})=\delta_G(e)$.

One integrates out $l_{1236}$ using $\delta_L(l_{12346})$ and obtains
\begin{equation}\label{l:1236''}
     l_{1236} = (h_{126}\rhd'l_{2346}) l_{1246} h_{146}\rhd'(g_{46}\rhd l_{1234})l_{1346}^{-1}\,h_{136}\rhd'\{h_{346}, (g_{46}g_{34})\rhd h_{123}\}_{\mathrm{p}}\,,
\end{equation}
$l_{1246}$ using $\delta_L(l_{12456})$ and obtains
\begin{equation}\label{l:1246}
    l_{1246}= (h_{126}\rhd'l_{2456}) l_{1256} h_{156}\rhd'(g_{56}\rhd l_{1245})l_{1456}{}^{-1}\,h_{146}\rhd'\{h_{456}, (g_{56}g_{45})\rhd h_{124}\}_{\mathrm{p}}\,,
\end{equation}
and $l_{1346}$ using $\delta_L(l_{13456})$ and obtains
\begin{equation}\label{l:1346}
    l_{1346}= (h_{136}\rhd'l_{3456}) l_{1356} h_{156}\rhd'(g_{56}\rhd l_{1345})l_{1456}{}^{-1}\,h_{146}\rhd'\{h_{456}, (g_{56}g_{45})\rhd h_{134}\}_{\mathrm{p}}\,.
\end{equation}
The remaining $\delta$-functions on $H$ reduce on $\delta_H(e)^3$, similarly as in the case of $1 \leftrightarrow 5$ Pachner move, \ie\,, one obtains $\delta_H(h_{1256})=\delta_H(h_{1356})=\delta_H(h_{1456})=\delta_H(e)$.
For the remaining $\delta$-function $\delta_L(l_{12356})$,
\begin{equation}
\begin{aligned}
\delta_L(l_{12356})&=&\delta_L\Big( l_{1236}{}^{-1}(h_{126}\rhd'l_{2356}) l_{1256} h_{156}\rhd'(g_{56}\rhd l_{1235})l_{1356}{}^{-1}h_{136}\rhd ' \{h_{356}, (g_{56}g_{35})\rhd h_{123}\}_{\mathrm{p}}\Big)\,,
\end{aligned}
\end{equation}
one obtains, after substituting the equations (\ref{l:1236''}), (\ref{l:1246}), and (\ref{l:1346}), the following
\begin{equation}
 \begin{aligned}
 \delta_L(l_{12356})&=&\delta_L\Big(h_{136}\rhd'\{h_{346}, (g_{46}g_{34})\rhd h_{123}\}_{\mathrm{p}}{}^{-1}l_{1346}h_{146}\rhd'(g_{46}\rhd l_{1234})^{-1}l_{1246}^{-1} (h_{126}\rhd'l_{2346})^{-1} \,\\&&(h_{126}\rhd'l_{2356}) l_{1256} h_{156}\rhd'(g_{56}\rhd l_{1235})l_{1356}{}^{-1}h_{136}\rhd ' \{h_{356}, (g_{56}g_{35})\rhd h_{123}\}_{\mathrm{p}}\Big)\\&=&\delta_L\Big((h_{126}\rhd'l_{2456})^{-1}(h_{126}\rhd'l_{2346})^{-1}(h_{126}\rhd'l_{2356}) (h_{256}g_{56}\rhd h_{125}{}^{-1})\rhd'l_{1235} \\&& \delta(l_{1256})\rhd'\Big( \delta(l_{1356}){}^{-1}\rhd'\big(h_{136}\rhd ' \{h_{356}, (g_{56}g_{35})\rhd h_{123}\}_{\mathrm{p}}(h_{136}h_{346})\rhd'\{h_{346}^{-1}, g_{36}\rhd h_{123}\}_{\mathrm{p}}\\&&(h_{136}\rhd'l_{3456}) \big)h_{156}\rhd'(g_{56}\rhd l_{1345})l_{1456}{}^{-1}\,h_{146}\rhd'\{h_{456}, (g_{56}g_{45})\rhd h_{134}\}_{\mathrm{p}}h_{146}\rhd'(g_{46}\rhd l_{1234})^{-1}\\&&h_{146}\rhd'\{h_{456}, (g_{56}g_{45})\rhd h_{124}\}_{\mathrm{p}}^{-1}l_{1456}h_{156}\rhd'(g_{56}\rhd l_{1245})^{-1}\Big)\Big)\,.
 \end{aligned}
 \end{equation}
 Commuting the elements in order to match the l.h.s. of the move, \ie\,, the $\delta$-function given by the equation (\ref{l23456}), and using the identity (\ref{eq:id01}), \ie\,, 
 \begin{equation}
     \{h_{346}^{-1}h_{356}, (g_{56}g_{35})\rhd h_{123}\}_{\mathrm{p}}=h_{346}^{-1}\rhd' \{h_{356}, (g_{56}g_{35})\rhd h_{123}\}_{\mathrm{p}} \{h_{346}^{-1}, g_{36}\rhd h_{123}\}_{\mathrm{p}}\,,
 \end{equation}
 one obtains
 \begin{equation}
     \begin{aligned}\label{l12356'}
\delta_L(l_{12356})&=&\delta_L\Big((h_{126}\rhd'l_{2456})^{-1}(h_{126}\rhd'l_{2346})^{-1}(h_{126}\rhd'l_{2356}) (h_{126}h_{256}g_{56}\rhd h_{125}{}^{-1})\rhd'l_{1235} \\&& \delta(l_{1256})\rhd'\Big( \delta(l_{1356}){}^{-1}\rhd'\big((h_{136}h_{346})\rhd ' \{h_{346}^{-1}h_{356}, (g_{56}g_{35})\rhd h_{123}\}_{\mathrm{p}}(h_{136}\rhd'l_{3456}) \big)\\&&h_{156}\rhd'(g_{56}\rhd l_{1345})(\delta(l_{1456}){}^{-1}\,\,h_{146})\rhd'\big(\{h_{456}, (g_{56}g_{45})\rhd h_{134}\}_{\mathrm{p}}(g_{46}\rhd l_{1234})^{-1}\\&&\{h_{456}, (g_{56}g_{45})\rhd h_{124}\}_{\mathrm{p}}^{-1}\big)h_{156}\rhd'(g_{56}\rhd l_{1245})^{-1}\Big)\Big)\,.\end{aligned}
\end{equation}
Using the identity (\ref{eq:id01}) again one rewrites the following term as
\begin{equation}
\begin{aligned}
  &&(h_{136}h_{346})\rhd ' \{h_{346}^{-1}h_{356}, (g_{56}g_{35})\rhd h_{123}\}_{\mathrm{p}}(h_{136}\rhd'l_{3456})=\\&&(h_{136}h_{346})\rhd ' \{h_{346}^{-1}\rhd' \delta(l_{3456})^{-1}h_{456} g_{56}\rhd h_{345}^{-1}, (g_{56}g_{35})\rhd h_{123}\}_{\mathrm{p}}(h_{136}\rhd'l_{3456})=\\&&(h_{136}\rhd' \delta(l_{3456})^{-1}h_{136}h_{346})\rhd'\big(\{h_{456} g_{56}\rhd h_{345}^{-1}, (g_{56}g_{35})\rhd h_{123}\}_{\mathrm{p}} ((g_{46}g_{34})\rhd h_{123} h_{346}^{-1})\rhd' l_{3456}^{-1}\big)\,,
\end{aligned}
 \end{equation}
 and substituting it in the equation (\ref{l12356'}) the $\delta$-function becomes:
\begin{equation}
    \begin{aligned}
\delta_L(l_{12356})&=&\delta_L\Big((h_{126}\rhd'l_{2456})^{-1}(h_{126}\rhd'l_{2346})^{-1}(h_{126}\rhd'l_{2356}) (h_{126}h_{256}g_{56}\rhd h_{125}{}^{-1})\rhd'l_{1235} \\&& \delta(l_{1256})\rhd'\Big( (\delta(l_{1356}){}^{-1}h_{136}\rhd'\delta(l_{3456})^{-1}h_{136}h_{346}) \rhd' \\&&\big(\{h_{456}g_{56}\rhd h_{345}^{-1}, (g_{56}g_{35})\rhd h_{123}\}_{\mathrm{p}}((g_{46}g_{34})\rhd h_{123}h_{346}^{-1})\rhd'l_{3456})\big)\\&& (h_{156}g_{56}\rhd h_{135} g_{56}\rhd (h_{345}g_{45}\rhd h_{134}^{-1})h_{456}^{-1})\rhd'\big(\{h_{456}, (g_{56}g_{45})\rhd h_{134}\}_{\mathrm{p}}(g_{46}\rhd l_{1234})^{-1}\\&&\{h_{456}, (g_{56}g_{45})\rhd h_{124}\}_{\mathrm{p}}^{-1}\big)\Big)(h_{126}h_{256}g_{56}\rhd h_{125}{}^{-1})\rhd'\big(h_{156}\rhd'(g_{56}\rhd l_{1345})(g_{56}\rhd l_{1245})^{-1}\big)\Big)\,.
\end{aligned}
\end{equation}
Commuting the elements $l_{3456}$ and $\{h_{456}g_{56}\rhd h_{345}, (g_{56}g_{35})\rhd h_{123}\}_{\mathrm{p}}$, and using the identity (\ref{eq:id01}) to rewrite this Peiffer lifting, one obtains
\begin{equation}\label{eq:14lll}
    \begin{aligned}
    \delta_L(l_{12356})&=&\delta_L\Big((h_{126}\rhd'l_{2456})^{-1}(h_{126}\rhd'l_{2346})^{-1}(h_{126}\rhd'l_{2356}) (h_{126}h_{256}g_{56}\rhd h_{125}{}^{-1})\rhd'l_{1235}\\&&
    \big(h_{126}h_{256}g_{56}\rhd h_{125}{}^{-1}h_{135}(g_{56}g_{35})\rhd h_{123}g_{56}\rhd h_{356}^{-1})\rhd'g_{56}\rhd l_{3456}\\&& (h_{126}h_{256}g_{56}\rhd h_{125}^{-1}g_{56}\rhd h_{135}g_{56}\rhd h_{345}) \rhd' \Big( \{g_{56}\rhd h_{345}^{-1},(g_{56}g_{35})\rhd h_{123}\}_{\mathrm{p}}\\&&h_{456}^{-1}\rhd' \{h_{456},(g_{56}g_{45}g_{34})\rhd h_{123}\}_{\mathrm{p}} ((g_{56}g_{45})\rhd h_{134}^{-1}h_{456}^{-1})\rhd'\big(\{h_{456}, (g_{56}g_{45})\rhd h_{134}\}_{\mathrm{p}}(g_{46}\rhd l_{1234})^{-1}\\&&\{h_{456}, (g_{56}g_{45})\rhd h_{124}\}_{\mathrm{p}}^{-1}\big)\Big)(h_{126}h_{256}g_{56}\rhd h_{125}{}^{-1})\rhd'\big(h_{156}\rhd'(g_{56}\rhd l_{1345})(g_{56}\rhd l_{1245})^{-1}\big)\Big)\,.
    \end{aligned}
\end{equation}
After the similar transformations as in the case of $1\leftrightarrow 5$ move, commuting the element $l_{1234}$ so that the order of the elements matches the order in the expression (\ref{l23456}), and acting to the whole expression with $h_{126}^{-1}$ one obtains
\begin{equation}\label{l12356}
\begin{aligned}\delta_L(l_{12356})&=&\delta_L\Big(l_{2456}{}^{-1} l_{2346}{}^{-1} l_{2356} (h_{256}g_{56}\rhd h_{125}{}^{-1})\rhd'g_{56}\rhd l_{1235}\big(h_{256}g_{56}\rhd h_{125}{}^{-1}g_{56}\rhd h_{135}\big)\rhd'\\&&\Big((g_{35}\rhd h_{123}h_{356}^{-1})\rhd'l_{3456})\{g_{56}\rhd h_{345}, (g_{56}g_{45}g_{34})\rhd h_{123}\}_{\mathrm{p}}^{-1}(g_{56}\rhd h_{345}(g_{56}g_{45})\rhd (h_{123}h_{234}^{-1})h_{456}^{-1})\rhd'\\&&\{h_{456}, (g_{56}g_{45})\rhd h_{234}\}_{\mathrm{p}}\Big)(h_{256}g_{56}\rhd h_{125}{}^{-1})\rhd'g_{56}\rhd l_{1345}\\&&(h_{256}g_{56}\rhd h_{125}{}^{-1}g_{56}\rhd h_{145})\rhd'((g_{56}g_{45})\rhd l_{1234})^{-1}(h_{256}g_{56}\rhd h_{125}{}^{-1})\rhd'g_{56}\rhd l_{1245}^{-1}\Big) \,.
\end{aligned}
\end{equation}
which is precisely the equation (\ref{l23456}).
The remaining integration over the element $h_{156}$ of the group $H$ and remaining integration over the three elements of the group $L$, $l_{1246}$, $l_{1256}$, and $l_{1356}$, are trivial, yielding the result on the r.h.s. to:
\begin{equation}
  r.h.s.= \delta_G(e)^3\,\delta_H(e)^3\,\delta_L(l_{12356})=|G|^3\,|H|^3\,\delta_L(l_{12356})\,.
\end{equation}
The prefactors are $  |G|^{-8}|H|^{-1}|L|^{-1}$ on the l.h.s., and $  |G|^{-11}|H|^{-3}|L|^{-1}$ on the r.h.s. compensate for the left-over factors.
\subsection{Pachner move \texorpdfstring{$3\leftrightarrow 3$}{3-3}}
Let us first investigate the r.h.s. of the move. First, one integrates out the $l_{1235}$, exploiting $ \delta_L(l_{12345})$ and obtains
\begin{equation}\label{eq:l1235}
l_{1235} = (h_{125}\rhd'l_{2345}) l_{1245} h_{145}\rhd'(g_{45}\rhd l_{1234})l_{1345}{}^{-1}\,h_{135}\rhd'\{h_{345}, (g_{45}g_{34})\rhd h_{123}\}_{\mathrm{p}}\,,
\end{equation}
and one integrates out $l_{1236}$, exploiting $ \delta_L(l_{12356})$ and obtains
\begin{equation}\label{eq:l1236}
    l_{1236}=(h_{126}\rhd'l_{2356}) l_{1256} h_{156}\rhd'(g_{56}\rhd l_{1235})l_{1356}{}^{-1}h_{136}'\rhd\{h_{356}, (g_{56}g_{35})\rhd h_{123}\}_{\mathrm{p}}\,.
\end{equation}
Next, one integrates out $h_{123}$, exploiting $ \delta_H(l_{1234})$ and obtains:
\begin{equation}\label{eq:h123}
    h_{123} =  g_{34}^{-1}\rhd h_{134}^{-1}\,g_{34}^{-1}\rhd \delta(l_{1234})^{-1}\,g_{34}^{-1}\rhd h_{124} \,g_{34}^{-1}\rhd h_{234}.
\end{equation}
The $\delta$-function $\delta_G(g_{123})$, when using the equation (\ref{eq:h123}) reads
\begin{equation}
\begin{aligned}
 \delta_G(g_{123})=\delta_G\big(g_{34}^{-1}\rhd \partial(h_{134})^{-1}\,g_{34}^{-1}\rhd\partial( \delta(l_{1234}))^{-1}\,g_{34}^{-1}\rhd \partial(h_{124}) \,g_{34}^{-1}\rhd \partial(h_{234})\,g_{23}\,g_{12}\,g_{13}^{-1}\big)\,,
\end{aligned}
\end{equation}
which then using the condition $\partial\delta=0$, reduces to
\begin{equation}
    \delta_G(g_{123})=\delta_G\big( \partial(h_{134})^{-1}\, \partial(h_{124}) \, \partial(h_{234})\,g_{34}^{-1}g_{23}\,g_{12}\,g_{13}^{-1}g_{34}\big)\,.
\end{equation}
Using the condition (\ref{id:g}) for the triangles $({134})$, $({124})$, and $({234})$, it finally reduces to 
\begin{equation}
    \delta_G(g_{123})=\delta_G\big(e\big)\,.
\end{equation}
For the $\delta$-function $\delta_H(h_{1235})$, one obtains, after using the equation (\ref{eq:l1235}):
\begin{equation}
\begin{aligned}
      \delta_H(h_{1235})&=&\delta_H\Big((h_{125}\delta(l_{2345})h_{125}^{-1}) \delta(l_{1245}) (h_{145}(g_{45}\rhd \delta(l_{1234}))h_{145}^{-1})\delta(l_{1345}){}^{-1}\,h_{135}\rhd'\{h_{345}, g_{35}\rhd h_{123}\}_{\mathrm{p}}\,h_{135}\,\\&&((g_{35}g_{34}{}^{-1})\rhd (h_{134}^{-1}\,\delta(l_{1234})^{-1}\,h_{124} \,h_{234}))\,h_{235}^{-1}\,h_{125}^{-1}\Big)\,.
\end{aligned}
\end{equation}
Using the $\delta$-functions $\delta_L(h_{2345})$, $\delta_L(h_{1245})$, and $\delta_L(h_{1345})$, that appear on both sides of the move, and are thus part of the integrand,
\begin{equation}
\begin{aligned}
\delta(l_{2345})&=&h_{235}\,h_{345}\,(g_{45}\rhd h_{234}^{-1})\, h_{245}^{-1}\,,\\
\delta(l_{1245})&=&h_{125}\,h_{245}\,(g_{45}\rhd h_{124}^{-1})\, h_{145}^{-1}\,,\\
\delta(l_{1345})^{-1}&=&h_{145}\,(g_{45}\rhd h_{134})\,h_{345}^{-1}\,h_{135} ^{-1}\,,
\end{aligned}
\end{equation}
one obtains:
\begin{equation}
\begin{aligned}
    \delta_H(h_{1235})&=&\delta_H\Big(h_{125}h_{235}\,h_{345}\,(g_{45}\rhd h_{234}^{-1})\, h_{245}^{-1}h_{125}^{-1} h_{125}\,h_{245}\,(g_{45}\rhd h_{124}^{-1})\, h_{145}^{-1} h_{145}(g_{45}\rhd \delta(l_{1234}))h_{145}^{-1}\\&&h_{145}\,(g_{45}\rhd h_{134})\,h_{345}^{-1}\,h_{135} ^{-1}h_{135}\rhd\delta(\{h_{345}, (g_{45}g_{34})\rhd h_{123}\}_{\mathrm{p}})\\&&h_{135}\,((g_{35}g_{34}{}^{-1})\rhd (h_{134}^{-1}\delta(l_{1234})^{-1}\,h_{124} \,h_{234}))\,h_{235}^{-1}\,h_{125}^{-1}\Big)\\&=&\delta_H\Big(h_{345}(g_{45}g_{34})\rhd h_{123}^{-1} \,h_{345}^{-1}\delta(\{h_{345},(g_{45}g_{34})\rhd h_{123}\}_{\mathrm{p}})\,(g_{35}\rhd h_{123})\Big)\,.
    \end{aligned}
\end{equation}
Substituting $g_{35}=\partial(h_{345})g_{45}g_{34}$, and applying the identity 
\begin{equation}\label{eq:Pfajferovkomutator3}
   \delta\{ h_1\,,h_2 \}_{\mathrm{p}}(\partial(h_1) \rhd h_2)h_1h_2^{-1}h_1^{-1}=e \,,
\end{equation}
for $h_1=h_{345}$ and $h_2=(g_{45}g_{34})\rhd h_{123}$, one obtains
\begin{equation}
\begin{aligned}
    \delta_H(h_{1235})=\delta_H(e).
    \end{aligned}
\end{equation}
Similarly, one obtains that $\delta_H(h_{1236})=\delta_H(e)$.
The remaining $\delta$-function $\delta_H
(l_{12346})$ reads
\begin{equation}\label{eq:33rhs}
\begin{aligned}
\delta_L(l_{12346})&=&\delta_L\big( l_{1236}{}^{-1}(h_{126}\rhd'l_{2346}) l_{1246} h_{146}\rhd'(g_{46}\rhd l_{1234})l_{1346}{}^{-1}\,h_{136}\rhd'\{h_{346}, (g_{46}g_{34})\rhd h_{123}\}_{\mathrm{p}}\big)\,.
\end{aligned}
\end{equation}
After substituting the equation (\ref{eq:l1236}), and then the equation (\ref{eq:l1235}), one obtains:
\begin{equation}
    \begin{aligned}
   \delta_L(l_{12346})&=&\delta_L\big(h_{136}\rhd'\{h_{356}, (g_{56}g_{35})\rhd h_{123}\}^{-1}_{\mathrm{p}}l_{1356}h_{156}\rhd'(g_{56}\rhd l_{1235})^{-1}l_{1256}^{-1} h_{126}\rhd'l_{2356}^{-1}\\&&(h_{126}\rhd'l_{2346}) l_{1246} h_{146}\rhd'(g_{46}\rhd l_{1234})l_{1346}{}^{-1}h_{136}\rhd'\{h_{346}, (g_{46}g_{34})\rhd h_{123}\}_{\mathrm{p}}\big)\\&=&\delta_L\big(h_{136}\rhd'\{h_{356}, (g_{56}g_{35})\rhd h_{123}\}^{-1}_{\mathrm{p}}l_{1356}\\&&h_{156}\rhd'(g_{56}\rhd ((h_{125}\rhd'l_{2345}) l_{1245} h_{145}\rhd'(g_{45}\rhd l_{1234})l_{1345}{}^{-1}\,h_{135}\rhd'\{h_{345}, (g_{45}g_{34})\rhd h_{123}\}_{\mathrm{p}}))^{-1}\\&&l_{1256}^{-1} h_{126}\rhd'l_{2356}^{-1}(h_{126}\rhd'l_{2346}) l_{1246} h_{146}\rhd'(g_{46}\rhd l_{1234})l_{1346}{}^{-1}h_{136}\rhd'\{h_{346}, (g_{46}g_{34})\rhd h_{123}\}_{\mathrm{p}}\big)\,.
   \end{aligned}
   \end{equation}
   After commuting the elements, \ie\,, using the Peiffer identity for the crossed module $(L\overset{\delta}{\rightarrow}H, \rhd')$, one obtains
   \begin{equation}\resizebox{0.96\hsize}{!}{$\begin{aligned}\label{eq:33move11}
       \delta_L(l_{12346})&=&\delta_L\big(h_{136}\rhd'\{h_{356}, (g_{56}g_{35})\rhd h_{123}\}^{-1}_{\mathrm{p}}\\&&(\delta(l_{1356})h_{156}g_{56}\rhd  h_{135})\rhd' g_{56}\rhd\{h_{345}, (g_{45}g_{34})\rhd h_{123}\}^{-1}_{\mathrm{p}} l_{1356}h_{156}\rhd'(g_{56}\rhd l_{1345})\\&& (h_{156}g_{56}\rhd h_{145})\rhd'((g_{56}g_{45})\rhd l_{1234})^{-1}h_{156}\rhd'(g_{56}\rhd l_{1245})^{-1}(h_{156}g_{56}\rhd h_{125})\rhd'(g_{56}\rhd l_{2345}^{-1})l_{1256}^{-1}\\&& h_{126}\rhd'l_{2356}^{-1}(h_{126}\rhd'l_{2346}) l_{1246} h_{146}\rhd'(g_{46}\rhd l_{1234})l_{1346}{}^{-1}h_{136}\rhd'\{h_{346}, (g_{46}g_{34})\rhd h_{123}\}_{\mathrm{p}}\big)\\
      &=&\delta_L\big((\delta(l_{1346})^{-1}h_{136})\rhd'\{h_{346}, (g_{46}g_{34})\rhd h_{123}\}_{\mathrm{p}}(\delta(l_{1346})^{-1}h_{136})\rhd'\{h_{356}, (g_{56}g_{35})\rhd h_{123}\}^{-1}_{\mathrm{p}}\\&&((\delta(l_{1346})^{-1}\delta(l_{1356})h_{156}g_{56}\rhd  h_{135})\rhd' g_{56}\rhd\{h_{345}, (g_{45}g_{34})\rhd h_{123}\}^{-1}_{\mathrm{p}} \\&& (\delta(l_{1346})^{-1} \delta(l_{1356}) h_{156}\rhd'(g_{56}\rhd \delta(l_{1345}))h_{156}g_{56}\rhd h_{145})\rhd'((g_{56}g_{45})\rhd l_{1234})^{-1}l_{1346}^{-1} l_{1356}h_{156}\rhd'(g_{56}\rhd l_{1345})\\&&h_{156}\rhd'(g_{56}\rhd l_{1245})^{-1}(h_{156}g_{56}\rhd h_{125})\rhd'(g_{56}\rhd l_{2345}^{-1})l_{1256}^{-1} h_{126}\rhd'l_{2356}^{-1}(h_{126}\rhd'l_{2346}) l_{1246} h_{146}\rhd'(g_{46}\rhd l_{1234})\big)\,.
\end{aligned}$}
\end{equation}
Using the identity (\ref{eq:id03}) one obtains that   
\begin{equation}
\{h_{346}, (g_{46}g_{34})\rhd h_{123}\}_{\mathrm{p}}=h_{346}\rhd'\{h_{346}^{-1},g_{36}\rhd h_{123}\}_{\mathrm{p}}^{-1}\,.
\end{equation}
Using a variant of the identity (\ref{eq:id01}), \ie\,, that
\begin{equation}
    \{h_1 h_2 h_3, h_4\}_{\mathrm{p}}^{-1}=\{h_1,\partial(h_2 h_3)\rhd h_4\}_{\mathrm{p}}^{-1}h_1\rhd'\{h_2,\partial(h_2)\rhd h_4\}_{\mathrm{p}}^{-1}(h_1h_2)\rhd' \{h_3, h_4\}_{\mathrm{p}}^{-1}\,,
    \end{equation}
one obtains that 
\begin{equation}
\begin{aligned}
    \{h_{346}^{-1}\,h_{356}\,(g_{56}\rhd h_{345}), (g_{56}g_{45}g_{34})\rhd h_{123}\}_{\mathrm{p}}^{-1}&=&\{h_{346}^{-1}, (g_{46}g_{34})\rhd h_{123}\}_{\mathrm{p}}^{-1}h_{346}^{-1}\rhd'\{h_{356}, (g_{56}g_{35})\rhd h_{123}\}^{-1}_{\mathrm{p}}\\&&(h_{346}^{-1}h_{356})\rhd'\{g_{56}\rhd h_{345}, (g_{56}g_{45}g_{34})\rhd h_{123}\}^{-1}_{\mathrm{p}}\,,
    \end{aligned}
\end{equation}
rendering the expression (\ref{eq:33move11}) to
  \begin{equation}\resizebox{0.99\hsize}{!}{$\begin{aligned}\label{eq:33move12}
       \delta_L(l_{12346})&=&\delta_L\big((h_{146}g_{46}\rhd h_{134})\rhd'\{h_{346}^{-1}\,h_{356}\,(g_{56}\rhd h_{345}), (g_{56}g_{45}g_{34})\rhd h_{123}\}_{\mathrm{p}}^{-1} \\&& (\delta(l_{1346})^{-1} \delta(l_{1356}) h_{156}\rhd'(g_{56}\rhd \delta(l_{1345}))h_{156}g_{56}\rhd h_{145})\rhd'((g_{56}g_{45})\rhd l_{1234})^{-1}l_{1346}^{-1} l_{1356}h_{156}\rhd'(g_{56}\rhd l_{1345})\\&&h_{156}\rhd'(g_{56}\rhd l_{1245})^{-1}(h_{156}g_{56}\rhd h_{125})\rhd'(g_{56}\rhd l_{2345}^{-1})l_{1256}^{-1} h_{126}\rhd'l_{2356}^{-1}(h_{126}\rhd'l_{2346}) l_{1246} h_{146}\rhd'(g_{46}\rhd l_{1234})\big)\,.
\end{aligned}$}
\end{equation}
Substituting the equation (\ref{eq:h123}), and using the identity (\ref{eq:id02}), one obtains that the expression,
\begin{equation}
\resizebox{0.99\hsize}{!}{$
\begin{aligned}
    \{h_{346}^{-1}\,h_{356}\,(g_{56}\rhd h_{345}), (g_{56}g_{45}g_{34})\rhd h_{123}\}_{\mathrm{p}}^{-1}&=&\{h_{346}^{-1}\,h_{356}\,(g_{56}\rhd h_{345}), (g_{56}g_{45})\rhd ((h_{134}^{-1}\rhd' \delta(l_{1234})^{-1}) h_{134}^{-1} h_{124} h_{234}\}_{\mathrm{p}}^{-1}\\&=&(g_{46}\rhd (h_{134}^{-1}\rhd' \delta(l_{1234})^{-1}))\rhd' \{h_{346}^{-1}\,h_{356}\,(g_{56}\rhd h_{345}),(g_{56}g_{45})\rhd\\&& (h_{134}^{-1}h_{124}h_{234})\}_{\mathrm{p}}^{-1}\{h_{346}^{-1}\,h_{356}\,(g_{56}\rhd h_{345}),(g_{56}g_{45})\rhd (h_{134}^{-1}\rhd' \delta(l_{1234})^{-1})\}_{\mathrm{p}}^{-1}\,,
    \end{aligned}$}
    \end{equation}
using the identity (\ref{eq:id05}), \ie\,, that 
\begin{equation}
\begin{aligned}
\{h_{346}^{-1}\,h_{356}\,(g_{56}\rhd h_{345}),(g_{56}g_{45})\rhd (h_{134}^{-1}\rhd' \delta(l_{1234})^{-1})\}_{\mathrm{p}}^{-1}&=&(g_{46}\rhd(h_{134}^{-1}\rhd' l_{1234}^{-1}) )\\&& (h_{346}^{-1}\,h_{356}\,(g_{56}\rhd h_{345}))\rhd'((g_{56}g_{45})\rhd(h_{134}^{-1}\rhd' l_{1234}))\,,
\end{aligned}
\end{equation}
reduces to
\begin{equation}
    \begin{aligned}\label{eq:f33}
     \{h_{346}^{-1}\,h_{356}\,(g_{56}\rhd h_{345}), (g_{56}g_{45}g_{34})\rhd h_{123}\}_{\mathrm{p}}^{-1}&=&g_{46}\rhd (h_{134}^{-1}\rhd' \delta(l_{1234})^{-1})\\&& \{h_{346}^{-1}\,h_{356}\,(g_{56}\rhd h_{345}),(g_{56}g_{45})\rhd (h_{134}^{-1}h_{124}h_{234})\}_{\mathrm{p}}^{-1}\\&& (h_{346}^{-1}\,h_{356}\,(g_{56}\rhd h_{345}))\rhd'((g_{56}g_{45})\rhd(h_{134}^{-1}\rhd' l_{1234}))\,.
    \end{aligned}
\end{equation}
Substituting this result in the expression (\ref{eq:33move12}) the terms featuring $l_{1234}$ cancel, and finally the delta function $\delta_L(l_{12346})$ reads:
\begin{equation}
\resizebox{0.99\hsize}{!}{$\begin{aligned}\label{eq:33move13}
       \delta_L(l_{12346})&=&\delta_L\big((h_{146}g_{46}\rhd h_{134})\rhd'\{h_{346}^{-1}\,h_{356}\,(g_{56}\rhd h_{345}),(g_{56}g_{45})\rhd (h_{134}^{-1}h_{124}h_{234})\}_{\mathrm{p}}^{-1} l_{1346}^{-1} l_{1356} \\&&h_{156}\rhd'(g_{56}\rhd l_{1345})h_{156}\rhd'(g_{56}\rhd l_{1245})^{-1}(h_{156}g_{56}\rhd h_{125})\rhd'(g_{56}\rhd l_{2345}^{-1})l_{1256}^{-1} h_{126}\rhd'l_{2356}^{-1}(h_{126}\rhd'l_{2346}) l_{1246}\big)\,.
\end{aligned}$}
\end{equation}
One obtains that the integration over $l_{1234}$ is trivial, and the r.h.s. of the move finally reads
\begin{equation}
\begin{aligned}\label{eq:rhs33}
r.h.s.&=& \delta_G(e)\delta_H(e)^{2}\delta_L\big(h_{156}\rhd'(g_{56}\rhd l_{1245})^{-1}\, h_{156}\rhd'(g_{56}\rhd(h_{125}\rhd'l_{2345}))^{-1} \,l_{1256}^{-1}\, h_{126}\rhd'l_{2356}^{-1}(h_{126}\rhd'l_{2346})\\&& l_{1246}(h_{146}g_{46}\rhd h_{134})\rhd'\{h_{346}^{-1}\,h_{356}\,(g_{56}\rhd h_{345}),(g_{56}g_{45})\rhd (h_{134}^{-1}h_{124}h_{234})\}_{\mathrm{p}}^{-1} l_{1346}^{-1}l_{1356}h_{156}\rhd'(g_{56}\rhd  l_{1345})\,.
 \end{aligned}
\end{equation}
\newline
The integral of the l.h.s.\ reads
\begin{equation}
  \int_Hdh_{456}\int_{L^3}dl_{1456}dl_{2456}dl_{3456}\delta_G(g_{456})\,\delta_H(h_{3456})\delta_H(h_{2456})\delta_H(h_{1456})\delta_L(l_{23456})\delta_L(l_{13456})\delta_L(l_{12456})\,.
\end{equation}
First, one integrates out the $l_{1456}$, exploiting $ \delta_L(l_{13456})$ and obtains
\begin{equation}\label{eq:l1456}
l_{1456} =\,h_{146}\rhd\{h_{456}, (g_{56}g_{45})\rhd h_{134}\} l_{1346}{}^{-1}(h_{136}\rhd'l_{3456}) l_{1356} h_{156}\rhd'(g_{56}\rhd l_{1345})\,.
\end{equation}
Next, one one integrates out the $l_{2456}$, exploiting $ \delta_L(l_{23456})$ and obtains
\begin{equation}\label{eq:l2456}
    l_{2456}= \,h_{246}\rhd\{h_{456},(g_{56}g_{45})\rhd h_{234}\} l_{2346}{}^{-1}(h_{236}\rhd'l_{3456}) l_{2356} h_{256}\rhd'(g_{56}\rhd l_{2345})\,.
\end{equation}
Next, one integrates out $h_{456}$, exploiting $ \delta_H(h_{3456})$ and obtains
\begin{equation}\label{eq:h456}
    h_{456} = h_{346}^{-1}\, \delta(l_{3456})h_{356}\,(g_{56}\rhd h_{345})\,.
\end{equation}
Using the equation (\ref{eq:h456}), one obtains that
\begin{equation}
    \delta_G(g_{456})=\delta_G\big(\partial(h_{346})^{-1}\, \partial(h_{356})\,g_{56}\rhd \partial(h_{345})\,g_{56}\,g_{45}\,g_{46}^{-1}\big)\,,
\end{equation}
which, using the identity \eqref{id:g} for triangles $({346})$, $({356})$, and $({345})$, reduces to:
\begin{equation}
    \delta_G(g_{456})=\delta_G\big(e\big)\,.
\end{equation}
Similarly as done for the right-hand side of the move, one shows that $\delta_H(h_{1456})$, when using the equation (\ref{eq:l1456}), and $\delta_H(h_{2456})$, when using the equation (\ref{eq:l2456}), reduce to $\delta_H(e)^2$.
The remaining $\delta_L(l_{12456})$ now reads
\begin{equation}\label{eq:33lhs}
\delta_L(l_{12456})=\delta_L\big( l_{1246}{}^{-1}(h_{126}\rhd'l_{2456}) l_{1256} h_{156}\rhd'(g_{56}\rhd l_{1245})l_{1456}{}^{-1}\,h_{146}\rhd\{h_{456}, (g_{56}g_{45})\rhd h_{124}\}_{\mathrm{p}}\big)\,.
\end{equation}
Substituting the equations (\ref{eq:l1456}) and (\ref{eq:l2456}), one obtains
\begin{equation}
 \label{eq:33lhs01}
 \resizebox{0.94\hsize}{!}{$
\begin{aligned}
\delta_L(l_{12456})&=&\delta_L\big( l_{1246}{}^{-1}(h_{126}\rhd'(h_{246}\rhd\{h_{456},( g_{56}g_{45})\rhd h_{234}\}_{\mathrm{p}}{l_{2346}{}^{-1}(h_{236}\rhd'l_{3456}) l_{2356} h_{256}\rhd'(g_{56}\rhd l_{2345})\,}))\\&& l_{1256} h_{156}\rhd'(g_{56}\rhd l_{1245})  h_{156}\rhd'(g_{56}\rhd l_{1345}){}^{-1}l_{1356}^{-1}(h_{136}\rhd'l_{3456}) {}^{-1} l_{1346}h_{146}\rhd\{h_{456}, (g_{56}g_{45})\rhd h_{134}\}_{\mathrm{p}}^{-1}\,\\&&h_{146}\rhd\{h_{456}, (g_{56}g_{45})\rhd h_{124}\}_{\mathrm{p}}\big)\,.
\end{aligned}$}
\end{equation}
After commuting the elements, \ie\,, using the Peiffer identity for the crossed module $(L\overset{\delta}{\rightarrow}H, \rhd')$, one obtains
\begin{equation}\label{eq:33lhs03}
\resizebox{0.94\hsize}{!}{$
\begin{aligned}
\delta_L(l_{12456})&=&\delta_L\big( (\delta(l_{1246}){}^{-1}h_{126}h_{246})\rhd\{h_{456},( g_{56}g_{45})\rhd h_{234}\}_{\mathrm{p}}(\delta(l_{1246})^{-1}h_{126}\rhd \delta(l_{2346})^{-1}h_{126}h_{236})\rhd'l_{3456}\\&& l_{1246}^{-1} h_{126}\rhd'l_{2346}{}^{-1}h_{126}\rhd'l_{2356} (h_{126}h_{256})\rhd'(g_{56}\rhd l_{2345})\,)\\&& l_{1256} h_{156}\rhd'(g_{56}\rhd l_{1245})  h_{156}\rhd'(g_{56}\rhd l_{1345}){}^{-1}l_{1356}^{-1}l_{1346}(\delta(l_{1346})^{-1}h_{136})\rhd'l_{3456} {}^{-1}\\&& h_{146}\rhd\{h_{456}, (g_{56}g_{45})\rhd h_{134}\}_{\mathrm{p}}^{-1}\,h_{146}\rhd\{h_{456}, (g_{56}g_{45})\rhd h_{124}\}_{\mathrm{p}}\big)\,.
\end{aligned}$}
\end{equation}
Using the identity (\ref{eq:id06}) for the inverse of the element $\{h_{456}, (g_{56}g_{45})\rhd h_{134}\}_{\mathrm{p}}^{-1}$, and then the variant of the identity (\ref{eq:id02}), \ie\,, that is,
\begin{equation}
    \{h_1, h_2 h_3 h_4\}_{\mathrm{p}}= \{h_1, h_2\}_{\mathrm{p}}(\partial(h_1)\rhd h_2)\rhd' \{h_1, h_3\}_{\mathrm{p}} (\partial(h_1)\rhd (h_2 h_3))\rhd' \{h_1, h_4\}_{\mathrm{p}}\,,
\end{equation}
one obtains
\begin{equation}
\begin{aligned}
\{h_{456}, (g_{56}g_{45})\rhd(h_{134}^{-1}h_{124}h_{234})\}_{\mathrm{p}}&=&\{h_{456}, (g_{56}g_{45})\rhd h_{134}^{-1}\}_{\mathrm{p}}(g_{46}\rhd h_{134}^{-1})\rhd' \{h_{456}, (g_{56}g_{45})\rhd h_{124}\}_{\mathrm{p}}\\&& (g_{46} \rhd (h_{134}^{-1}h_{124}))\rhd' \{h_{456}, (g_{56}g_{45})\rhd h_{124}\}_{\mathrm{p}}\,,
\end{aligned}
\end{equation}
rendering the equation (\ref{eq:33lhs03}) to
\begin{equation}\label{eq:33lhs04}
\begin{aligned}
\delta_L(l_{12456})&=&\delta_L\big( (\delta(l_{1246})^{-1}h_{126}\rhd \delta(l_{2346})^{-1}h_{126}h_{236})\rhd'l_{3456}\\&& l_{1246}^{-1} h_{126}\rhd'l_{2346}{}^{-1}h_{126}\rhd'l_{2356} (h_{126}h_{256})\rhd'(g_{56}\rhd l_{2345})\,)\\&& l_{1256} h_{156}\rhd'(g_{56}\rhd l_{1245})  h_{156}\rhd'(g_{56}\rhd l_{1345}){}^{-1}l_{1356}^{-1}l_{1346}(\delta(l_{1346})^{-1}h_{136})\rhd'l_{3456} {}^{-1}\\&&(h_{146} g_{46}\rhd h_{134})\rhd' \{h_{456}, (g_{56}g_{45})\rhd(h_{134}^{-1}h_{124}h_{234})\}_{\mathrm{p}}\big)\,.
\end{aligned}
\end{equation}
Using the equation (\ref{eq:h456}), and the identities (\ref{eq:id01}) and (\ref{identitet}), similarly as for the r.h.s. of the move, one obtains that the terms featuring $l_{3456}$ cancel, \ie\,, the delta function $\delta_L(l_{12456})$ reads
\begin{equation}\label{eq:33lhs05}
\begin{aligned}
\delta_L(l_{12456})&=&\delta_L\big( l_{1246}^{-1} h_{126}\rhd'l_{2346}{}^{-1}h_{126}\rhd'l_{2356} (h_{126}h_{256})\rhd'(g_{56}\rhd l_{2345}))l_{1256} h_{156}\rhd'(g_{56}\rhd l_{1245}) \\&&  h_{156}\rhd'(g_{56}\rhd l_{1345}){}^{-1}l_{1356}^{-1}l_{1346} (h_{146} g_{46}\rhd h_{134})\rhd' \{h_{456}, (g_{56}g_{45})\rhd(h_{134}^{-1}h_{124}h_{234})\}_{\mathrm{p}}\big)\,.
\end{aligned}
\end{equation}
It follows that the integral over $l_{3456}$ is now trivial and l.h.s. of the move finally reduces to:
\begin{equation}
\begin{aligned}\label{eq:lhs33}
l.h.s. &=&\delta_G(e)\delta_H(e)^{2}\delta_L\big(h_{126}\rhd'l_{2346} l_{1246}(h_{146} g_{46}\rhd h_{134})\rhd' \{h_{456}, (g_{56}g_{45})\rhd(h_{134}^{-1}h_{124}h_{234})\}_{\mathrm{p}}^{-1}l_{1346}^{-1}\,\\&&l_{1356}\,h_{156}\rhd'(g_{56}\rhd l_{1345})\,h_{156}\rhd'(g_{56}\rhd l_{1245})^{-1}(h_{156} g_{56}\rhd h_{125})\rhd'(g_{56}\rhd l_{2345})^{-1}\,l_{1256}^{-1}\, h_{126}\rhd'l_{2356}^{-1}\,\big)\,.
\end{aligned}
\end{equation}
The expressions (\ref{eq:rhs33}) and (\ref{eq:33lhs}) are the same, which proves the invariance of the state sum (\ref{def_statesum}) under the Pachner move $3\leftrightarrow 3$. The numbers of
$k$-simplices agree on both sides of the $3\leftrightarrow 3$ move for all
$k$, and the prefactors play no role in this case.

%

\newenvironment{hpabstract}{%
  \renewcommand{\baselinestretch}{0.2}
  \begin{footnotesize}%
}{\end{footnotesize}}%
\newcommand{\hpeprint}[2]{%
  \href{http://www.arxiv.org/abs/#1}{\texttt{arxiv:#1#2}}}%
\newcommand{\hpspires}[1]{%
  \href{http://www.slac.stanford.edu/spires/find/hep/www?#1}{SPIRES Link}}%
\newcommand{\hpmathsci}[1]{%
  \href{http://www.ams.org/mathscinet-getitem?mr=#1}{\texttt{MR #1}}}%
\newcommand{\hpdoi}[1]{%
  \href{http://dx.doi.org/#1}{\ Journal Link}}%

\end{document}